\documentclass{aastex631}
\pdfoutput=1 

\usepackage{ulem}

\usepackage{rotating}
\usepackage{tikz}
\usepackage{booktabs}
\usepackage{amssymb}
\usepackage{graphicx}
\begin{document}

\title{A Stochastic Approach To Reconstruct Gamma Ray Burst Lightcurves}

\author{Maria G. Dainotti}
\affiliation{National Astronomical Observatory of Japan,2 Chome-21-1 Osawa, Mitaka, Tokyo 181-8588, Japan.}

\author{Ritwik Sharma}
\affiliation{Department of Physics, Deshbandhu College, University of Delhi,
New Delhi, India.}


\author{Aditya Narendra}
\affiliation{Astronomical Observatory of Jagiellonian University, Krakow.}

\author{Delina Levine}
\affiliation{Department of Astronomy, University of Maryland, College Park, MD 20742, USA}
\author{Enrico Rinaldi}
\affiliation{Interdisciplinary Theoretical \& Mathematical Science Program, RIKEN (iTHEMS),\\ 2-1 Hirosawa, Wako, Saitama, Japan 351-0198.}

\author{Agnieszka Pollo}
\affiliation{Astronomical Observatory of Jagiellonian University, Krakow.}
\affiliation{National Centre for Nuclear Research, Warsaw.}

\author{Gopal Bhatta}
\affiliation{Institute of Nuclear Physics Polish Academy of Sciences, PL-31342 Krakow, Poland.}




\begin{abstract}
Gamma-Ray Bursts (GRBs), being observed at high redshift ($z=9.4$), vital to cosmological studies and investigating Population III stars. To tackle these studies, we need correlations among relevant GRB variables with the requirement of small uncertainties on their variables. Thus, we must have good coverage of GRB light curves (LCs). However, gaps in the LC hinder the precise determination of GRB properties and are often unavoidable. Therefore, extensive categorization of GRB LCs remains a hurdle. We address LC gaps using a “stochastic reconstruction,” wherein we fit two pre-existing models (Willingale 2007; W07 and Broken Power Law; BPL) to the observed LC, then use the distribution of flux residuals from the original data to generate data to fill in the temporal gaps. We also demonstrate a model-independent LC reconstruction via Gaussian Processes.  
At 10\% noise, the uncertainty of the end time of the plateau, its correspondent flux, and the temporal decay index after the plateau decreases, on average, by 33.3\% 35.03\%, and 43.32\%, respectively for the W07, and by 33.3\%, 30.78\%, 43.9\% for the BPL. The slope of the plateau decreases by 14.76\% in the BPL. After using the Gaussian Process technique, we see similar trends of a decrease in uncertainty for all model parameters for both the W07 and BPL models. These improvements are essential for the application of GRBs as standard candles in cosmology, for the investigation of theoretical models and for inferring the redshift of GRBs with future machine learning analysis.
\end{abstract}



\section{Introduction} \label{sec:intro}

Gamma-Ray Bursts (GRBs) are transient astrophysical events that can be observed up to redshift $z=8.2$ \citep{Tanvir2009Nature,salvaterra2009grb} and $z=9.4$ \citep{Cucchiara2011}. Thus, they are excellent candidates for cosmological tools that can be utilized to probe the early universe. Moreover, a comprehensive characterization of GRBs can provide insight into Population III stars, the most ancient stars observed at the epoch of re-ionization. Having a reliable taxonomy of GRB classes and good data coverage will favor future population studies and thus will enhance the determination of the cosmological evolution of GRB properties and the investigation of their emission mechanism and/or progenitors. All these are foundational topics for astrophysics. 

The lack of a robust GRB classification scheme, the incompleteness of redshift information in the existing sample of observed GRBs, gaps in the light curves (LCs), and the need for a unique database repository are many of the difficult challenges that astrophysicists need to overcome. In this paper, we attempt to solve the issue related to the temporal gaps in the LCs.

Observationally, GRB emission can be divided into two phases: the prompt and the afterglow. 
The prompt is the primary phase observed from high-energy $\gamma$-rays to X-rays and sometimes in optical bands \citep{Vestrand2005Natur,Blake2005Natur,Beskin2010ApJ,2012MNRAS.421.1874G,2014Sci...343...38V}. The prompt is followed by an afterglow phase \citep{costa1997,vanParadijs1997,Piro1998,Gehrels2009ARA&A}, observed in multiple wavelengths such as X-rays, optical, and sometimes radio bands. 

GRBs have, historically, been divided into two classes depending on their prompt duration, $T_{90}$, the time interval during which a burst releases $90\%$ of its total background-subtracted counts, beginning after $5\%$ of the total counts have been measured \citep{mazets1981catalog, kouveliotou1993classification}. Short GRBs (SGRBs), with $T_{90}\leq 2$s, are produced by the merging of compact objects \citep{duncan1992,Narayan1992,usov1992,thompson1994MNRAS.270..480T,levan2008intrinsic,metzger2011MNRAS.413.2031M,Bucciantini2012MNRAS.419.1537B,perna2016ApJ...821L..18P}, and long GRBs (LGRBs), with $T_{90} \ge 2$s, result from collapsing massive stars \citep{woosley1993ApJ...405..273W,paczynski1998ApJ...494L..45P,macfayden1999ApJ...524..262M,bloom2002AJ....123.1111B,hjorth2003Natur.423..847H,Woosley2006ARA&A,woosley2006ApJ...637..914W,bucciantini2008,kumar2008Sci...321..376K,hjorth2012grb..book..169H,cano2017AdAst2017E...5C,lyman2017MNRAS.467.1795L,perna2018ApJ...859...48P,aloy2021MNRAS.500.4365A,Ahumada2021NatAs...5..917A}. 
However, this historical classification in LGRBs and SGRBs simplifies a more complex and realistic picture where many more classes have been discovered in the literature since then (see Section \ref{section:classification} for more details). In the analysis proposed here, the reconstruction of GRB LCs will improve GRB characterization to cast further light between different classes.

The Neil Gehrels Swift Observatory (Swift, \citet{Gehrels2004ApJ...611.1005G}) is crucial for observing GRB temporal properties. The Swift Burst Alert Telescope (BAT, 15-150 keV, \citep{barthelmy2005burst}) enables rapid detection of the prompt emission and a fast follow-up of the afterglow by the X-ray (XRT, 0.3-10 keV, \citep{burrows2005swift}) and Ultra-Violet telescopes (UVOT 170 - 600 nm, \citep{Roming2005}). Furthermore, due to rapid afterglow follow-up in several wavelengths, Swift data has shown new features in the GRB LCs \citep{Tagliaferri2005,Nousek2006,Troja2007}. 

Many X-ray LCs have a steep flux decay after the prompt emission ends, sometimes followed by flares and a plateau \citep{Zhang2006,OBrien2006,Nousek2006,Sakamoto2007,Liang2007,Willingale2007,Dainotti2008,dainotti2010discovery,Dainotti2016,dainotti2017a}. 
Specifically, XRT detections are available for 81\% of Swift GRBs: 42\% with X-ray plateaus \citep{Evans2009,Li2018b}.
The X-ray plateau generally lasts $10^2 - 10^5$ s and has a subsequent phase characterized by a power-law (PL) decay. 
In addition, about 30\% of optical LCs observed by UVOT and ground-based facilities also exhibit a shallow decay phase
\citep{Vestrand2005Natur,Kann2006,Zeh2006,panaitescu2008taxonomy,Kann2010,panaitescu2011optical,Kann2011,Li2012,Oates2012,Zaninoni2013,margutti2013MNRAS.428..729M,melandri2014A&A...565A..72M,Li2015,Li2018a,Si2018,dainotti2020b}. 
The plateau can be fitted with a broken power-law (BPL) \citep{Zhang2006,Racusin2009}, a smoothly broken power-law (SBPL), or the Willingale et al. (2007) phenomenological model (W07, \citet{Willingale2007}). The W07 model determines the time at the end of the plateau ($T_a$), its corresponding flux ($F_a$), and the temporal index after the plateau ($\alpha_a$). The BPL determines $T_a$, $F_a$, and the slope of the LC during the plateau ($\alpha_1$) and after the plateau ($\alpha_2$). These models are briefly described in Section \ref{section:sampling}.

The plateau is usually explained via the magnetar model, based on the dipole radiation emitted by the rotational energy of a newly-born neutron star (NS) 
In this model, the plateau ends when the NS reaches its critical spin-down timescale; the uncertainties on $T_a$ can be ascribed to the uncertainties on the magnetar spin period and magnetic field.

However, the characterization of the plateau emission can be hindered by temporal gaps which can occur in the beginning, during, and at the end of the plateau. These may arise from the orbital period of satellites, lack of fast follow-up studies, atmospheric turbulence, and instrumental errors or failures. Thus, extensive characterization of GRB LCs remains a bottleneck.

Therefore, we propose a method to reconstruct LCs starting from the plateau emission, which due to its theoretical interpretation within the magnetar model, is grounded in fundamental physics. Morphologically, the plateau has more standard features among diverse GRBs (e.g., its length and flatness) than the prompt properties. The plateau features have attracted attention due to their use in building relevant correlations with the plateau parameters and their application as cosmological tools. 
Specifically, \citet{Dainotti2008,Dainotti2010, dainotti2011a,dainotti2013determination,dainotti2015,dainotti2017a} and \citet{Li2018b} explored the luminosity at the end of the plateau, $L_{X,a}$ vs. its rest-frame time $T^{*}_{X,a}$ (known as the Dainotti relation or 2D L-T relation) \footnote{the rest-frame time is denoted with an asterisk}. The 2D relation has also been discovered in optical plateau emissions \citep{dainotti2020b,2022ApJS..261...25D}. Within the theoretical magnetar scenario, \citet{Rowlinson2014} showed that the X-ray Dainotti relation is reproduced with a slope for $L_{a,X}$-$T^{*}_{a, X}$ of $-1$. This correlation has already been applied in the cosmological framework to construct a GRB Hubble diagram out to $z>8$ \citep{cardone2009updated,cardone2010constraining,postnikov2014nonparametric,dainotti2013}. 

An extension of the 2D L-T relation, obtained by adding the peak prompt luminosity, $L_{X, peak}$, has led to the Dainotti 3D relation \citep{Dainotti2016,dainotti2017a,dainotti2020a, 2022ApJS..261...25D}. This 3D relation has also been successfully applied to constrain cosmological parameters \citep{dainotti2022MNRAS,2022MNRAS.514.1828D, 2022PASJ...74.1095D, 2022MNRAS.512..439C, 2022MNRAS.510.2928C,2023MNRAS.518.2201D}. 
Importantly, \citet{2022MNRAS.514.1828D} have shown that if we reduce the uncertainties on the parameters of the plateau emission by 47.5\%, we will reach the same precision on the cosmological value of $\Omega_M$ quoted in \citet{Conley2011} even now compared to the same accuracy which we would reach in 2037 based on current observation rates and parameter uncertainties. 
This shows how appealing a more robust LC reconstruction (LCR) can be, given that it would save us 15 years of observations to reduce the uncertainties on cosmological parameters to reach the same precision achieved by SNe Ia \citep[for details see][] {dainotti2020x}. 

In addition, the plateau emission has been the object of investigation concerning closure relations (relations between the temporal and spectral index during or after the plateau region), which allows us to test the viability of the standard fireball model for the observations carrying plateau emission in high-energy $\gamma$-rays, X-rays, or optical wavelengths \citep[][]{2019ApJ...883..134T, 2020ApJ...903...18S, 2021ApJS..255...13D, 2021PASJ...73..970D, 2022ApJ...940..169D, Willingale2007,Evans2009,Racusin2009,Kumar2010MNRAS.409..226K,Oates2012,Gaob2013NewAR..57..141G,Wang2015,Misra2021MNRAS.504.5685M,fraija2020closure,ryan2020ApJ...896..166R, 2022MNRAS.tmp.3552L, 2023Galax..11...25D}. Although closure relations are a quick way to test the standard fireball model, a more precise characterization of the decay index after the plateau emission would allow for better precision in determining these relations.

Lastly, another relevant application is to use the reconstructed LCs to train machine-learning (ML) models for redshift estimation and classification of all redshifts, especially of high-${z}$ GRBs.

Because of the many applications, both from a theoretical and a cosmological perspective, the plateau parameters must be well-constrained. In this regard reconstructing LCs with plateaus in the plateau region is extremely important. Indeed, LCs with gaps are, in many circumstances, not usable for cosmological applications. 
In addition, LCs with temporal gaps cannot be reliably used to test theoretical models that attempt to explain the GRB emission (e.g., the standard fireball model, \citet{cavallo1978MNRAS.183..359C,meszaros1993ApJ...405..278M,meszaros1997RvMA...10..127M,piran1999PhR...314..575P,Panaitescu2000,meszaros2001PThPS.143...33M,Panaitescu2001,2002ZhangMeszaros,Piran2004,Zhang2004IJMPA..19.2385Z,Zhang2006,scargle2020}). 
Thus, LCs with better coverage will facilitate modeling and determining correlations among critical physical properties.

Our reconstruction method gives us a glimpse into what the data likely would have been in these gaps and increases the overall density distribution of the LCs over time. As a result, we can improve how well LCs can be used as standard candles and for theoretical modeling. 

In this paper, we take the first step toward full LCR: We develop a reconstruction technique and test its performance on our dataset. This paper is divided into the following sections: in \S\ref{LCR} we provide the definition of the LCR, in \S\ref{sec:advantage} scientific motivation for the LCR, and in \S\ref{section:classification} we outline the GRB classification. Then, in \S\ref{section:methods}, we outline the data sample, Willingale and BPL models (\S\ref{section:sampling}), and in detail, the algorithms used to obtain the reconstructed LCs, including the functional form toy model (\S\ref{section:reconstruction}) and the Gaussian Processes (\S\ref{sec:GP}). The results are summarized in \S\ref{section:results}, of the functional form toy model approach (\S\ref{sec:funcresults}) and the Gaussian Processes (\S\ref{sec:GPresults}). We draw our conclusions in \S\ref{section:conclusion}.

\subsection{LC Reconstruction}\label{LCR}

LCR offers an innovative solution to the problem of temporal gaps in LCs. In several domains of astronomy, statistical LCR methods have been used to account for missing data. Such methods have also been applied to Cepheids using simulated annealing and Fourier decomposition techniques \citep{ngeow2003reconstructing}. A similar application of reconstruction methods on LCs can be seen in \citet{huber2010planetary} for planetary eclipse mapping and \citet{geiger1996light} for measurement of time delay in gravitational lens systems. However, an application of reconstruction techniques on GRB LCs has yet to be undertaken because GRB LCs have very variable features. They present a wide range of shapes, including smooth single pulses and multiple peaks with different widths occurring at different times, while others exhibit more complex behavior.

As a first step to address this problem and evaluate the performance of LCR, we have used a relatively simple methodology that chooses a model to fit the LCs. Many groups have classified LC morphology with a simple PL, BPL, smoothly BPL, or W07 model; some with plateaus, and some with double breaks
\citep{giblin2003,Nousek2006,pasquale2008,Evans2009,Racusin2009,liang2009LCR,cenko2010ApJ...711..641C}. 
Although our results depend on the underlying model chosen for the reconstruction, we can give a first estimate of how the LCR can enhance the determination of the plateau parameters.
Thus, we show how to enable more accurate morphological LC classification and determination of GRB parameters. Our LCR analysis reduced the uncertainties on the parameters of the plateau by 37.22\% for 10\% noise level on average among all plateau parameters for the reconstruction with the W07 functional form toy model and 30.69\% for 10\% noise level for the BPL functional form toy model. Furthermore, using the Gaussian Processes, the uncertainties on all plateau parameters are decreased, on average, by 31.43\% for the W07 and 21.99\% for the BPL (see \S\ref{section:results}).

\subsection{The advantage of the LCR in theoretical models}\label{sec:advantage}

We can anticipate that with the new light curves reconstructed, we can use them directly to test theoretical models. We here limit ourselves to provide two examples because this topic is far beyond the current focus of the paper, and it will be investigated in a forthcoming paper. One of the most immediate advantages is the use of the parameters of the BPL obtained with the reconstructed lightcurves with Gaussian processes or the other functional toy model LCR modeling to test the standard fireball model via the closure relationships, which are the relationships among the temporal, $\alpha$, and spectral indices, $\beta$, of the lightcurves. For a precise new estimate of the modeling one could check the differences between the results obtained in X-rays from previous papers \citep{Srinivasaragavan2020, dainotti2021b} and the ones from the LCR. We envision that the prediction on the closure relationships will be around 40\% more accurate compared to the current estimate in the literature. Indeed, 43.32\% is the percentage reduction on the uncertainty on the temporal index, $\alpha$,  after the plateau phase if we consider, for example, the W07 functional toy model for the reconstruction, and for the BPL, the uncertainty in the slope of the LC after the plateau phase, $\alpha_2$, decreases by 43.9\%. For the Gaussian Processes, the uncertainty in W07 $\alpha$ parameter decreases by 41.5\%, and the uncertainty in the $\alpha_2$ parameter decreases by 35.92\%. There are cases for which it is not possible to uniquely determine the outcome of the closure relationships, namely whether the closure relationship favors a scenario with a fast cooling or slow cooling or if a constant medium or a wind medium is preferred.  We can anticipate that a fraction of cases with the LCR will be able to discern better the interstellar medium and the regime and will be able to remove part of the degeneracy in the theoretical model investigation. \citet{Srinivasaragavan2020} found that $> 50\%$ follow 11 out of the 16 closure relationships tested. It is interesting to check how the exact percentage of fulfillment could change when LCR is adopted in this scenario.
Another model that can benefit from the LCR is the magnetar model. Indeed, it was shown in \citet{stratta2018} that due to the temporal gaps in some of the LC, the magnetic field, the spin period parameters, or the electron energy fraction could not be determined very precisely; namely, their uncertainties are so large that the estimate cannot be considered reliable. In this scenario, it is important to have full coverage of the lightcurve so that a combination of the magnetic field, the spin period and the electron energy fraction, could solve cases in which, due to these uncertainties on the parameters, it is even hard to distinguish between models (magnetars vs. black hole). The LCR can alleviate the degeneracy of the parameters of the model by reducing the uncertainties of the parameters. However, some degree of degeneracy inherent to the model will still remain unavoidable.
Here, we remind that this LCR will be approached as a first step, but it can also be extended in high-energy $\gamma$-rays, optical and radio wavelengths for the determination of the closure relationship in these respective energy bands \citep{dainotti2021c, 2023Galax..11...25D, 2022ApJ...940..169D, 2023MNRAS.519.4670L}.

\subsection{The connection of the LCR and the GRB classification and Properties}\label{section:classification}

Often the classification of GRB properties is challenging because of the general lack of data points in the LC or, more often, for the lack of data points in a crucial part of the LC for which a feature is expected, e.g., the beginning or end of the plateau emission, etc. 
This section aims to clarify how many GRBs classified in the literature as, for example, LGRBs, SGRBs, or the other classes defined below correspond to the morphological classes defined here. Indeed, this analysis is relevant because, with this study, we can check what percentage of the Good GRBs fall in a given class. 

Besides the Long and Short classification, we point out that other sub-classes have been discovered.
Intrinsically SGRBs (IS) have $T_{90}/(1+z)< 2s$ \citep{Levesque2010,Zhang2021NatAs,Rossi2021arXiv}. SGRBs with extended emission (SEE) \citep{Norris2006,Norris2010, Norris2000,Levan2007,dichiara2021} are SGRBs with mixed properties between LGRBs since they have $T_{90}>2$s, but they are harder in their spectral properties than the LGRBs, and in this regard, they are similar to SGRBs. From the observational point of view, they have a spike in the prompt followed by a long tail. The spike of the initial emission is followed by a brief pause in emission, lasting for $\simeq$10 seconds, after which the emission increases again on a timescale of 30-50 seconds \citep{Norris2006}. Ultra-long GRBs (ULGRBs) have $T_{90} \geq 1000$s \citep{stratta2013ultra,gendre2013ultra,Nakauchi2013ApJ...778...67N,Piro2014,Levan2014,Greiner2015Natur,Kann2018A&A,Gendre2019,ZhangBB2014}. These GRBs may originate from engine-driven explosions of stars much larger than LGRB progenitors \citep{Levan2014}. X-ray flashes (XRFs) \citep{soderberg2004redshift,chincarini2010MNRAS.406.2113C} are LGRBs with greater fluence in X-rays (2–30 keV) than in $\gamma$-rays (30-400 keV), while X-ray rich (XRR) GRBs are an intermediate class between XRFs and LGRBs that display very strong X-ray emission. SNe-GRBs are LGRBs associated with Supernovae of type Ic (SNe Ic) for which the SNe has been clearly observed \citep{2007A&A...471L..29D, 2022ApJ...938...41D}. These sub-classes may imply diverse progenitors or the same progenitor in different circumburst media.

\begin{figure}  
 \centering
\includegraphics[width=0.75\textwidth]{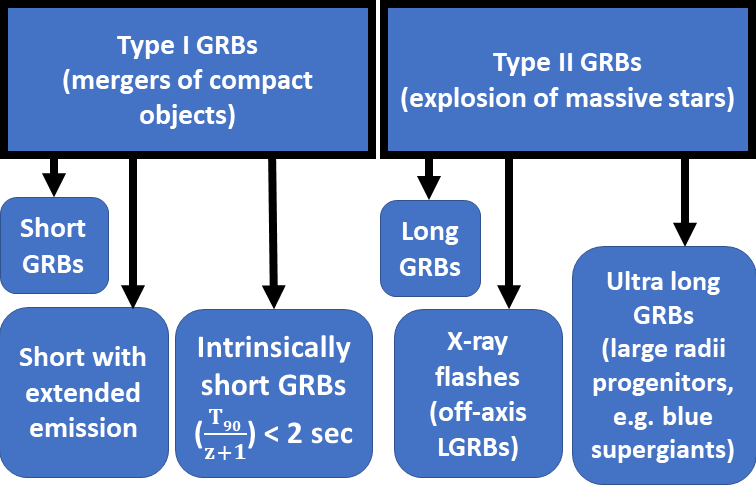}
    \caption{\small
   Schematic of GRB classes divided in Type I and Type II and their related sub-categories.
    }
      \label{fig:types}
\end{figure}

\citet{Zhang2007ApJ...655L..25Z,Zhang2007b,Zhang2009} unified the classification in terms of progenitors, with Type I and Type II GRBs classes. Type I include SGRBs, SEE, and IS; Type II includes LGRBs, XRFs, XRRs, GRB-SNe, and ULGRBs. For schematic pictures of the classes, see Fig. \ref{fig:types}.

\section{Methodology}\label{section:methods}

We aim to fill temporal gaps in GRB LCs by compiling all observations from Swift XRT taken from the BAT+XRT repository. To account for the missing data points, we adopt a stochastic approach where the reconstructed points are built from a chosen LC model.
To account for the realistic variation of the data, a noise function that considers the residuals from the initially observed flux data points is used to inject additive noise into the reconstructed points.

\subsection{Data Sampling, the Willingale Model and the Broken-Power law model}\label{section:sampling}

We take a sample of 455 GRBs (Table. \ref{tab:sample} from \citet{2020ApJ...903...18S}) with X-ray plateaus (222 with known redshift and 233 without known redshift), originally obtained from the Swift BAT-XRT repository \citep{Evans2009,Evans2007}. These GRBs are fitted with the W07 function, described in Eq. \ref{eqn1} as presented in \citet{2007ApJ...662.1093W}:

\begin{equation}
f(t) = \left \{
\begin{array}{ll}
\displaystyle{F_i \exp{\left ( \alpha_i \left( 1 - \frac{t}{T_i} \right) \right )} \exp{\left (
- \frac{t_i}{t} \right )}} {\hspace{1cm}\rm for} \ \ t < T_i \\
\\
\displaystyle{F_i \left ( \frac{t}{T_i} \right )^{-\alpha_i}
\exp{\left ( - \frac{t_i}{t} \right )}} {\hspace{1cm}\rm for} \ \ t \ge T_i, \\
\end{array}
\right .
\label{eqn1}
\end{equation}

Where $T_i$ and $F_i$ are the times and fluxes, respectively, either at the end of the prompt (denoted with $p$) or at the end of the plateau emission (denoted with $a$). The parameter $\alpha_i$ is the temporal index after $T_i$. The initial rise time is marked by the time $t_i$, which reads as $t_a$ for the afterglow emission. 
The maximum flux occurs at $t_i = \sqrt{tT_c/a_i}$.

We also fit the GRBs with the simple broken power law model (BPL):
\begin{equation}
f(t) = \left \{
\begin{array}{ll}
\displaystyle{F_i \left (\frac{t}{T_i} \right)^{-\alpha_1} 
} & {\rm for} \ \ t < T_i \\
\displaystyle{F_i \left ( \frac{t}{T_i} \right )^{-\alpha_2}
} & {\rm for} \ \ t \ge T_i, \\
\end{array}
\right .
\label{eq:simpleBPL}  
\end{equation}
Where $T_i$ and $F_i$ are the times and fluxes, respectively, at the break time (coincident with the end time of the plateau), $\alpha_1$ is the slope of the LC before the break, and $\alpha_2$ is the slope of the LC after the break.
In both our Willingale and BPL fit, as in W07, we take the logarithm in base 10 of these functions.

\begin{table*}
    \centering
    \begin{tabular}{lcccccr}
    \hline
       GRB NAME	&	$z$	&	$T_{90}$	&	$\frac{T_{90}}{(1+z)}$	&	LONG/SHORT	&	Type	&	Reference	\\
       \hline

GRB050128	&	-	&	28.00	&	-	&	L	&	SNe-GRB	&	[1] 	\\																				
GRB050315	&	1.95	&	95.40	&	32.35	&	L	&	XRR	&	[2]	\\																				
GRB050318	&	1.44	&	40.00	&	16.37	&	L	& XRR	&	[1] 	\\																				
GRB050319	&	3.24	&	151.74	&	35.79	&	L	&	XRF	&	[2]	\\																				
GRB050401	&	2.90	&	33.30	&	8.54	&	L	&	L	&	[1]	\\																				
GRB050416A	&	0.65	&	6.62	&	4.01	&	L	&
XRF/SNe-GRB, D	&	[1], [3] 	\\																				
GRB050502B	&	5.20	&	17.72	&	2.86	&	L	& XRR	&	[2]	\\																				
GRB050505	&	4.27	&	58.90	&	11.18	&	L	&	L	&  [1]		\\																				
GRB050607	&	-	&	48.00	&	-	&	L	&	XRR	&	[2]	\\																				
    \hline
    \end{tabular}
    \caption{Full sample of 455 GRBs, including redshift $z$, observer-frame burst duration $T_{90}$ and rest-frame duration $\frac{T_{90}}{(1+z)}$ long vs. short classification, GRB type, and reference. Full table is available online.
    References: 
    [1] \citet{2008ApJS..175..179S}; 
    [2] \citet{2018ApJ...866...97B};
    [3] \citet{2022ApJ...938...41D}
    }
    \label{tab:sample}
\end{table*}

We focus our reconstruction efforts on the plateau region of the GRB LCs, where our methodology works effectively. Before the plateau, the prompt emission has significant variability compared to the afterglow, and thus the prompt is difficult to model accurately. The stochastic approach used in this paper will also not account for flares and bumps in the afterglow since such variations are randomized to a very high degree. 
This is a first step towards understanding the feasibility of LCR. More complex methods that can generalize this first attempt to the prompt emission or flares in the afterglow phase will be explored in future works.

\begin{figure*}
\begin{center}
\includegraphics[width=.45\textwidth]{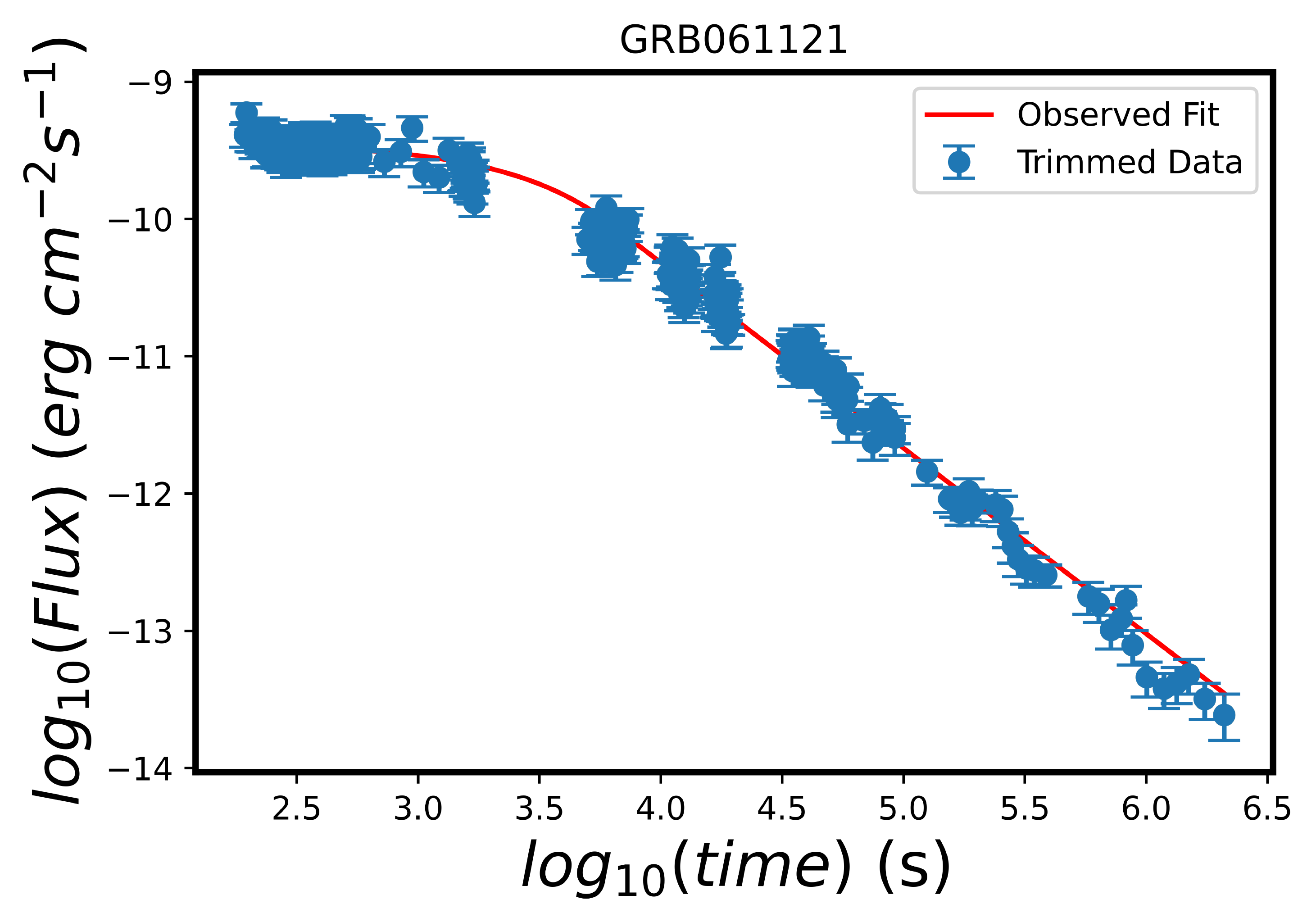}
\includegraphics[width=.45\textwidth]{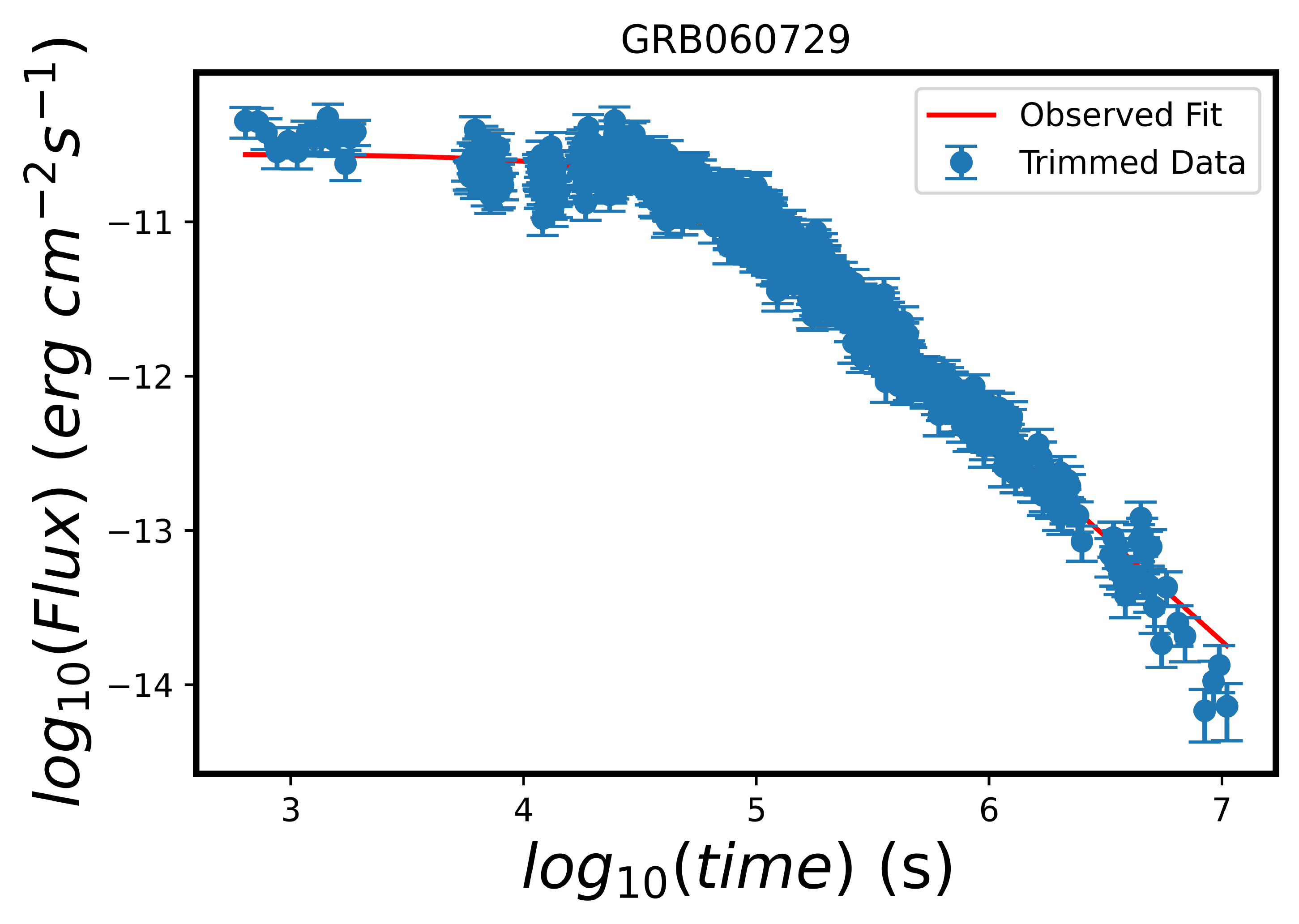}
\includegraphics[width=.45\textwidth]{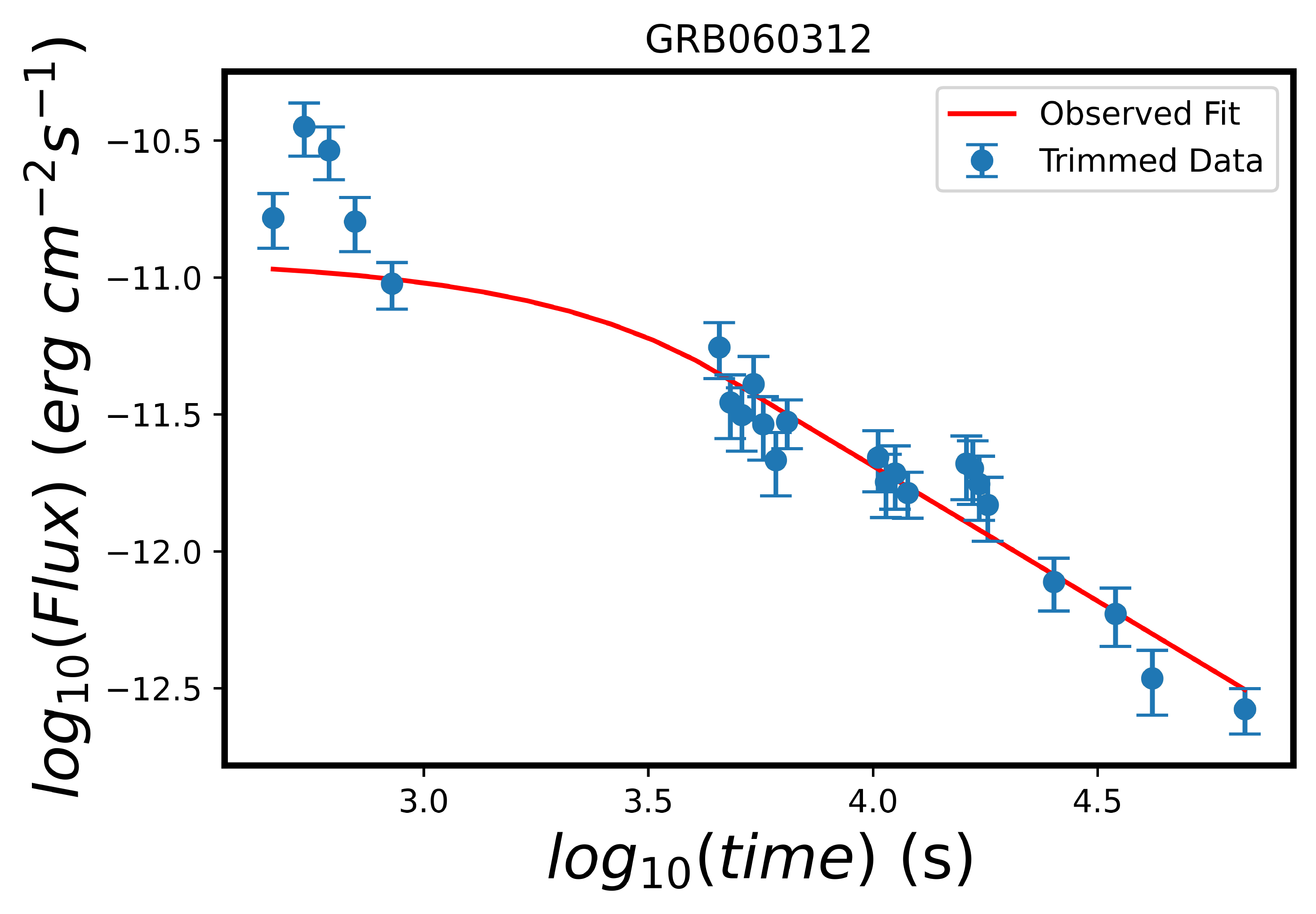}
\includegraphics[width=.45\textwidth]{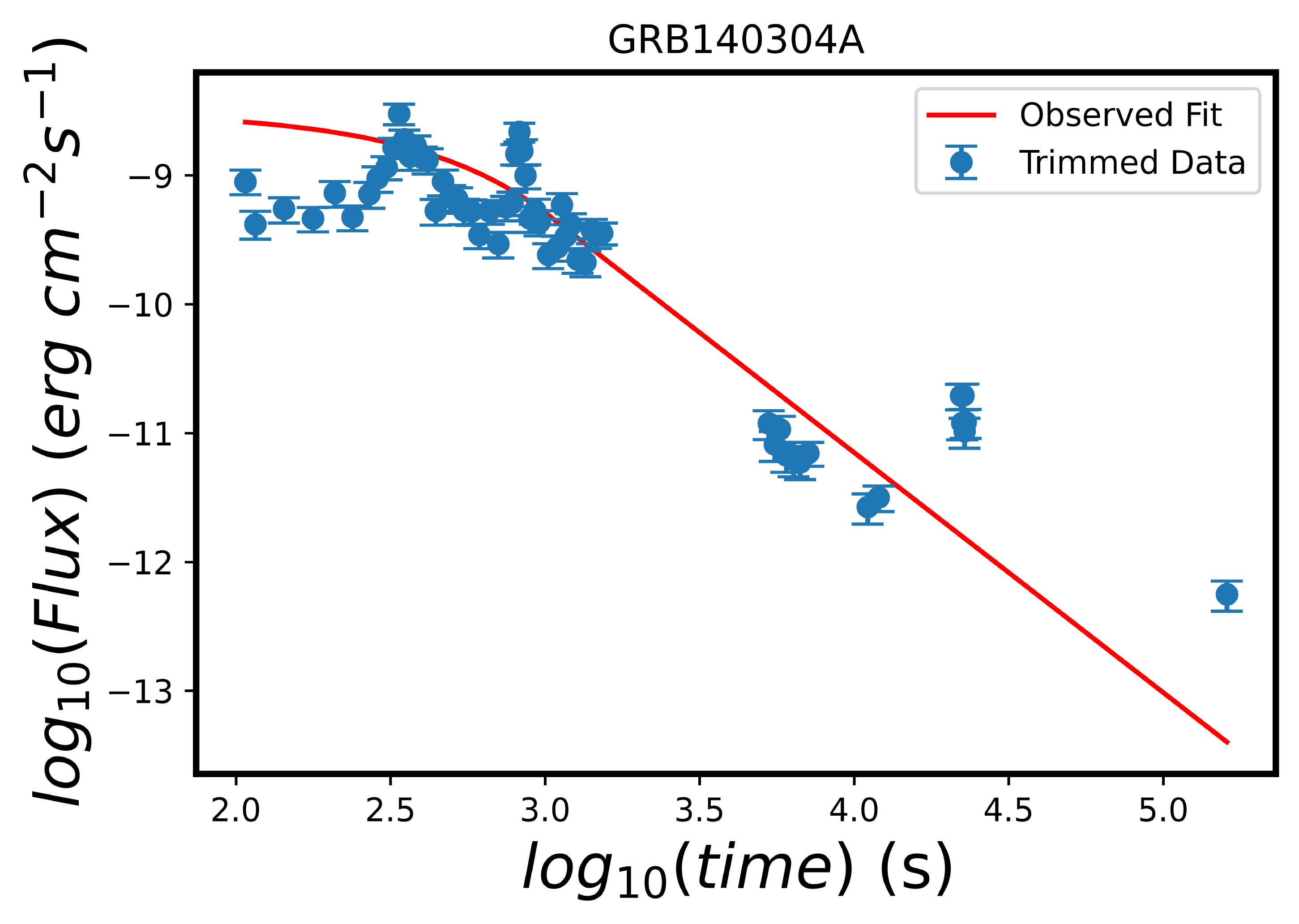}  
\end{center}
    \caption{LCs divided into four categories depending on the afterglow feature. i) Good GRBs (top left); ii) A break in the end of the LC (top right); iii) Flares/Bumps in the afterglow (bottom left); iv) Flares/Bumps with a Double Break at the end of the LC (bottom right).}
    \label{fig:categories}
\end{figure*}

The 455 LCs were then separated into categories based on the afterglow features. For subsequent analysis, we categorize the GRB afterglow LCs into those having i) good approximation with the W07 model (hereafter called Good GRBs), ii) flares or bumps throughout the afterglow region, iii) a double break at the end of the LC, iv) flares/bumps along with a double break. Fig. \ref{fig:categories} shows LCs belonging to each category with the W07 model fit superimposed. 
We take the Willingale function parameter values (for each GRB LC) from \citet{2020ApJ...903...18S}.  

\begin{figure*}
\begin{center}
\includegraphics[width=.45\textwidth]{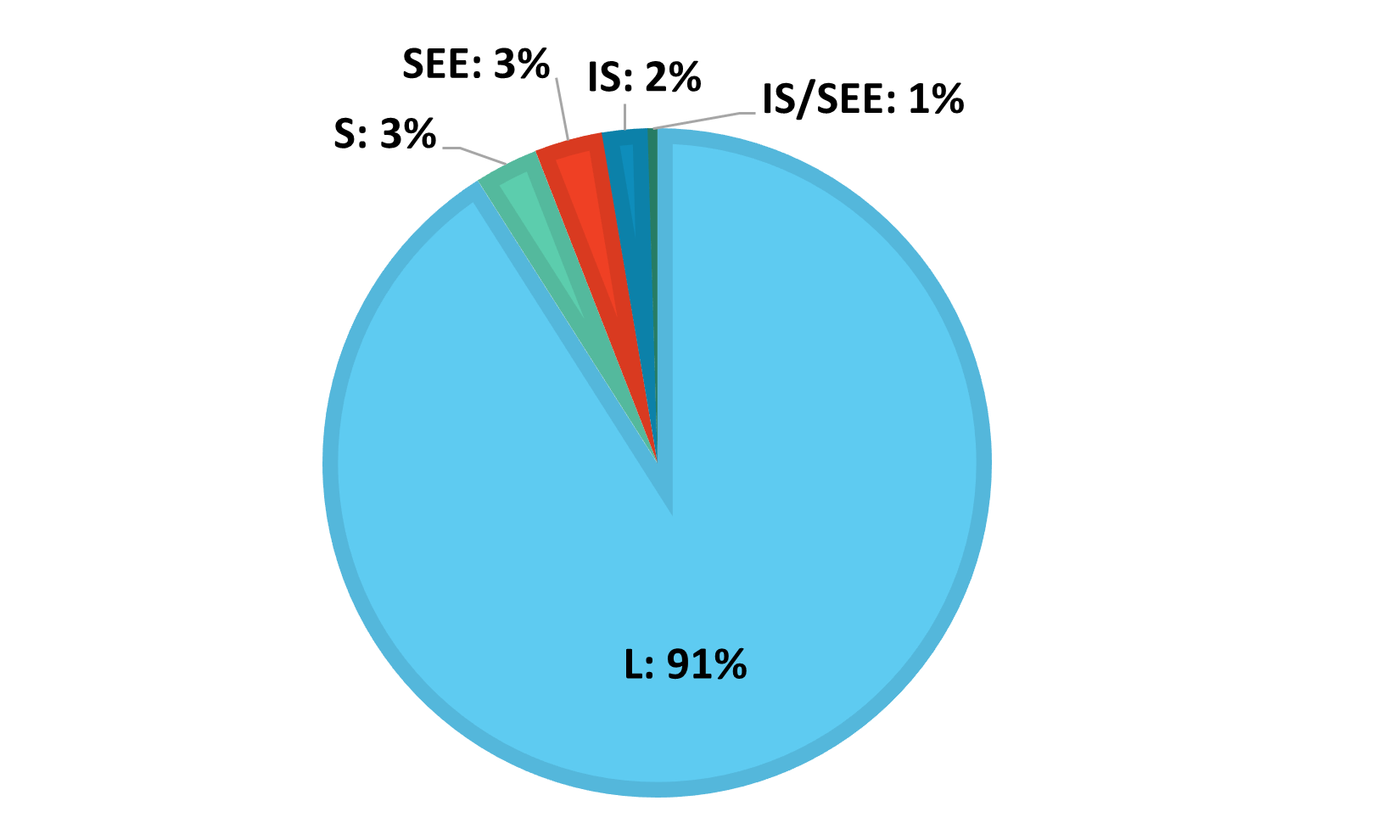}
\includegraphics[width=.45\textwidth]{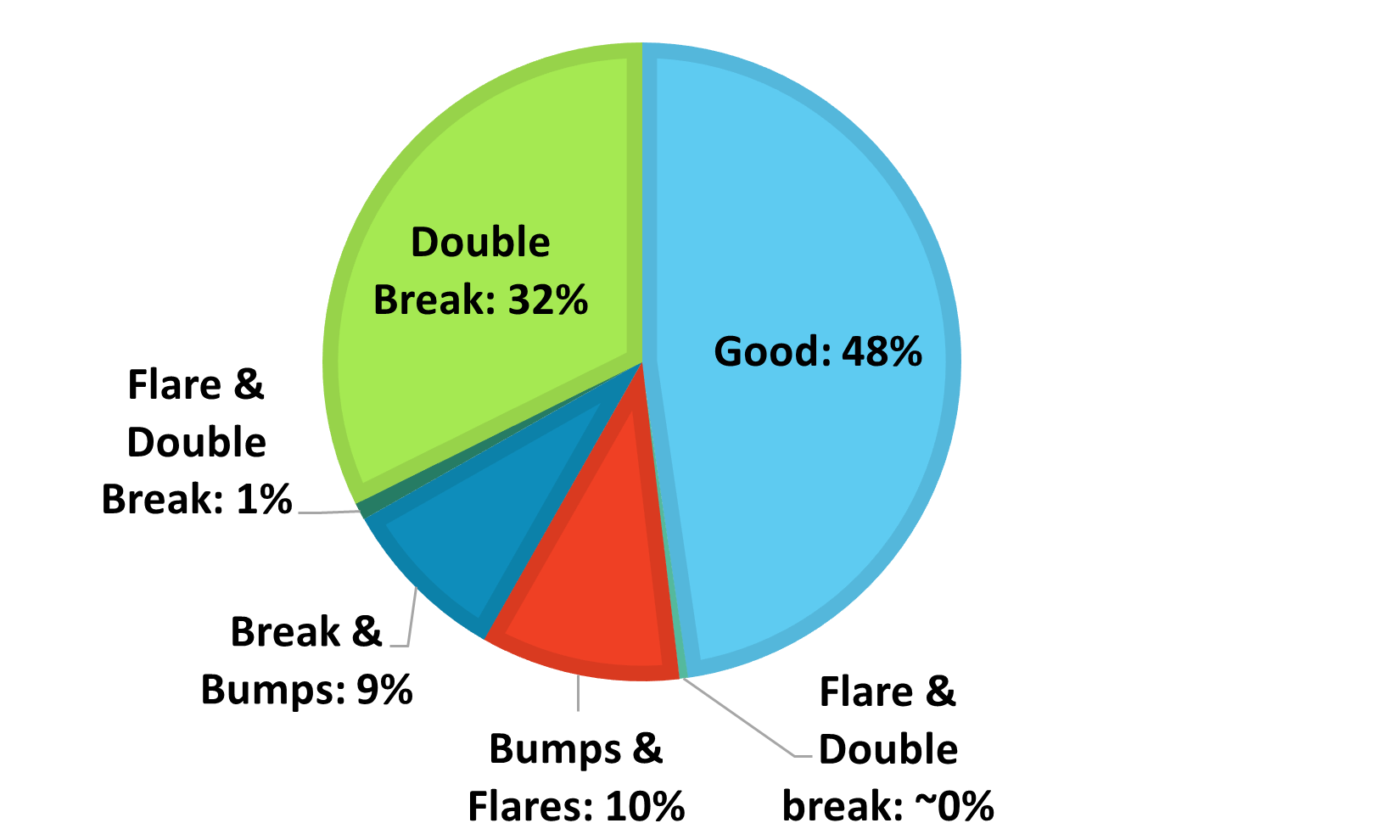}
\includegraphics[width=.45\textwidth]{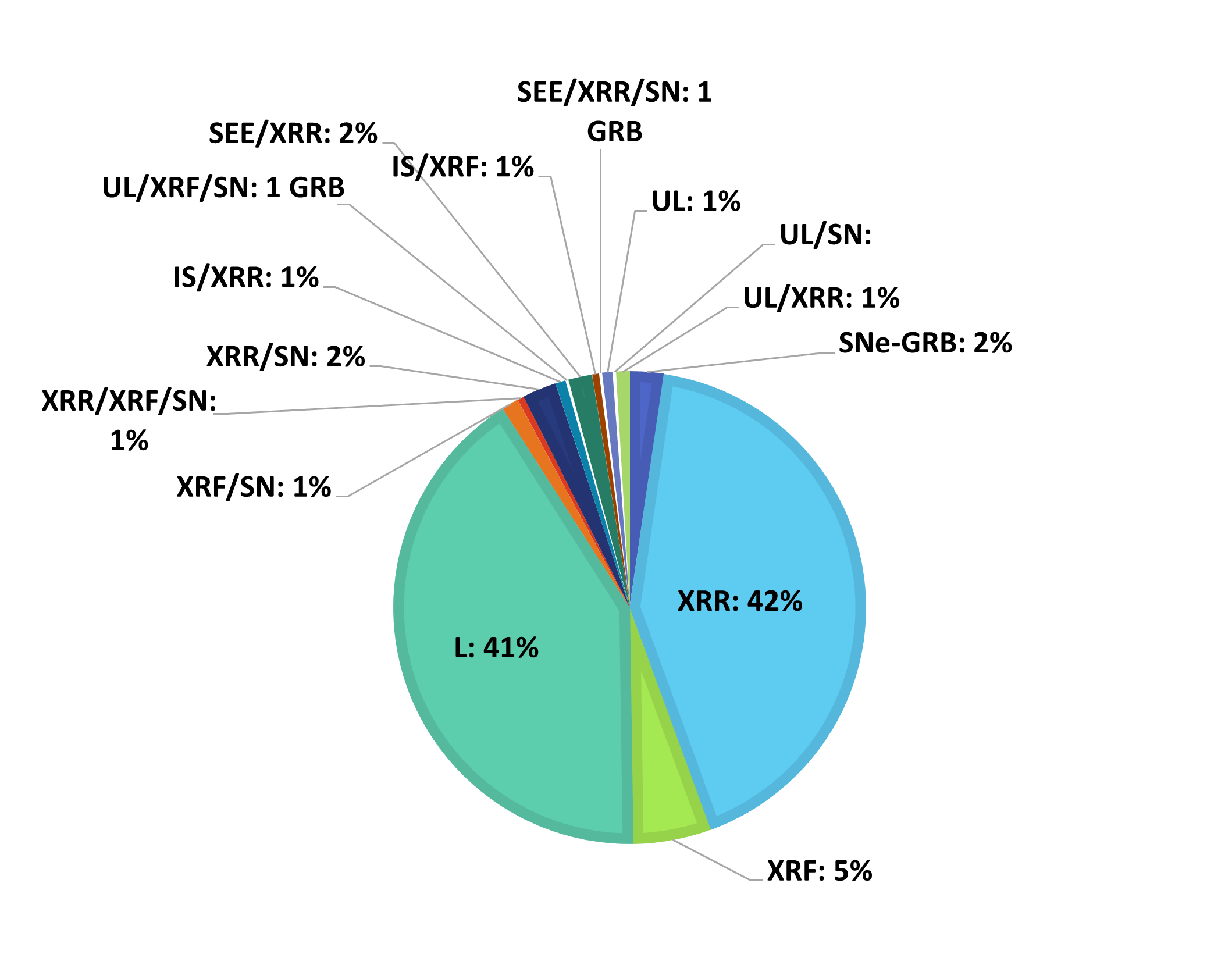}
\includegraphics[width=.45\textwidth]{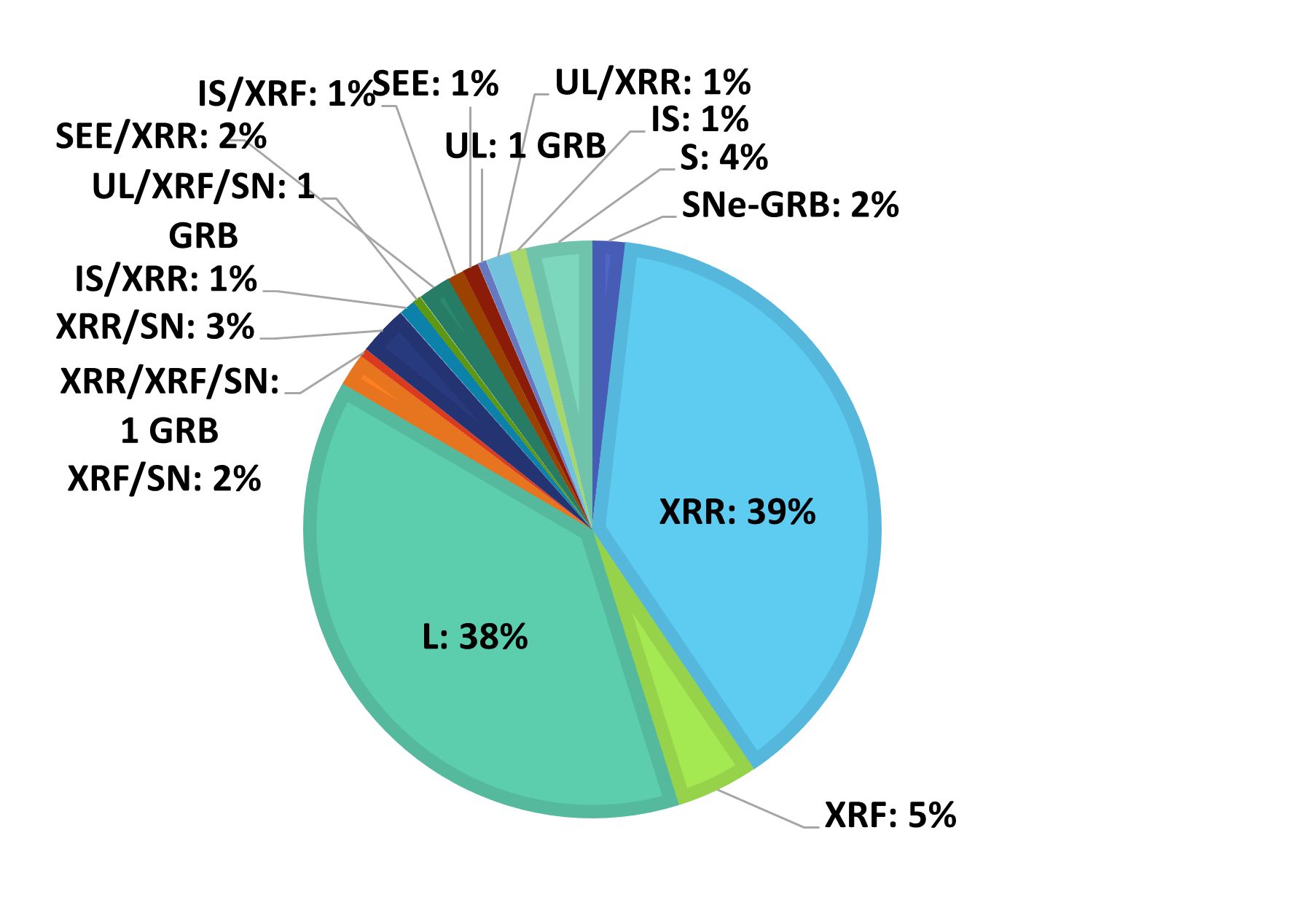}  
\end{center}

    \caption{Top left: breakdown of all 455 GRBs by duration: long (L), short (S), short with extended emission (SEE) and intrinsically short (IS); top right: breakdown of all 455 GRBs by morphological class; bottom left: breakdown of all Long GRBs into types: X-ray rich (XRR), X-ray flash (XRF), supernova-associated (SNe-GRB), ultra-long (UL), SEE, IS, and combinations thereof; bottom right: breakdown of only Good GRBs used for reconstruction into types.}
    \label{fig:piecharts}
\end{figure*}

We segregate the GRBs into classes to better understand the characterization of the Good LCs, which make up 48\% of our total sample of 455 GRBs. We show a breakdown of our sample into classes in Fig. \ref{fig:piecharts} - the top left chart shows classification by duration into LGRBs, SGRBs, SEEs and IS GRBs. The top right chart shows classification into morphological classes. The bottom two charts show classification by type (XRRs, XRFs, ULGRBs, SNe-GRBs) for all LGRBs (left) and all Good GRBs (right). For the Good GRBs, we see that the largest fraction of the GRBs are XRR GRBs (39\%), with the second-most frequent class being simply LGRBs (38\%). The third-most frequent class is the XRF (5\%). This would indicate that many LCs in our Good sample have very strong emissions in X-rays. There is no strong preference for any other class, with the other classes comprising $< 5\%$ of the Good sample.

\subsection{The Reconstruction method with the functional forms}\label{section:reconstruction}

We segregate the 455 GRBs into the previously-mentioned classes to understand how many Good GRBs belong to a given class. 
Then, for simplicity, we limit the application of the reconstruction technique to only the Good GRB LCs, which still constitute a large fraction (48\%; 218 GRBs) of our total sample. Indeed, we aim for a procedure that works for well-behaved morphological LCs and can be extended to more complex cases in the future.

We compute the flux residual for each LC. The flux residual is defined as the difference between the logarithm of the flux value of the original LC and the corresponding logarithmic value of the flux given by the W07 fit for a given instant of time $t$ (in $\log_{10}$ scale). This can be represented by: 

\begin{equation}
\log_{10} F_{\rm{res}}= \log_{10} F^{obs}_{t} - \log_{10} f(t) 
\label{eqn2}
\end{equation}

where $\log_{10} F_{\rm{res}}$ is the logarithm of the flux residual, $\log_{10} f(t)$ is the W07 model or the BPL model flux at time $t$ and $\log_{10} F^{\rm{obs}}_{t}$ is the observed log flux at time $t$.

\begin{figure}
\centering

W07 Fit and residuals

\includegraphics[width=.4\textwidth]{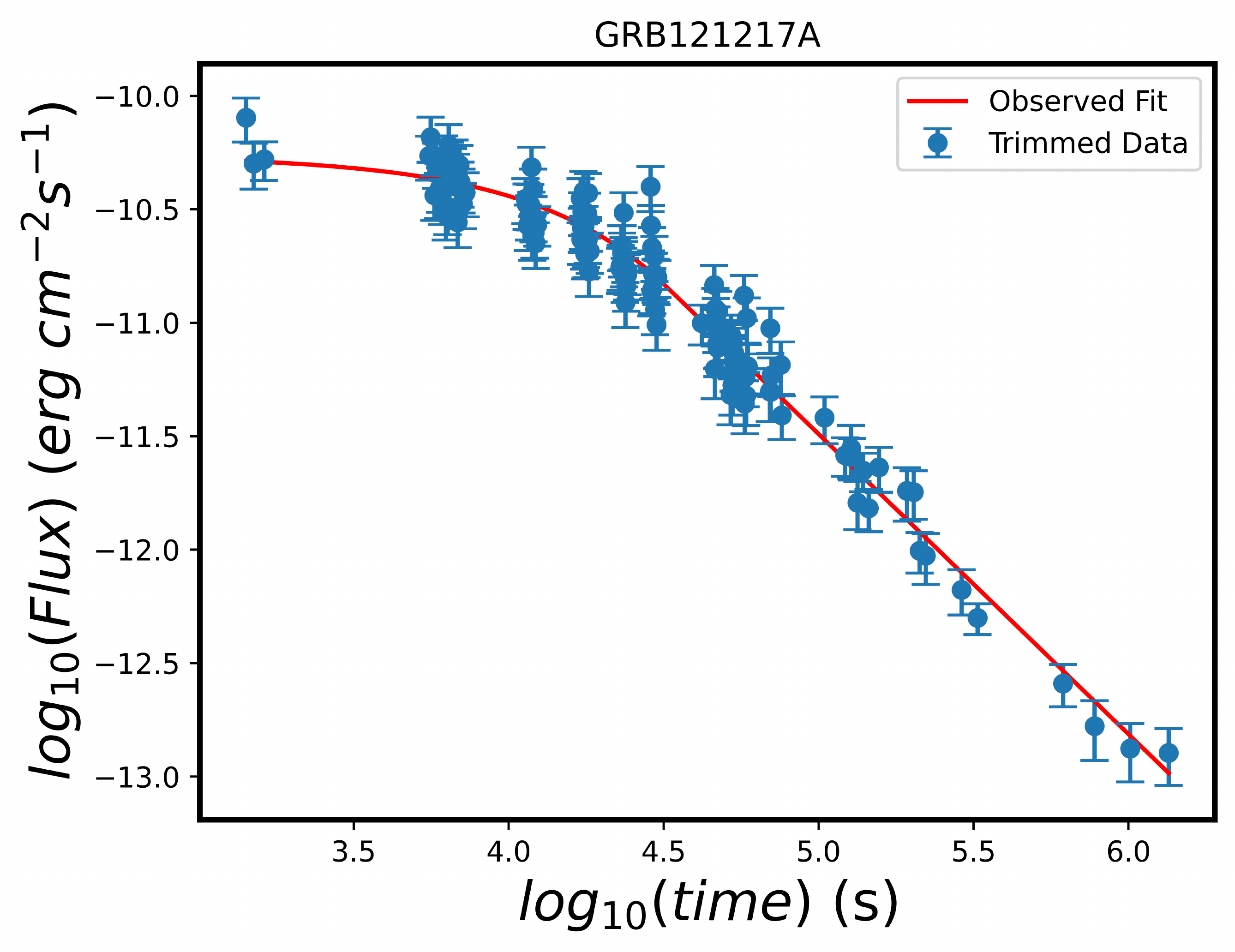}
\includegraphics[width=.4\textwidth]{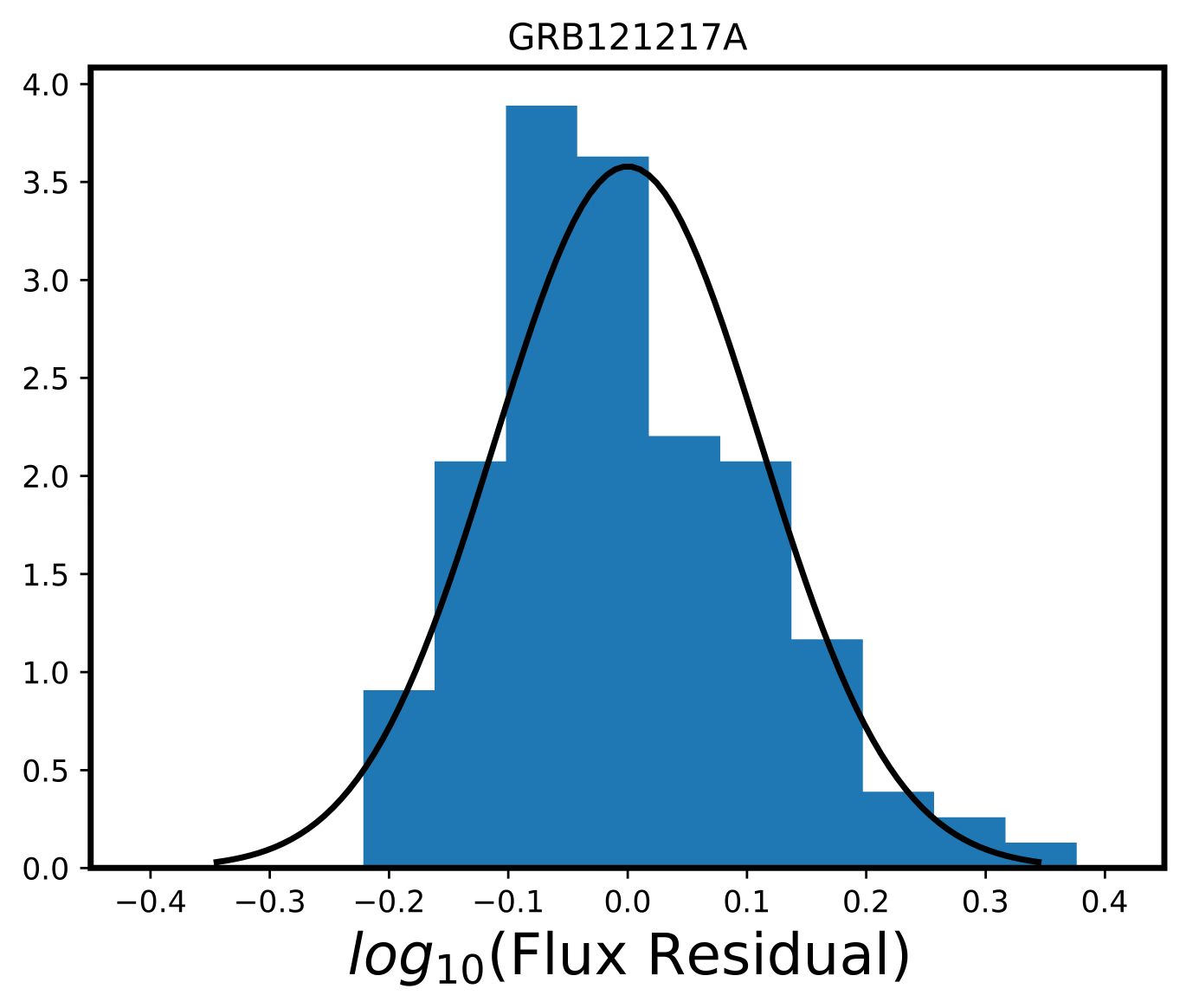}

BPL Fit and residuals

\includegraphics[width=.4\textwidth]{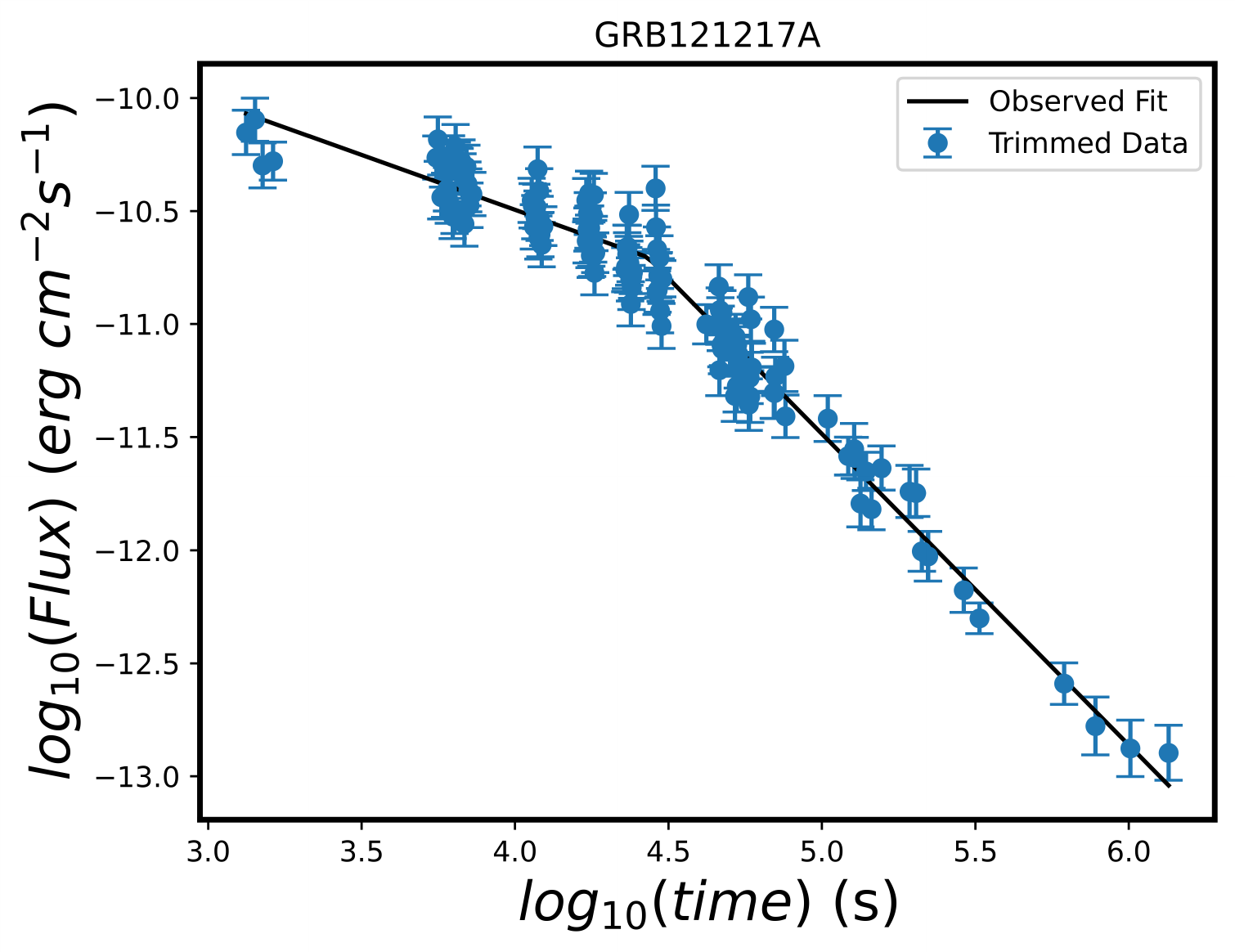}
\includegraphics[width=.4\textwidth]{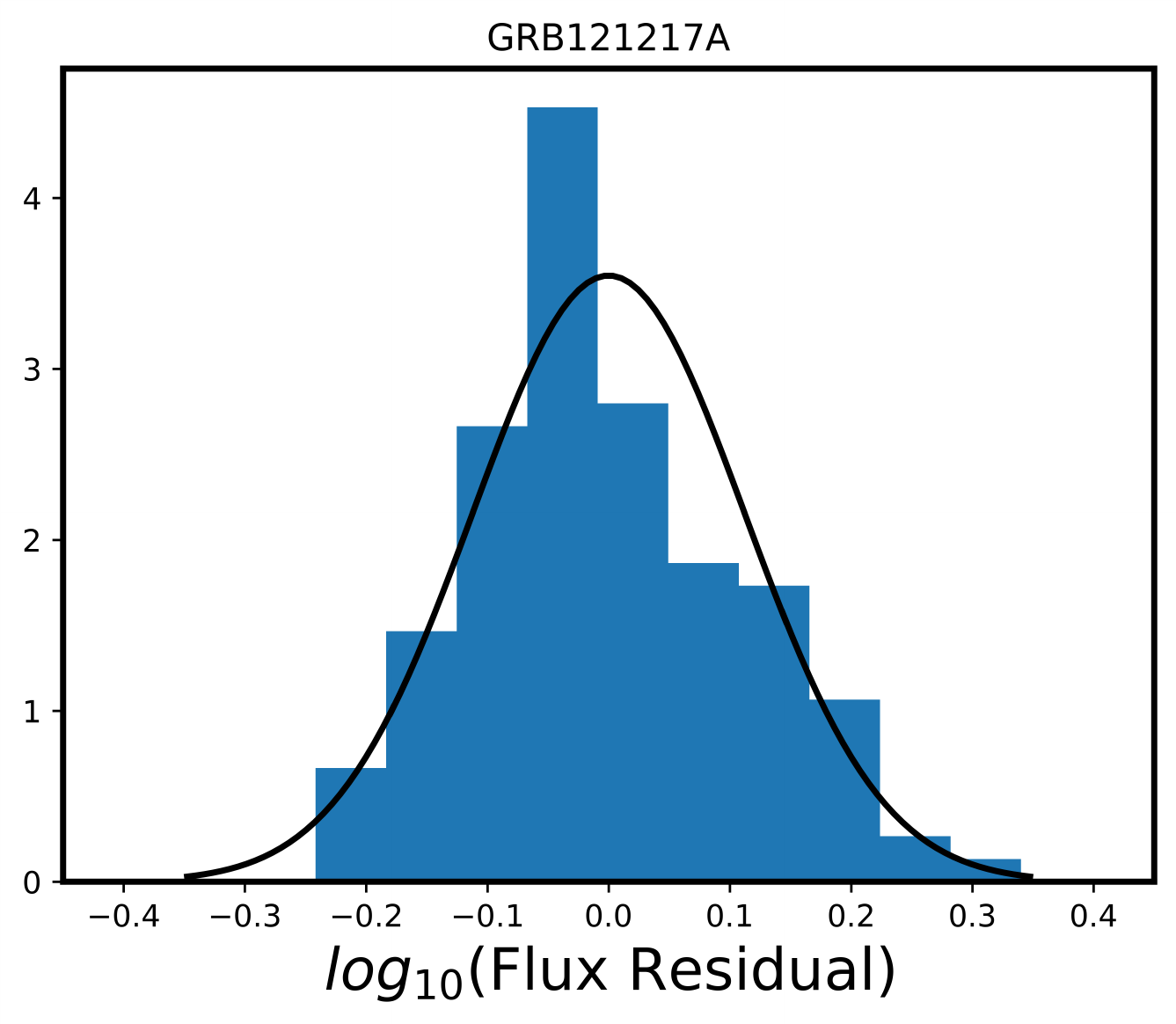}
    \caption{The left panels show the LCs of GRB GRB 121217A starting from the plateau emission, and the best-fit W07 model is shown in red on the upper panels and the BPL fit shown in black on the lower panels.
    The right panels show the log(flux) residual histogram and the best fit Gaussian distribution shown in black.}
    \label{fig:histograms}
\end{figure}

In the left panels of Fig. \ref{fig:histograms}, we show as an example the fitting of a Good LC, GRB 121217A with the W07 model (top; shown in red) beginning from the plateau emission and the BPL model (bottom; shown in black). 
In the right panel of Fig. \ref{fig:histograms}, we show the corresponding $\log_{10}$ flux residual histograms, or Normalized histogram of $F_{(res)}$ detailed in Eq. \ref{eqn2}, with the best-fit Gaussian overlaid for this GRBs.
The purpose of generating these histograms is to check the deviation of original flux values from the fitted W07 or BPL model. Based on these histograms, we performed the fitting using the best-fit distribution for each GRB, which resulted in the Gaussian/Normal distribution:

\begin{equation}
P(x) = \frac{1}{{\sigma \sqrt {2\pi } }}e^{{{ - \left( {x - \mu } \right)^2 } \mathord{\left/ {\vphantom {{ - \left( {x - \mu } \right)^2 } {2\sigma ^2 }}} \right. \kern-\nulldelimiterspace} {2\sigma ^2 }}}
\label{eqn3}
\end{equation}

where $\mu$ is the mean of the distribution, $\sigma$ is the standard deviation and $\frac{1}{\sigma\sqrt{2\pi}}$ is the normalization constant.

We draw random variate samples from the fitted Gaussian to generate our reconstructed data points based on our previous assumption that the flux has Gaussian-distributed errors. 
We checked that the Gaussian is indeed the best-fit distribution with the Mathematica 
\footnote{using Wolfram Mathematica 12.3} 
function \texttt{FindDistribution} and find that the values for normality for GRB121217A are $\mu=-0.0016$, and $\sigma=0.12$.

For a given time, $t$, the reconstructed flux at that time is defined in Eq. \ref{eqnreconstruction} as the sum of the fitted flux value and the noise based on the random variate sampled from the flux residual Gaussian distribution: 

\begin{equation}
\log_{10} F^{\rm{recon}}_{t} = \log_{10} f(t) + (1+n)\times RV_{\mathcal{N}}
\label{eqnreconstruction}
\end{equation}

Here, $F^{\rm{recon}}_{t}$ is the reconstructed value of flux at time $t$, $f(t)$ is the W07 or BPL flux value at time $t$, $n$ is the noise level, and $RV_{\mathcal{N}}$ is the random variate sampled from the Normal (Gaussian) distribution. 


We add the noise parameter to mimic the realistic variations seen in observed LCs. We assume the noise to be stationary, meaning that the noise on the flux does not change with time and is random without temporal autocorrelation.
Usually, it is common to encounter cases in which the noise level is increased by 10\% or 20\%. We acknowledge that the choice of noise level is arbitrary, but the purpose here is to show how LCR works with realistic cases. Anyone with the detailed procedure we have described can change the noise level for the particular LCs they are reconstructing. 

The $RV_{\mathcal{N}}$ for this noise is generated at every instant of time, starting from the beginning of the plateau from the distribution of the residuals fitted with a Gaussian. Thus, the reconstructed flux value is different for each point. In this manner, we can reconstruct data points within the temporal gaps of each LC. 

Applying this method will result in all the reconstructed data points in the afterglow LC being, statistically, the same distance from the best-fit of the W07 or BPL model as the initially-observed data points.
We choose the time of the reconstructed points along the fitting line with a time range distributed equally in the log scale. To this end, for practicality, we use the function in Python called \texttt{geomspace}, with the minimum and maximum time set at the beginning of the plateau and the last data point of the observed LC, respectively. 
Thus, each reconstructed LC is customized to each observed LC, which ensures the realistic nature of these reconstructed LCs for future use in cosmological studies. 

Due to the random noise added to the reconstructed data points, we perform the reconstruction steps mentioned above for each GRB 100 times. This helps in de-randomizing the results and stabilizing them. Then, the chosen function is fit on each iteration of the reconstructed LC, providing us with 100 new fit parameters and error estimates for each GRB. It should be noted that the new fit is performed on the LC, which contains the reconstructed flux and the original observations. So the combined LC is used here. The flux uncertainties for each reconstructed data point are generated randomly from a Gaussian distribution which is the best-fit distribution of the uncertainties of the observed fluxes.




Thus, with the LCR procedure, we combine the reconstructed and original data points to create a new enhanced LC, which we refit with the chosen model with a least-squares regression. For this, we use the \texttt{minimize()} method from the \texttt{lmfit} library (\cite{newville2016lmfit}), which is an interface for fitting curves in Python. It is a further extension of the optimization techniques available in the methods of \texttt{scipy.optimize}. The \texttt{minimize()} function is used as an optimization function that minimizes the residuals to the fit line while doing the refitting. We provide this function with the model we want to fit, the initial guess for the parameters, the data to fit (the reconstructed LCs), and the method to be used for fitting. This function returns the updated values of the chosen model parameters and their associated uncertainties.


\subsection{The Reconstruction method with the Gaussian processes}\label{sec:GP}

We here briefly describe Gaussian Processes and how we use them. Gaussian process (hereafter called GP) is a generic supervised learning method that can generate precise predictions, and is designed to solve regression problems. It is based on the mathematical properties of the Gaussian/Normal distribution.

Due to its probabilistic nature, a GP model does not produce merely a singular prediction but rather  calculates the likelihood for each possible prediction. 

A GP works on the postulates of Bayesian inference. It begins by inspecting the prior, which contains the leverage of knowledge of existing trends in the initial data, and using Bayesian inference to compute a posterior, a probabilistic description of the outcomes that agree with both the data and the prior. Each prediction has an associated confidence interval, which limits the region of the likelihood of the prediction.

The capacity to find confidence intervals/regions is the true utility of a GP model. The set of possible predictions might be a Normal distribution with the mean as the model’s prediction and a variance that characterizes the deviation of the predictions from the mean. A lower uncertainty corresponds to a thinner confidence region, and the model is said to be confident with its prediction.

Covariance functions, also known as Kernels, are a fundamental requirement to use a GP model. A Kernel can be defined as a measure of the degree of similarity between two given input space data points. This contains the assumption that two similar data points should yield two similar output values. Kernels can be of different types: those depending directly on the difference between the two data points, that is, $x-x'$, and those depending on the specific values of the data points $x$ and $x'$ themselves. Kernels depending upon just their difference, are translation-invariant in the input space. Furthermore, if the kernel in question depends merely on $|x-x'|$, then the kernel is said to be isotropic in the input space. An example of such a kernel would be a Radial Basis Function (RBF) kernel.

We here use the RBF plus the contribution of a White noise Kernel because we assume that the noises of the flux are independent and identically normally distributed. We use the built-in function \textit{GaussianProcessRegressor}, which enables prediction without prior fitting but only with prior knowledge of the GP. The prior in our analysis is not normalized, meaning that the mean is centered at zero. Here normalizing means that we would normalize the fluxes by removing the mean and scaling to unit-variance. We choose the default option as false. Then we fit the GP regression model with the built-in \textit{fit} function to our data. Next, we have used the in-built \textit{predict} function to obtain the reconstructed data points.

To reconstruct the LC with GP, we have chosen as the interval for the reconstruction 95\% confidence interval. We have populated the data points by using the Eq. \ref{eqnreconstruction} again, but here the function $f(t)$, instead of being the W07 or the BPL function, is the function found via the GP. To avoid the distribution of the data points in the Gaussian processes depending heavily on the shape of the function obtained with the Gaussian Regressor function, a built-in function in Python, for each different LC, we perform 100 MCMC simulations of the reconstructed LC. Then, we randomly pick one value for the data point and its associated uncertainty.

\section{Results}\label{section:results}

\subsection{Results from functional form reconstruction}\label{sec:funcresults}


Applying our LCR methodology to the 218 Good GRBs in our sample, we expect to see a reduction in the uncertainties on the chosen model parameters. To measure this, we compute the error fractions associated with each of the model parameters for the original and reconstructed fit.
The error fractions for the three Willingale parameters are given by:

\begin{equation}
 EF_{\log_{10}(T_{a})}=\left|\frac{\Delta{\log_{10}(T_{a})}}{\log_{10}(T_{a})}\right|,
 \label{eqn5}
\end{equation}

\begin{equation}
 EF_{\log_{10}(F_{a})}=\left|\frac{\Delta{\log_{10}(F_{a})}}{\log_{10}(F_{a})}\right|, 
 \label{eqn6}
\end{equation}

\begin{equation}
EF_{\alpha_{a}}=\left|\frac{\Delta{\alpha_{a}}}{\alpha_{a}}\right|.    
\label{eqn7}
\end{equation}

Here, $EF_{X}$ is the error fraction associated with the parameter $X$, and $\Delta{X}$ is the uncertainty associated with the parameter $X$. The $X$ and $\Delta{X}$ for $\log_{10}(T_a)$, $\log_{10}(F_a)$ and $\alpha_{a}$ before reconstruction are taken from \citet{2020ApJ...903...18S}, as previously indicated.

The computed error fractions before and after reconstruction for the W07 model is performed for the full sample of 218 Good GRBs for two noise levels, 10\% and 20\%, in Table \ref{tab:table2}. The first three columns show the error fraction for each W07 parameter from the original fitting. The second three columns show the error fraction for each W07 from the new fitting after reconstruction. These are averages of 100 iterations, as mentioned previously.

We calculate the percentage decrease in the error fractions for each noise level to analyze the improvement of the fit following the reconstruction. The formula used for the percentage decrease is given as: 

\begin{equation}
\%_{DEC}= \frac{\left|EF^{\rm{after}}_{X}\right|-\left|EF^{\rm{before}}_{X}\right|}{\left|EF^{\rm{before}}_{X}\right|}\times 100   
\label{eqn8}
\end{equation}

The last three columns of Table \ref{tab:table2} show the percent change in the error fraction after reconstruction for the two noise levels. From our analysis, we see that in all cases, the error fraction on the reconstructed parameters is lower than the error fraction on the original parameters. To better understand how this applies to the sample as a whole, we compute the average percentage decrease in error fraction for all Good GRBs tested at each noise level. For the 10\% noise level, we obtain a 33.33\% decrease in error fraction on the $\log(T_a)$ parameter, a 35.03\% decrease in error fraction on the $\log(F_a)$ parameter, and a 43.32\% decrease in error fraction on the $\alpha$ parameter. For the 20\% noise level, we obtain a 29.49\% decrease in error fraction on the $\log(T_a)$ parameter, a 31.24\% decrease in error fraction on the $\log(F_a)$ parameter, and a 40.57\% decrease in error fraction on the $\alpha$ parameter. As expected, when we introduce a larger noise level, the decrease on the uncertainty also decreases since the spread around the best-fit line increases.

\begin{figure*}
\begin{center}

W07 reconstruction

\includegraphics[width=.4\textwidth]{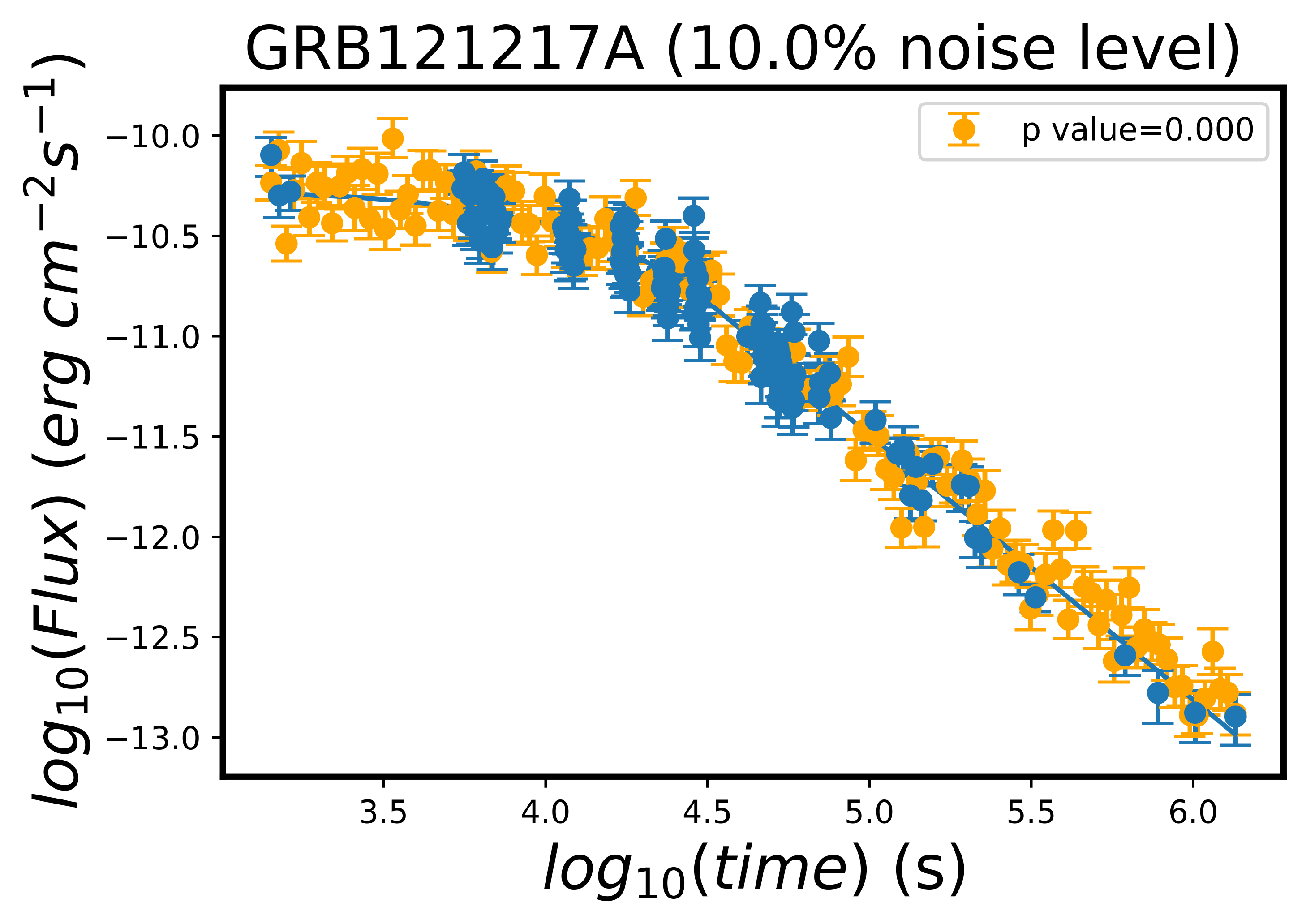}
\includegraphics[width=.4\textwidth]{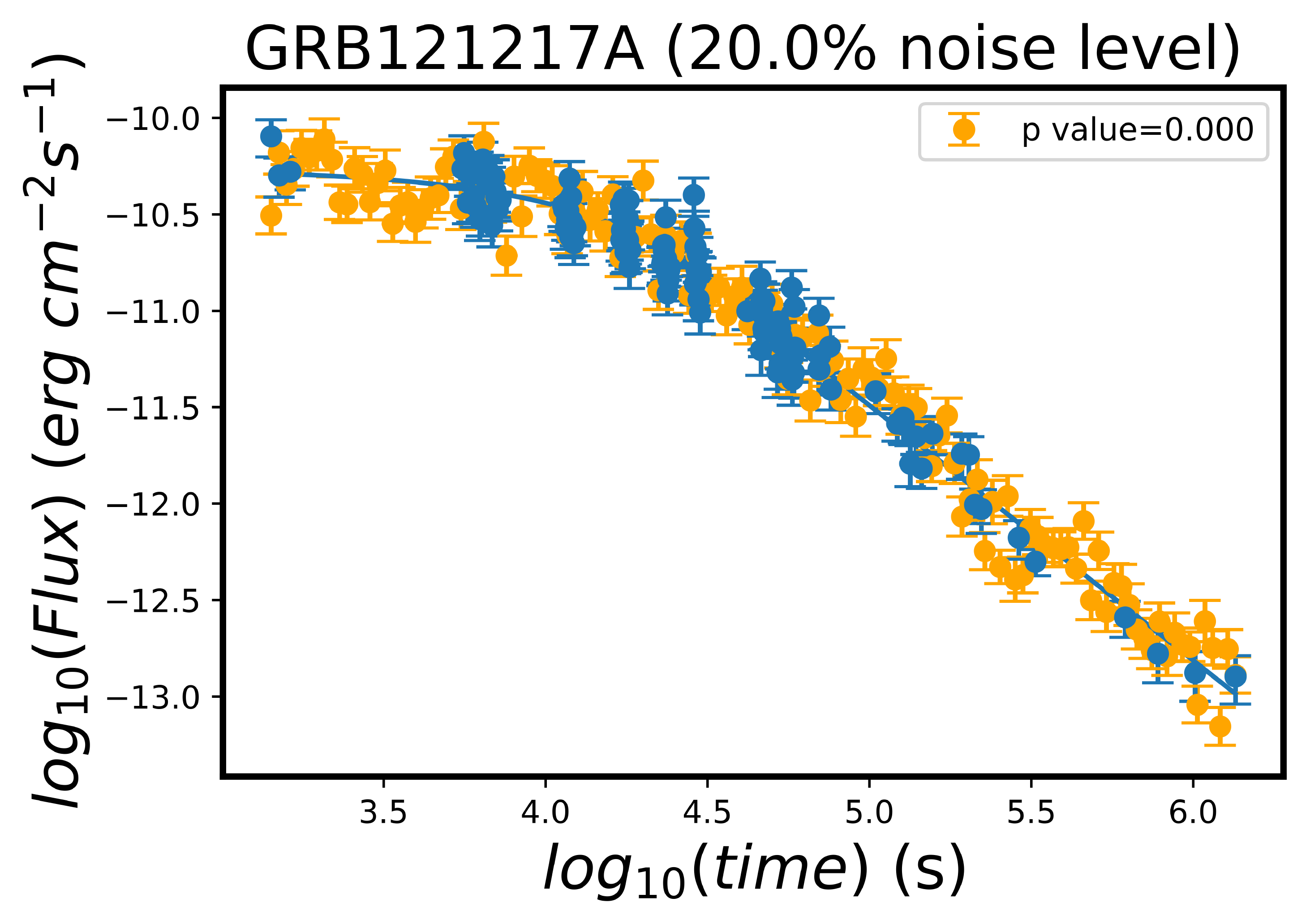}

BPL reconstruction

\includegraphics[width=.4\textwidth]{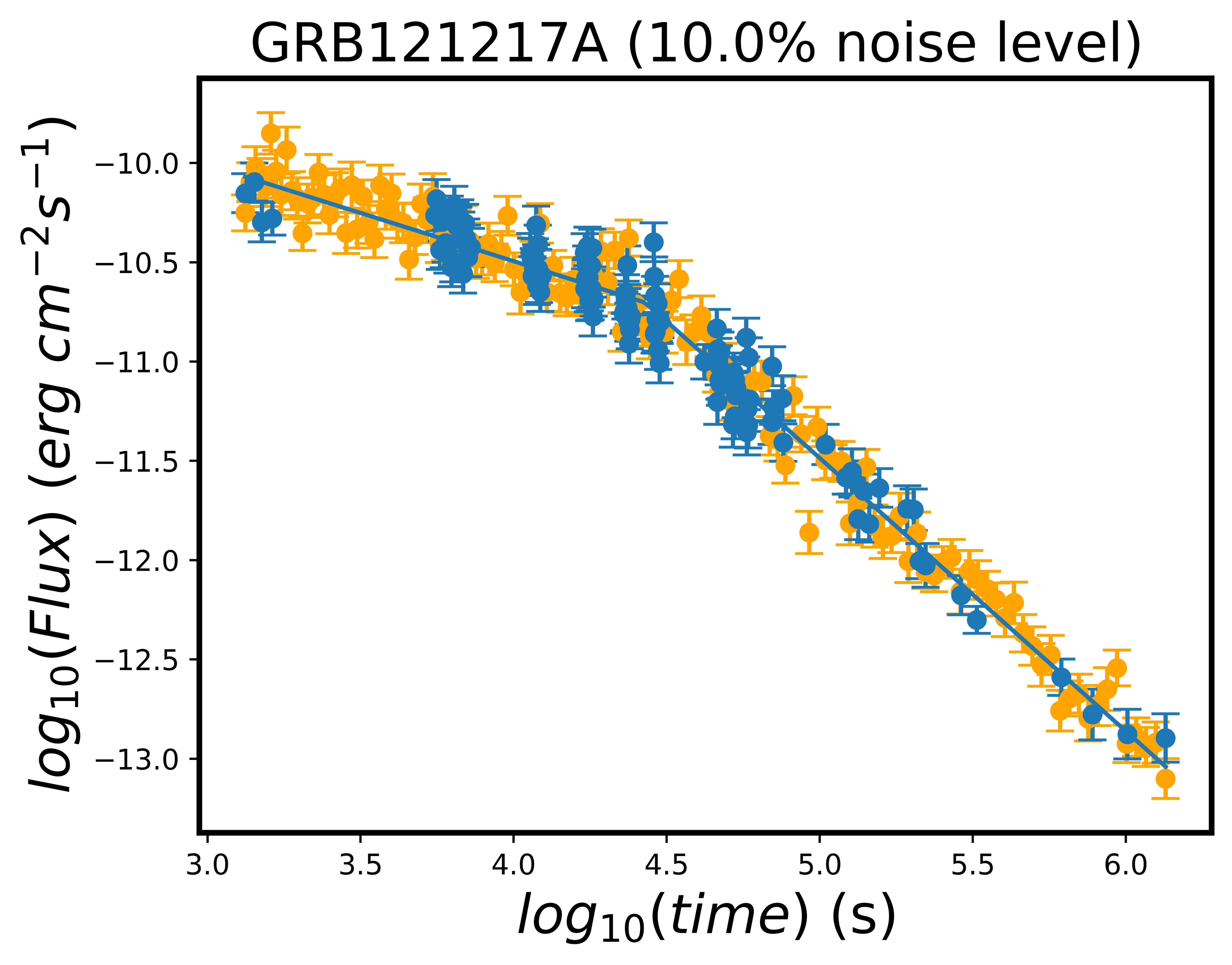}
\includegraphics[width=.4\textwidth]{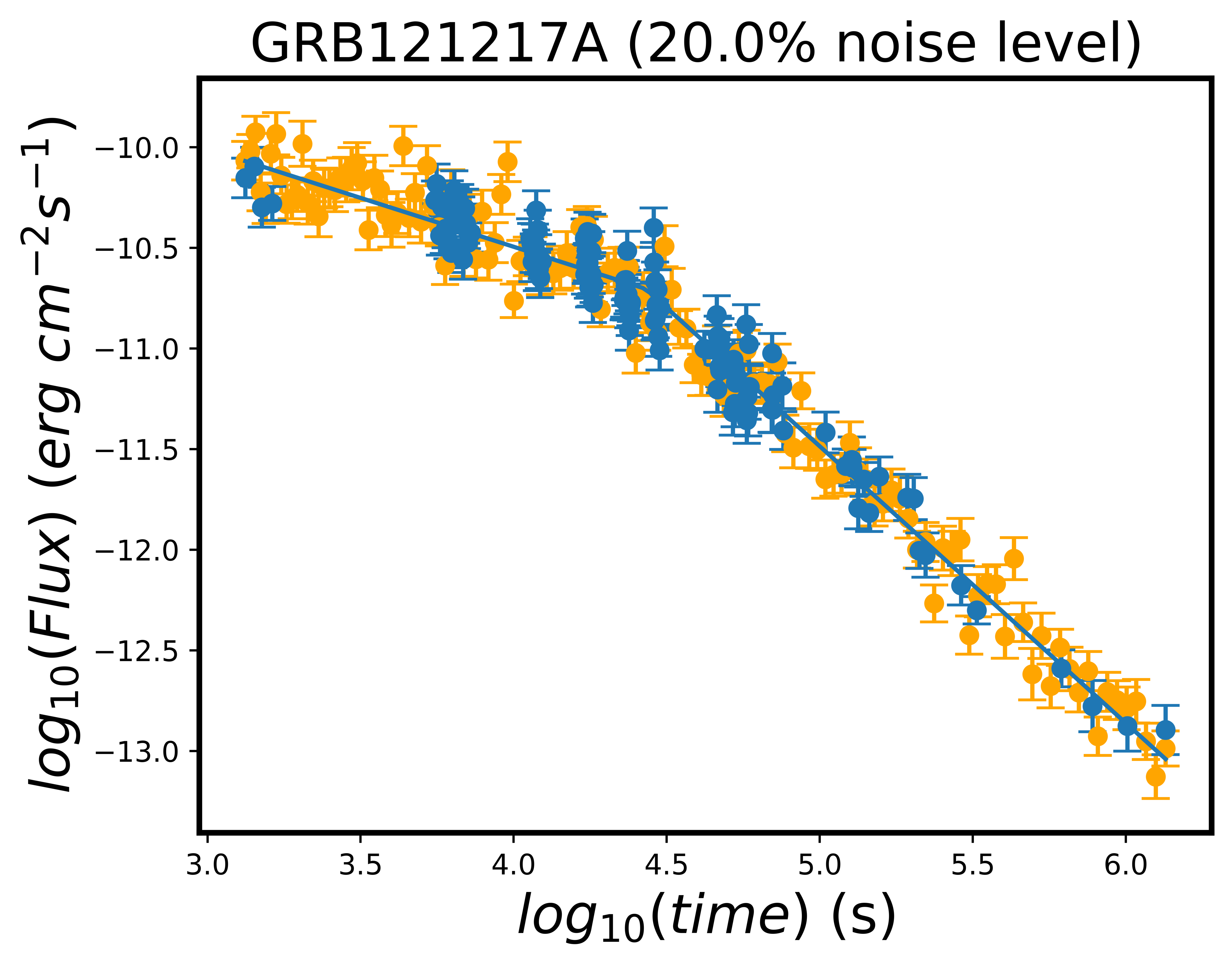}

\end{center}
    \caption{Reconstructed LCs for GRB121217A at two noise levels, using the Willingale (W07) and Broken Power Law (BPL) fits. The first row shows the reconstructed LCs using the W07 model at 10\% and 20\% noise, and the bottom row show the reconstructed LCs using the BPL model at 10\% and 20\% noise.
    These data points are generated from the noise distribution produced by fitting a Gaussian distribution on the flux residual histograms.}
   
    \label{fig:recon}
\end{figure*}

\begin{figure*}
\begin{center}
\includegraphics[width=.95\textwidth]{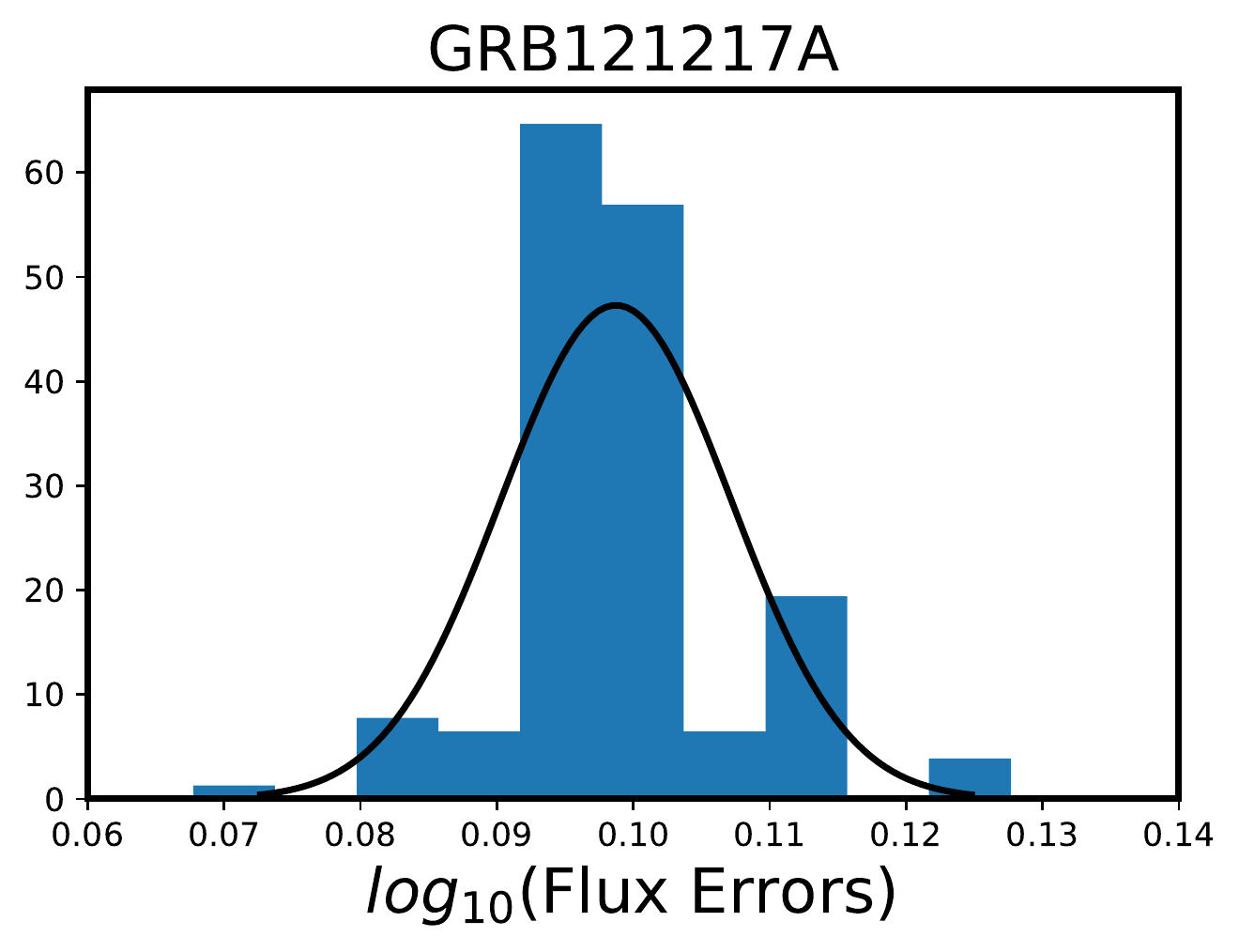}
\caption{Histogram of the error bars of the observed fluxes, shown to be approximated with a Gaussian distribution. }
\label{fig:reconres}
\end{center}
\end{figure*}

\begin{figure}
\fig{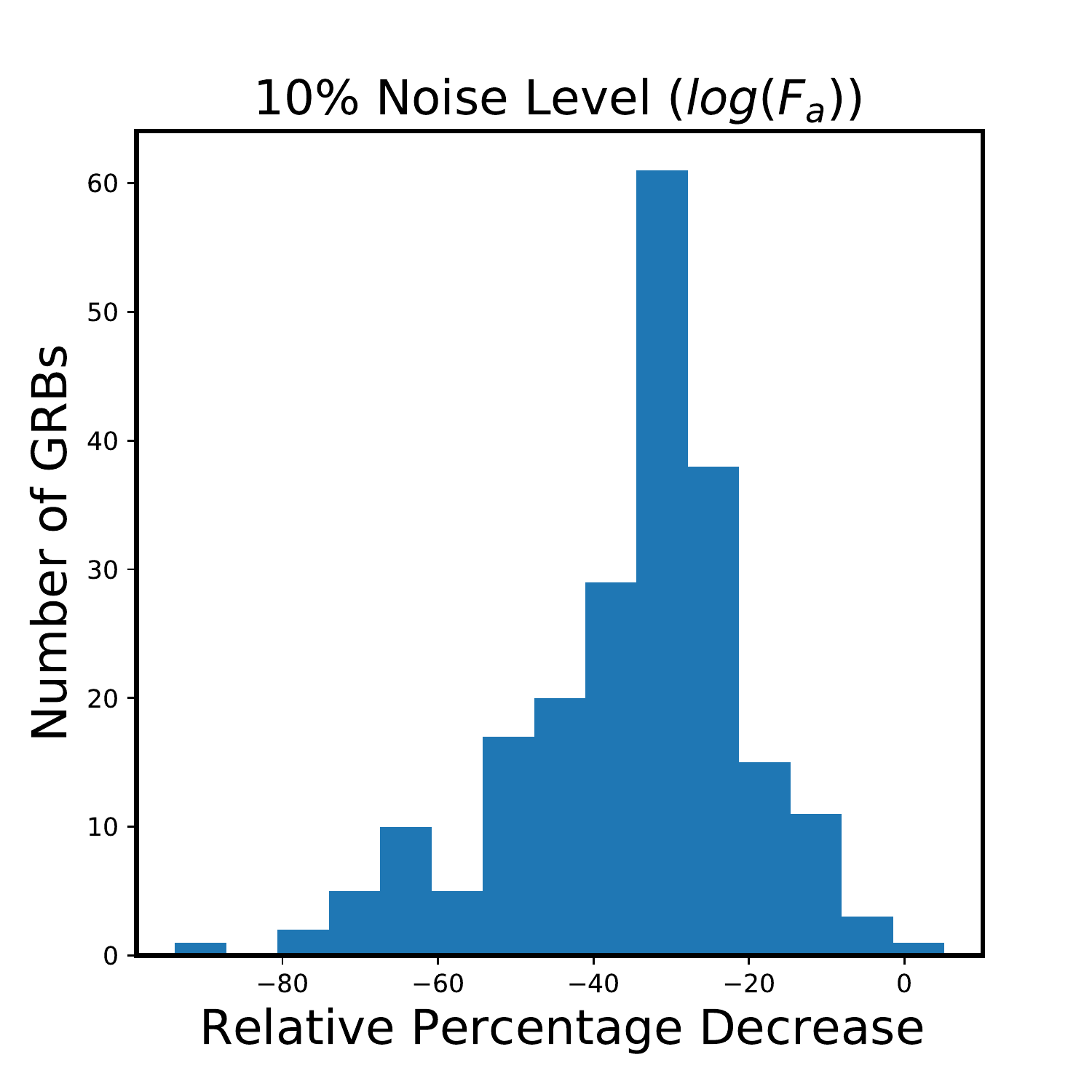}{0.27\textwidth}
{(a) Distribution of the 10\% noise level for the $\log F_a$ with W07.}
\fig{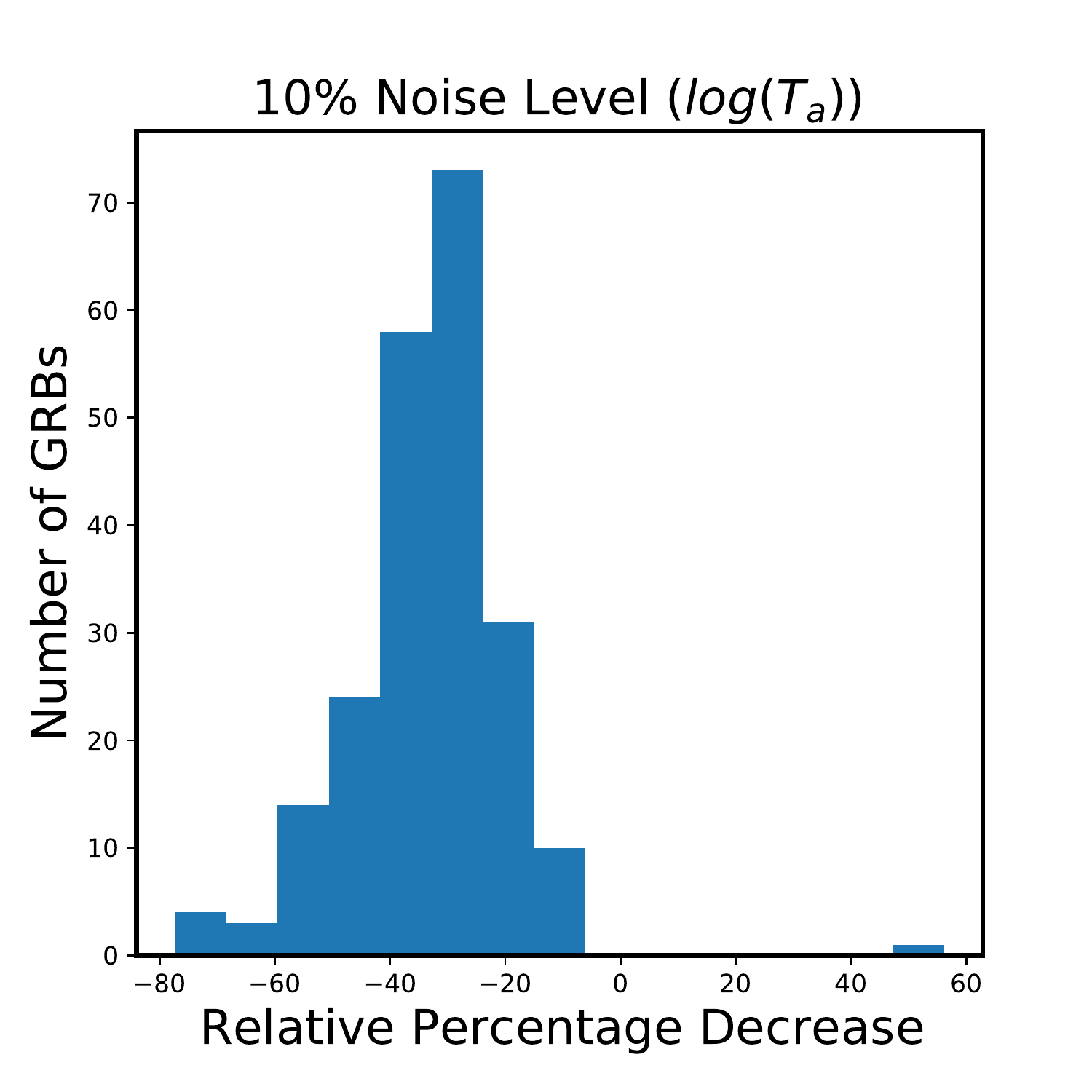}{0.27\textwidth}
{(b) Distribution of the 10\% noise level for the $\log T_a$ with W07}
\fig{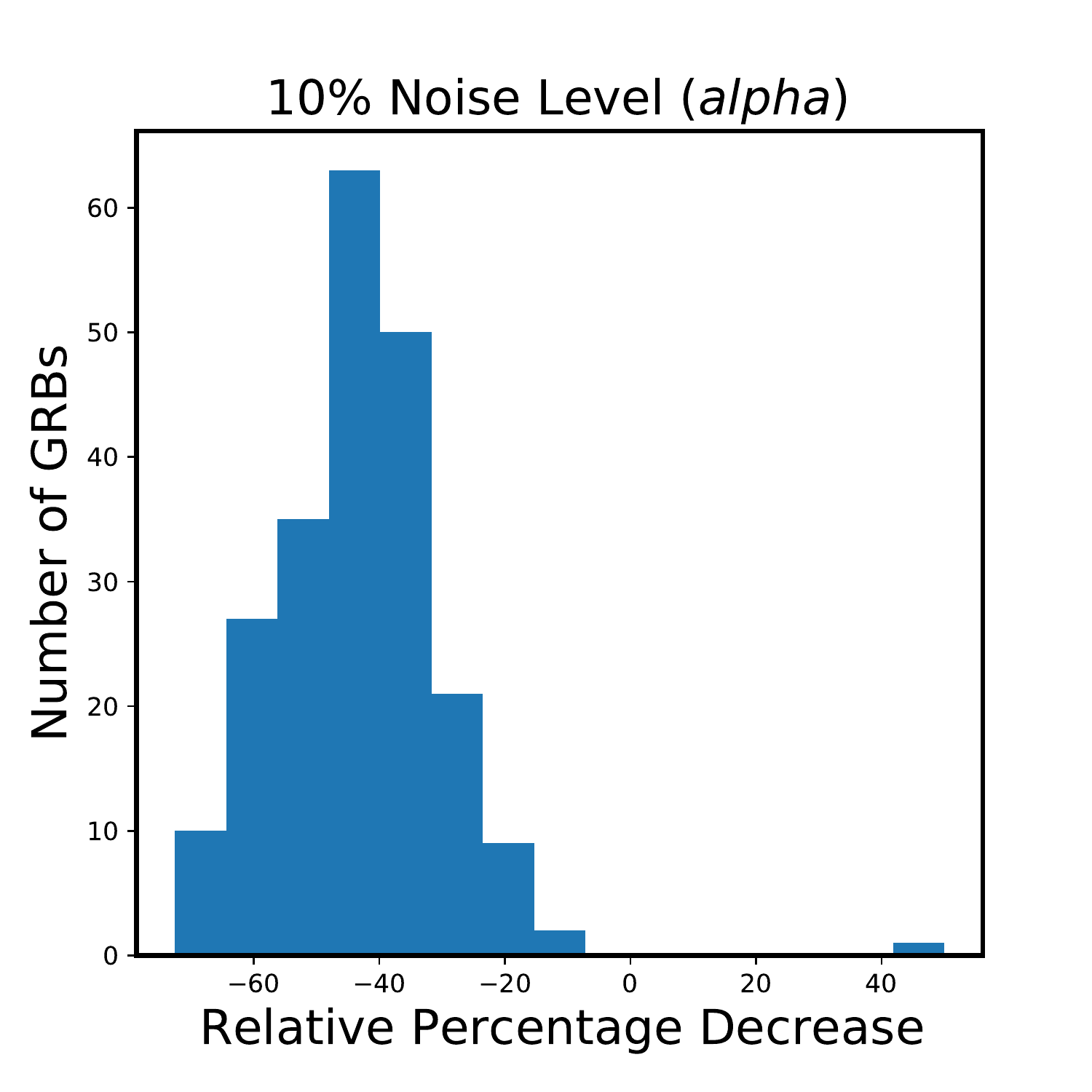}{0.27\textwidth}
{(c) Distribution of the 10\% noise level for the $\alpha_a$ with W07}

\fig{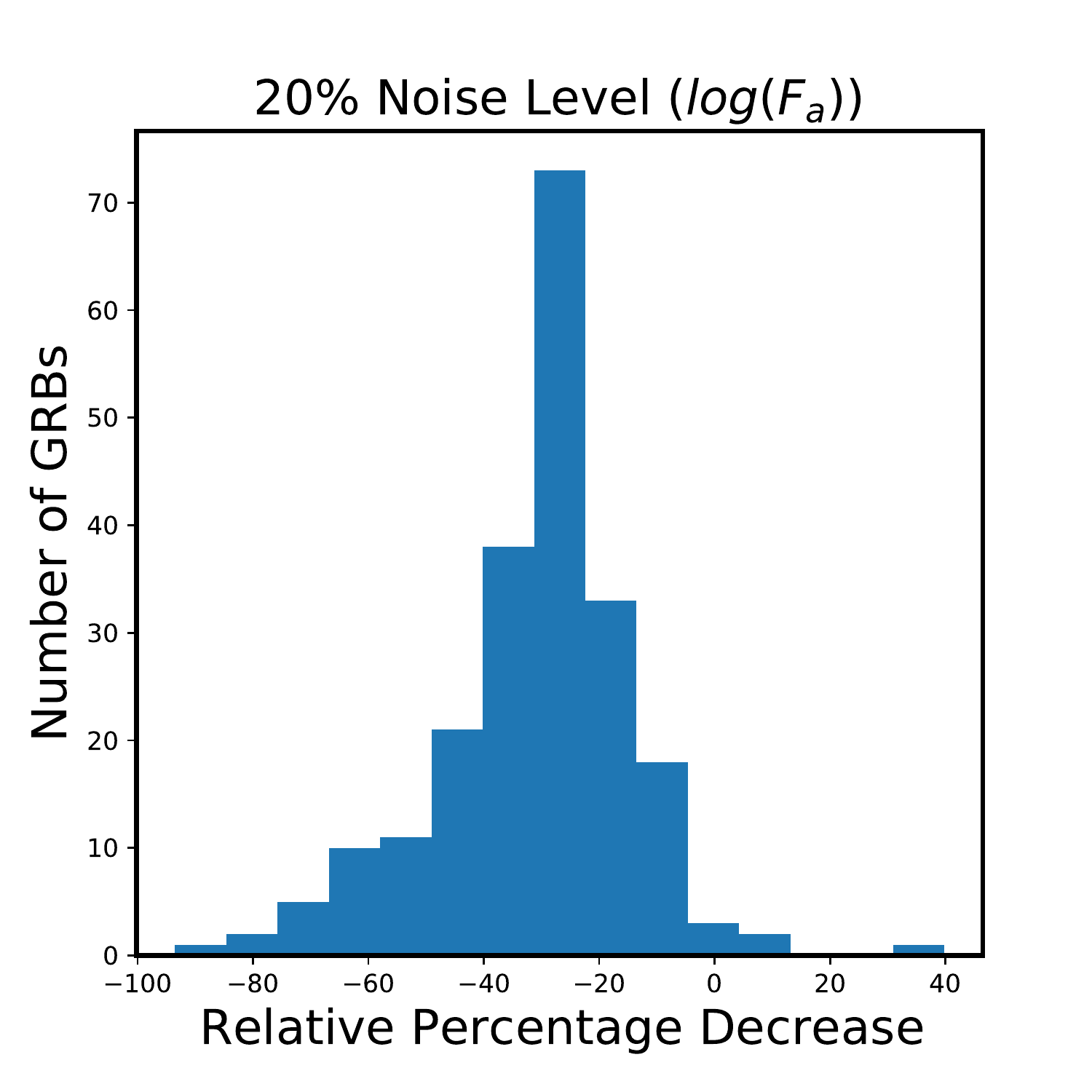}{0.27\textwidth}
{(d) Distribution of the 20\% noise level for the $\log F_a$ with the W07}
\fig{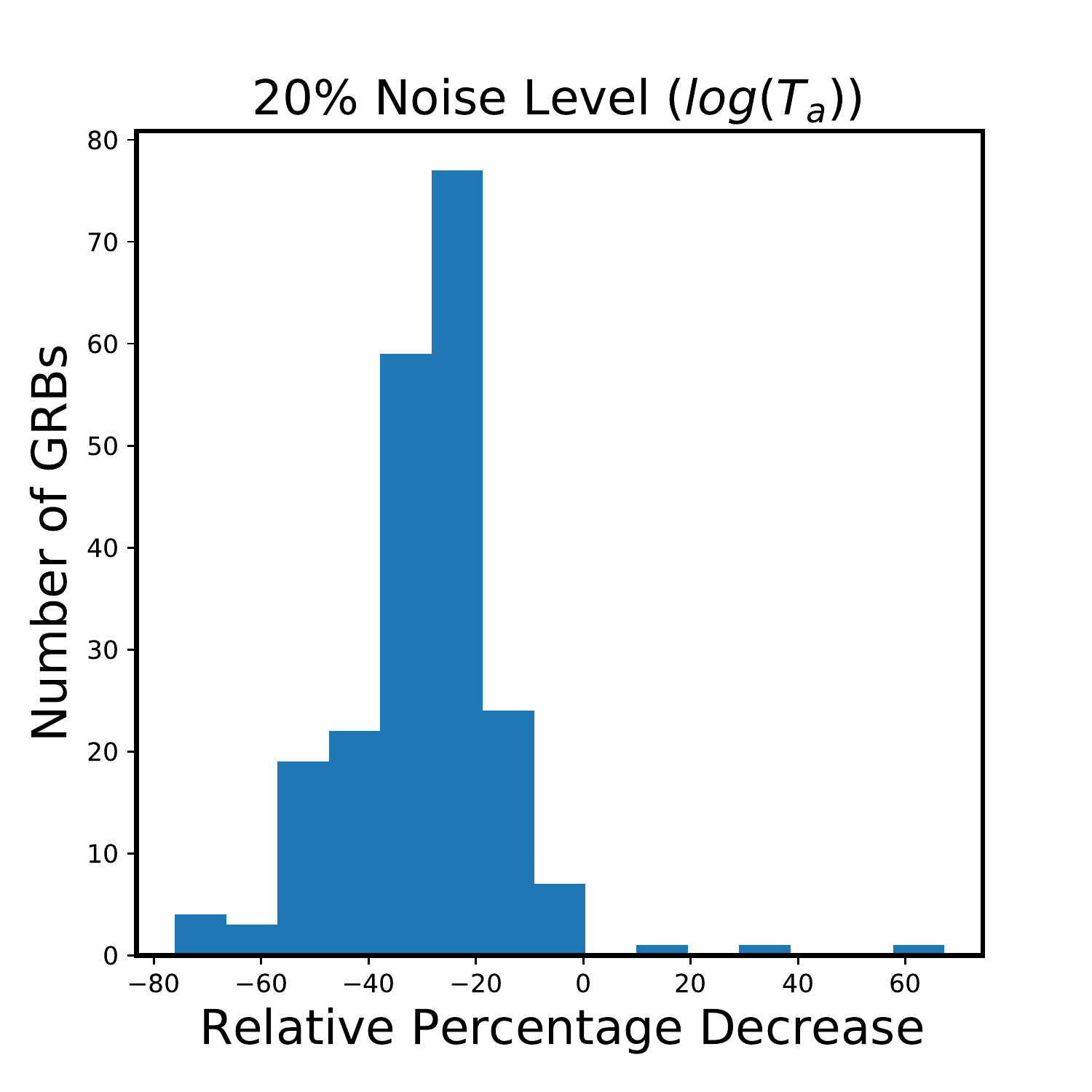}{0.27\textwidth}
{(e) Distribution of the 20\% noise level for the $\log T_a$ with W07}
\fig{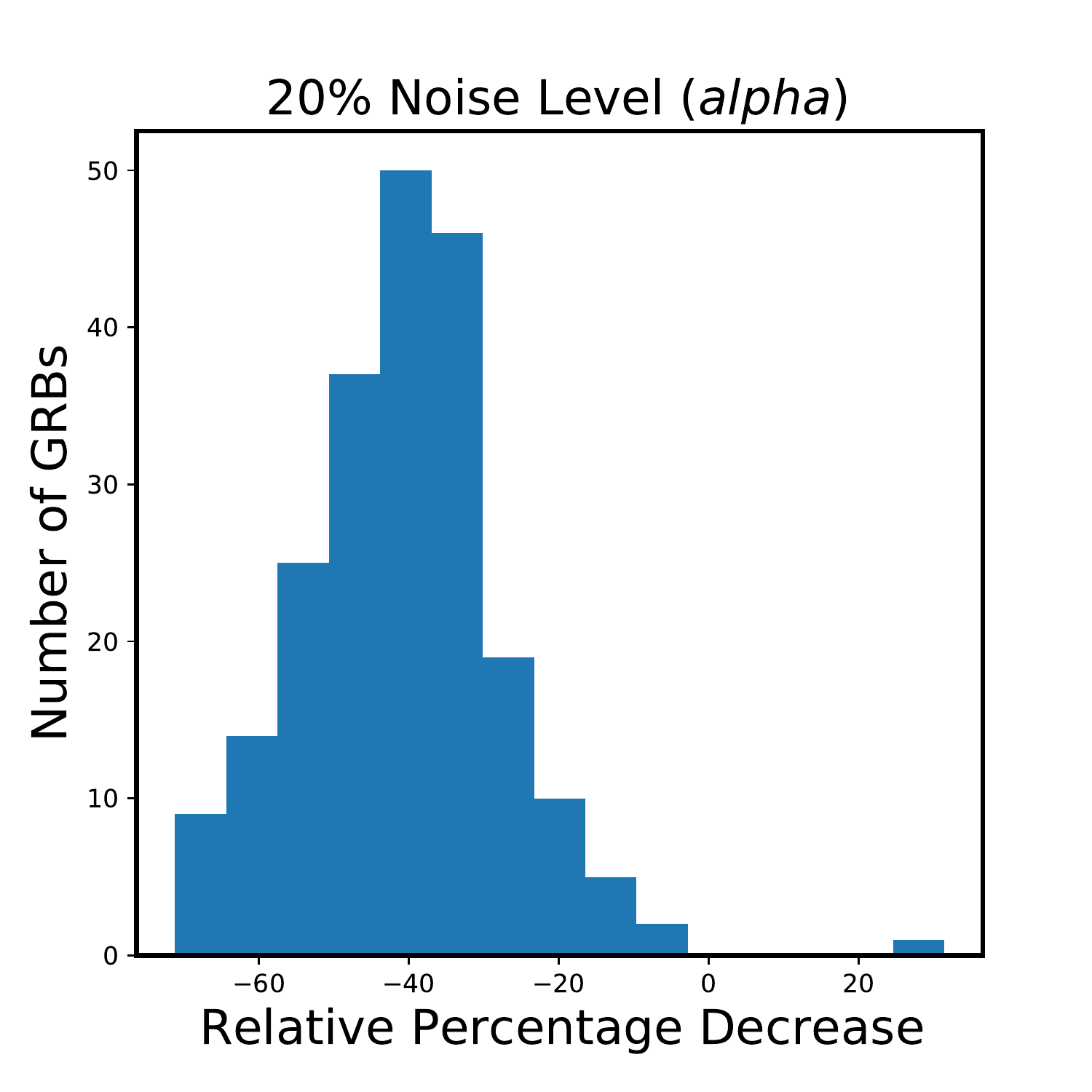}{0.27\textwidth}
{(f) Distribution of the 20\% noise level for the $\alpha$ with W07}

\caption{Distribution of Relative Percentage Decrease for the parameters of the W07 function when we apply the W07 for the reconstruction.}

\label{fig:W07reconstruction}
\end{figure}

A visual comparison between two noise levels for a sample GRB, GRB121217A, for 10\% and 20\% of the W07 fit is shown in the upper panels of Fig. \ref{fig:recon}. We observe that lower noise levels generate points somewhat close to the  W07 fit, and these points are further spread out as the noise level increases. The reconstructed LC plots are picked randomly from the 100 reconstructions for each GRB. As expected, we can approximate the flux uncertainties with a Gaussian distribution as shown in Fig. \ref{fig:reconres}. The histograms of Fig. \ref{fig:W07reconstruction} show the distribution of the percentage decrease (upper panel shows the 10\% noise level, while the bottom panel shows the 20\% noise level) for all three of the W07 parameters. As is evident, the reconstruction leads to a decrease in the errors on the W07 parameters for all the GRBs. After fitting the model, we expect a decrease in the uncertainty since we are refitting more points with the same model. However, the point of this approach is to provide a first simple solution to the lightcurve reconstruction. This toy model for the reconstruction is indeed model dependent, but in principle this method can work on any model.

To assess the method's dependency on a particular model, we also fit with the BPL model and check the differences regarding the decrease in the uncertainties. To test the BPL, we perform the same methodology as used on the W07 fits. Again, we see an overall decrease in error fraction on the reconstructed parameters as compared to the error fraction on the original parameters. We again compute the average percentage decrease in error fraction for all Good GRBs tested at each noise level. For the 10\% noise level, we obtain a 33.31\% decrease in error fraction on the $\log(T_a)$ parameter, a 30.79\% decrease in error fraction on the $\log(F_a)$ parameter, a 14.76\% decrease in error fraction on the $\alpha_1$ parameter, and a 43.98\% decrease in error fraction on the $\alpha_2$ parameter. For the 20\% noise level, we obtain a 29.88\% decrease in error fraction on the $\log(T_a)$ parameter, a 27.20\% decrease in error fraction on the $\log(F_a)$ parameter, a 1.78\% decrease in error fraction on the $\alpha_1$ parameter, and a 41.1\% decrease in error fraction on the $\alpha_2$ parameter.

\begin{figure}
 

\fig{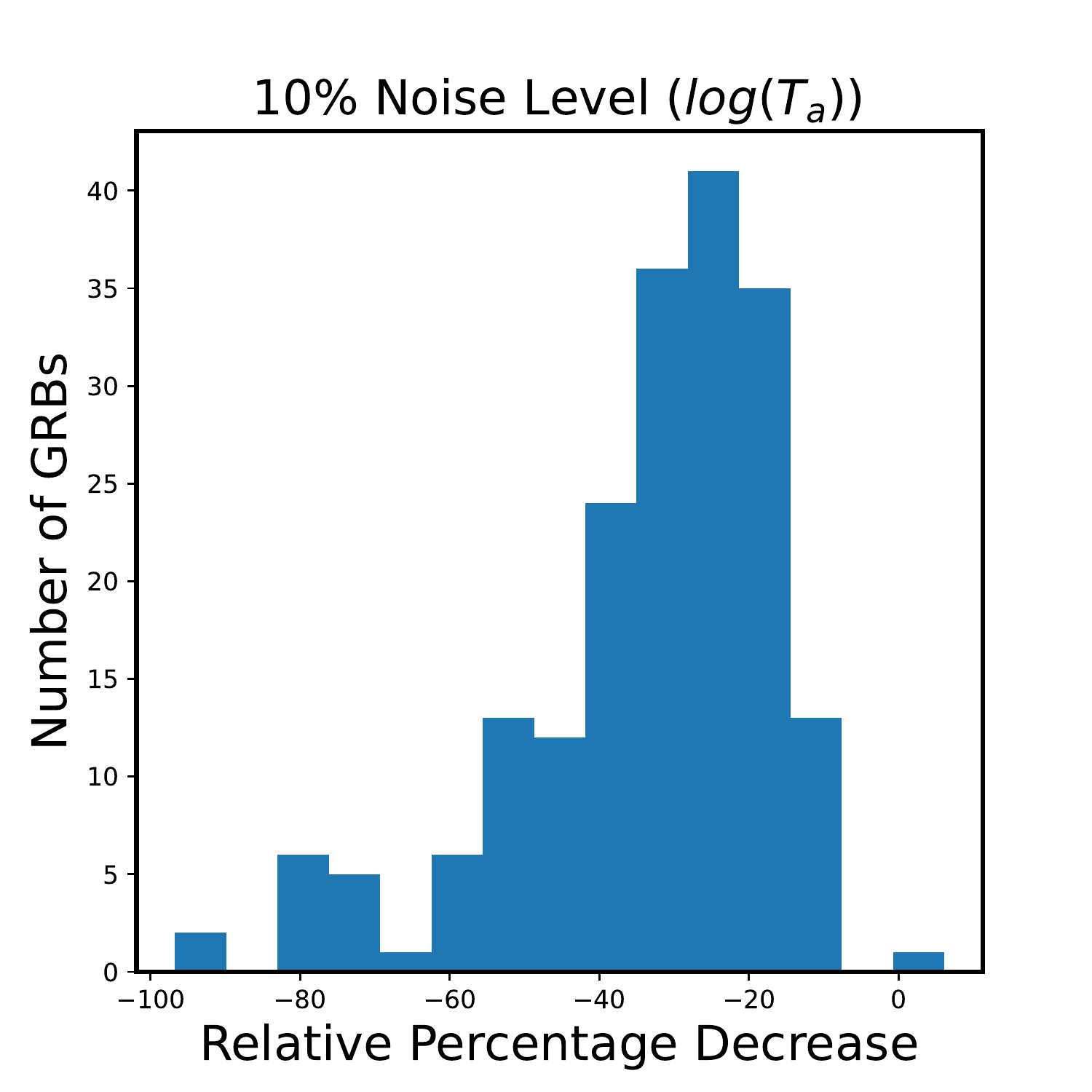}{0.22\textwidth}
{(a) Distribution of the 10\% noise level for the $\log(F_a)$ with BPL}
\fig{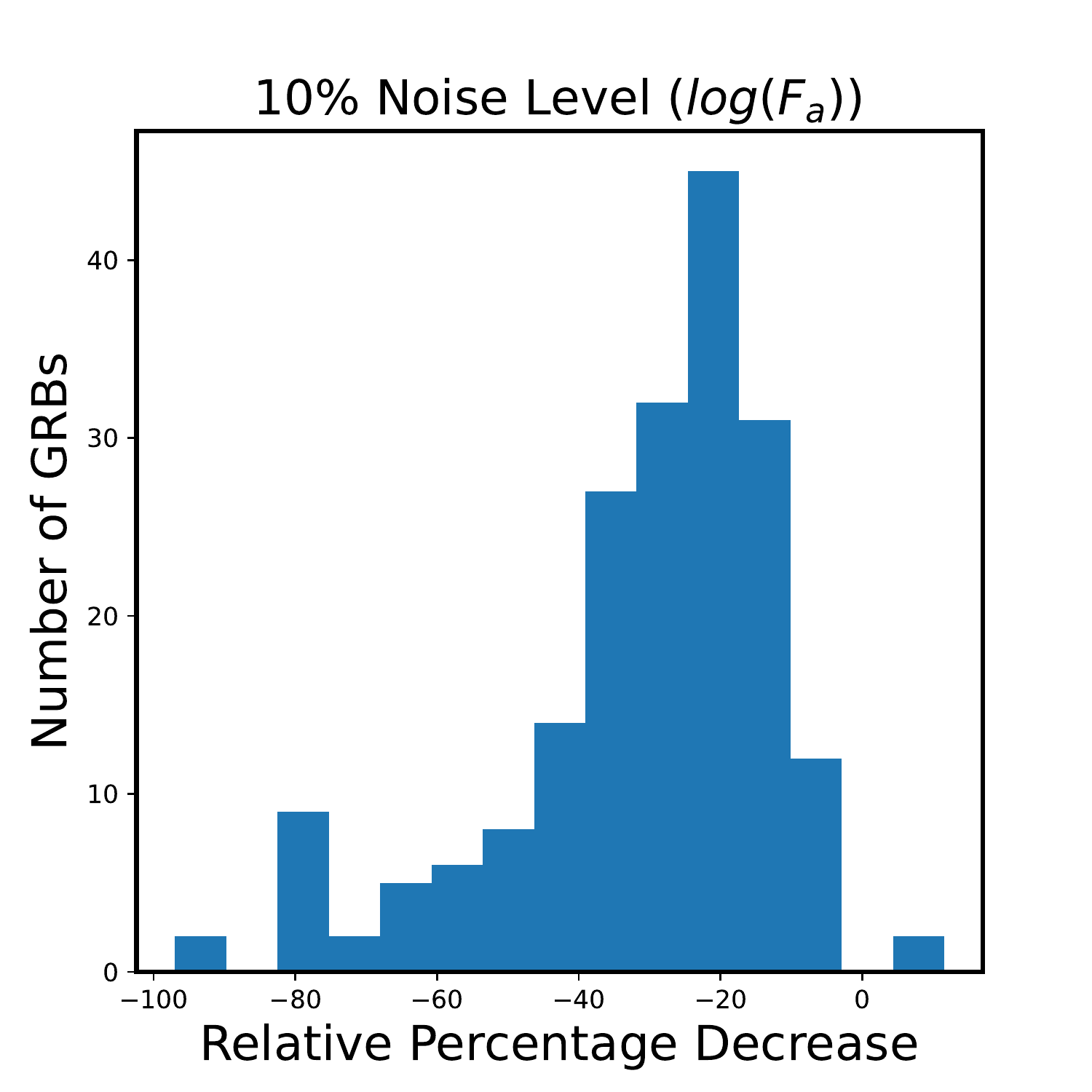}{0.22\textwidth}
{(b) Distribution of the 10\% noise level for the $\log(T_a)$ with BPL}
\fig{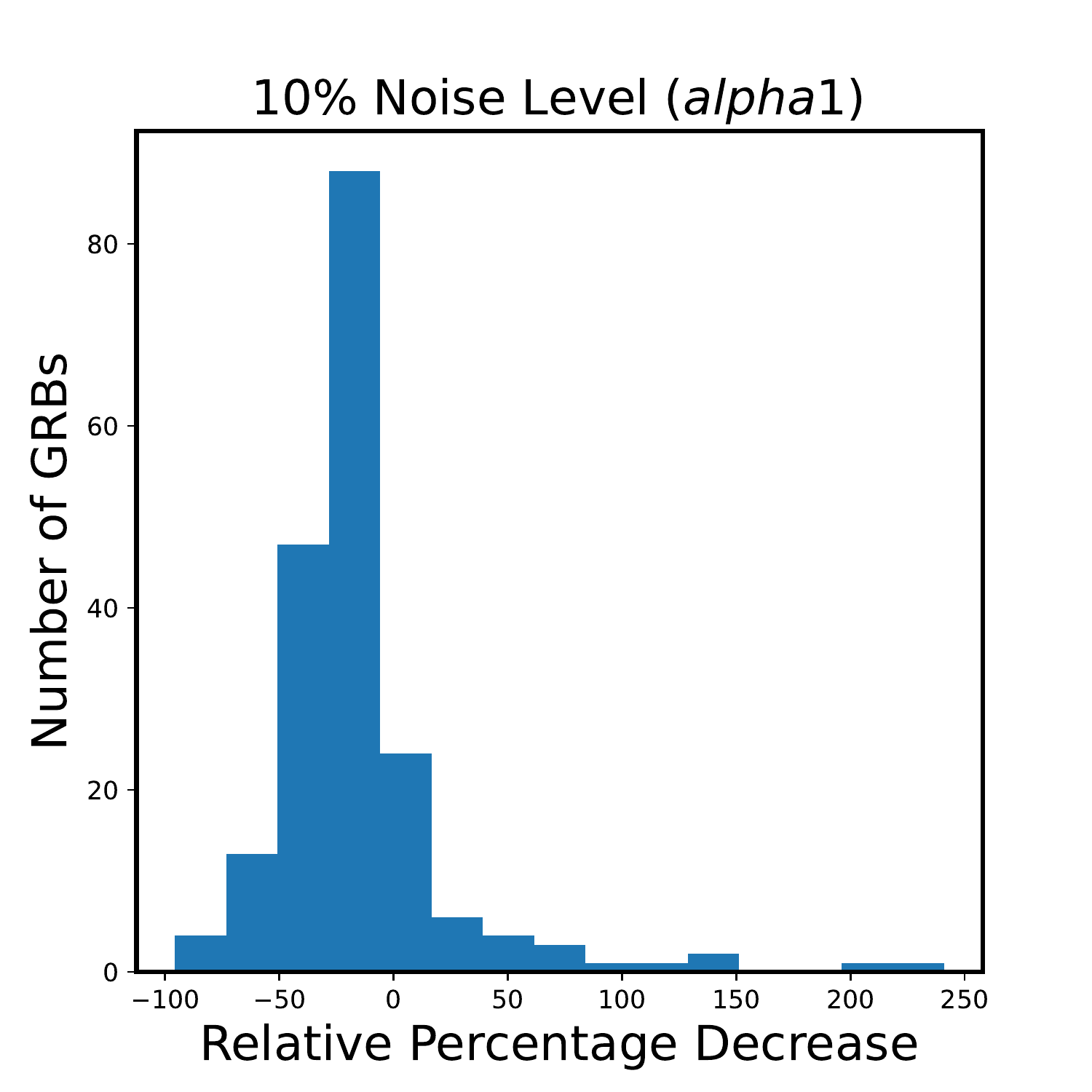}{0.22\textwidth}
{(c) Distribution of the 10\% noise level for the $\alpha_1$ with BPL }
\fig{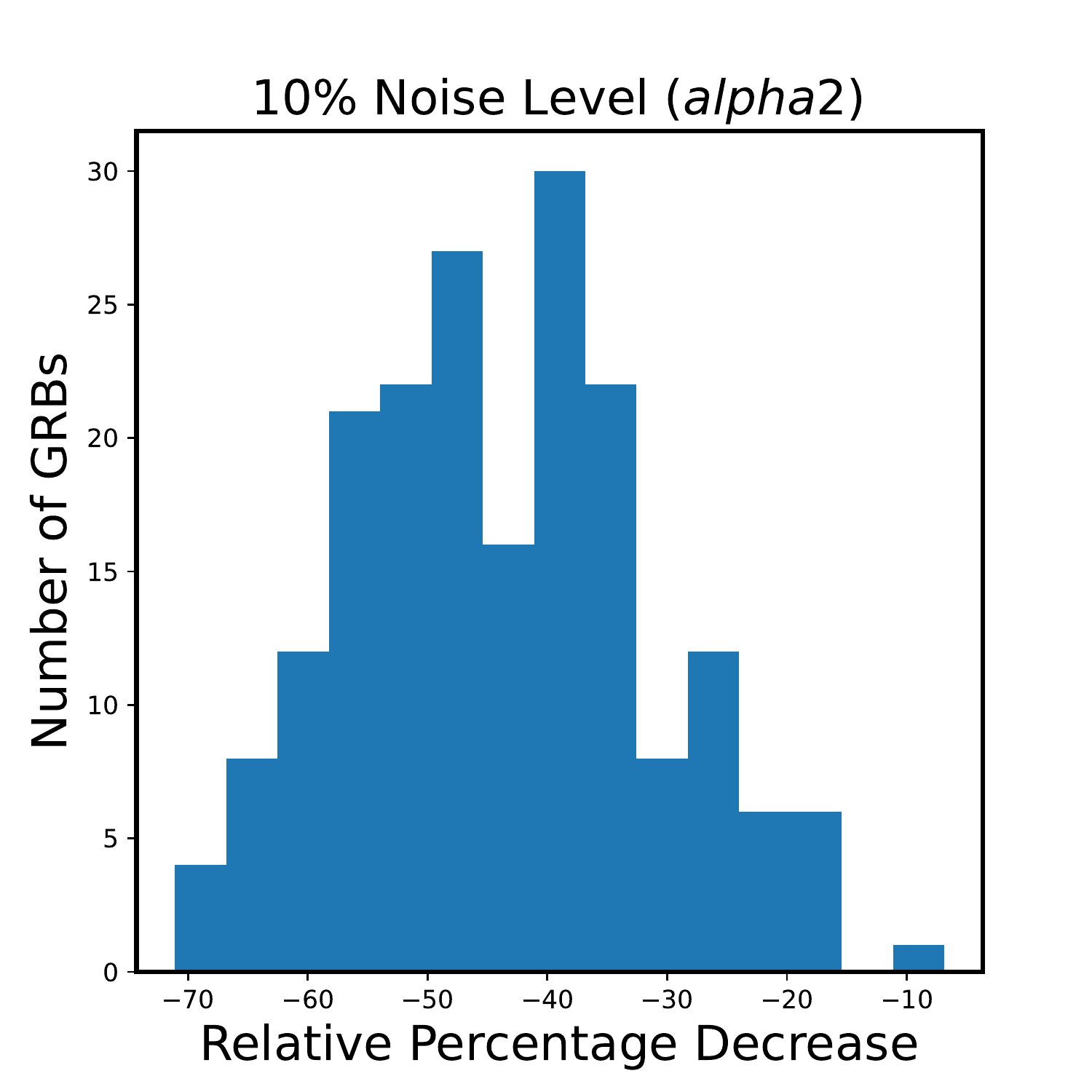}{0.22\textwidth}
{(d) Distribution of the 10\% noise level for the $\alpha_2$ with BPL}

\fig{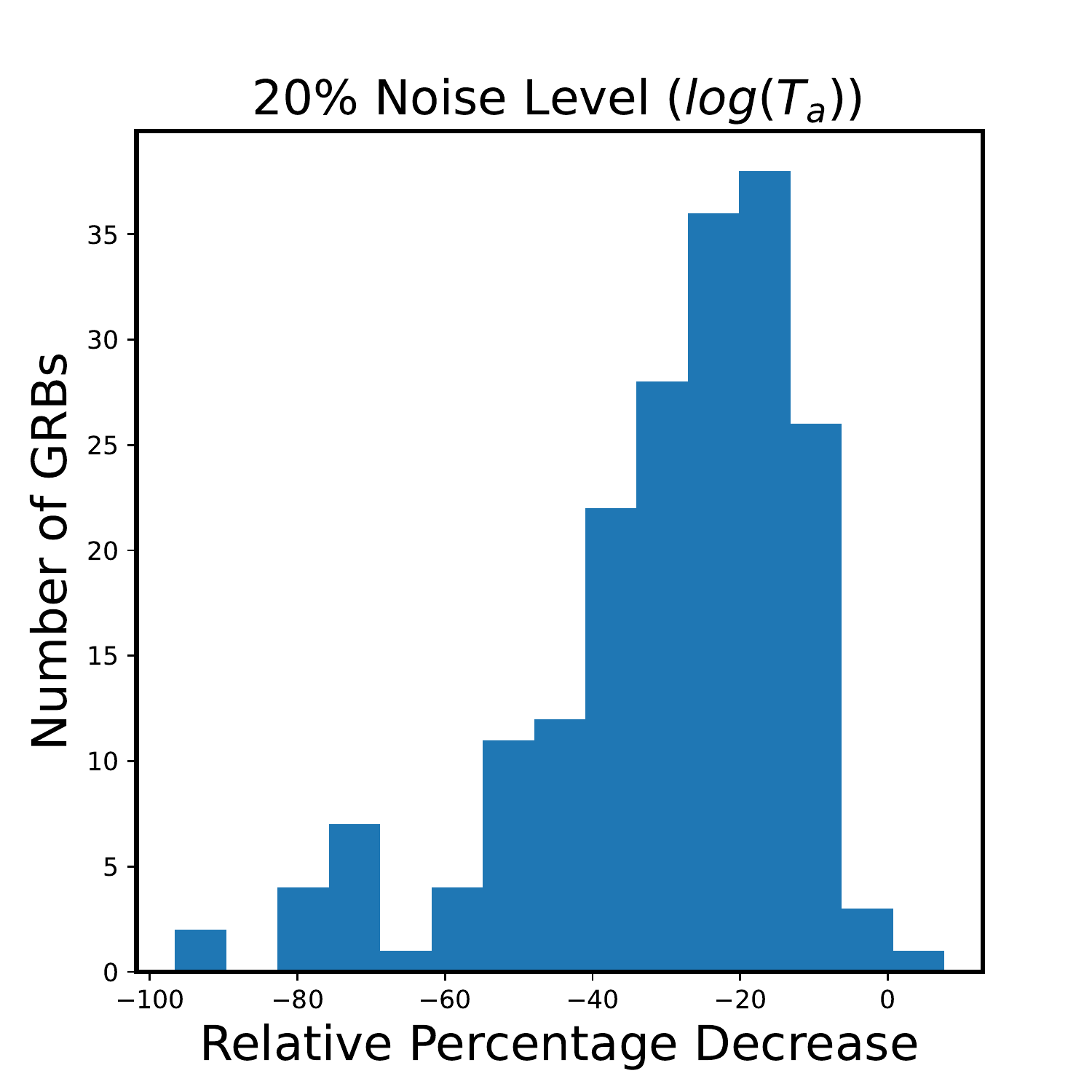}{0.22\textwidth}
{(e) Distribution of the 20\% noise level for the $\log(F_a)$ with BPL}
\fig{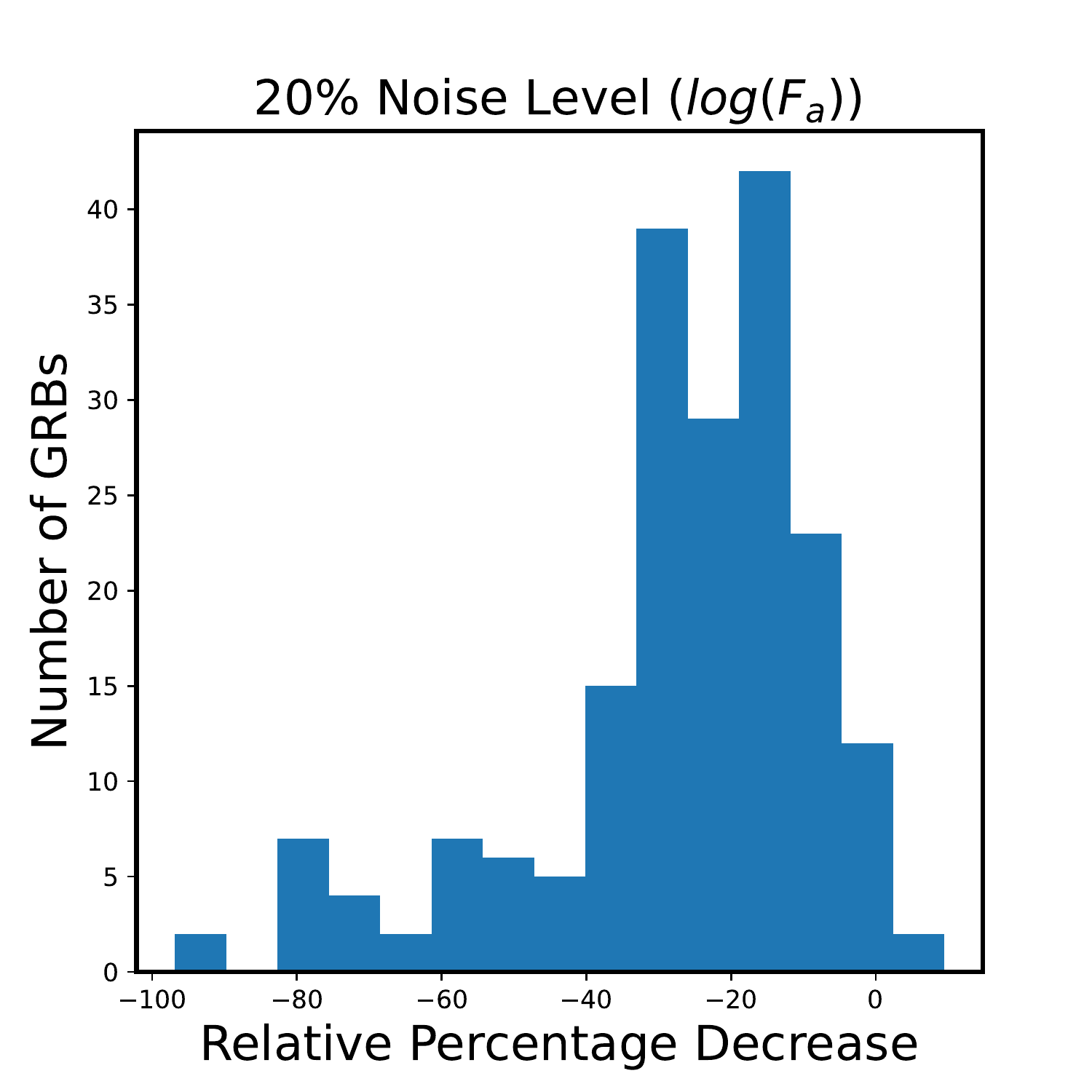}{0.22\textwidth}
{(f) Distribution of the 20\% noise level for the $\log(T_a)$ with BPL}
\fig{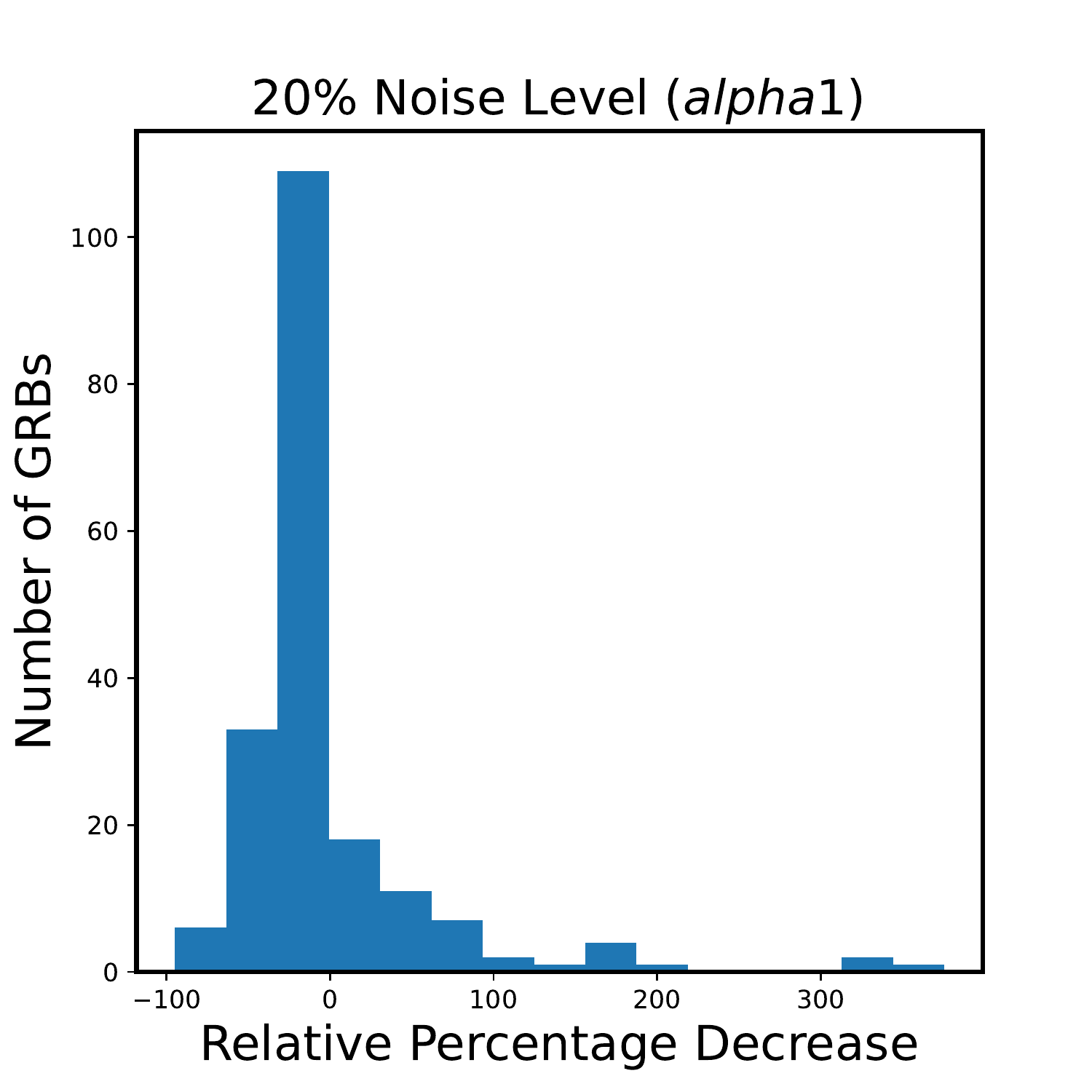}{0.22\textwidth}
{(g) Distribution of the 20\% noise level for the $\alpha_1$ with BPL}
\fig{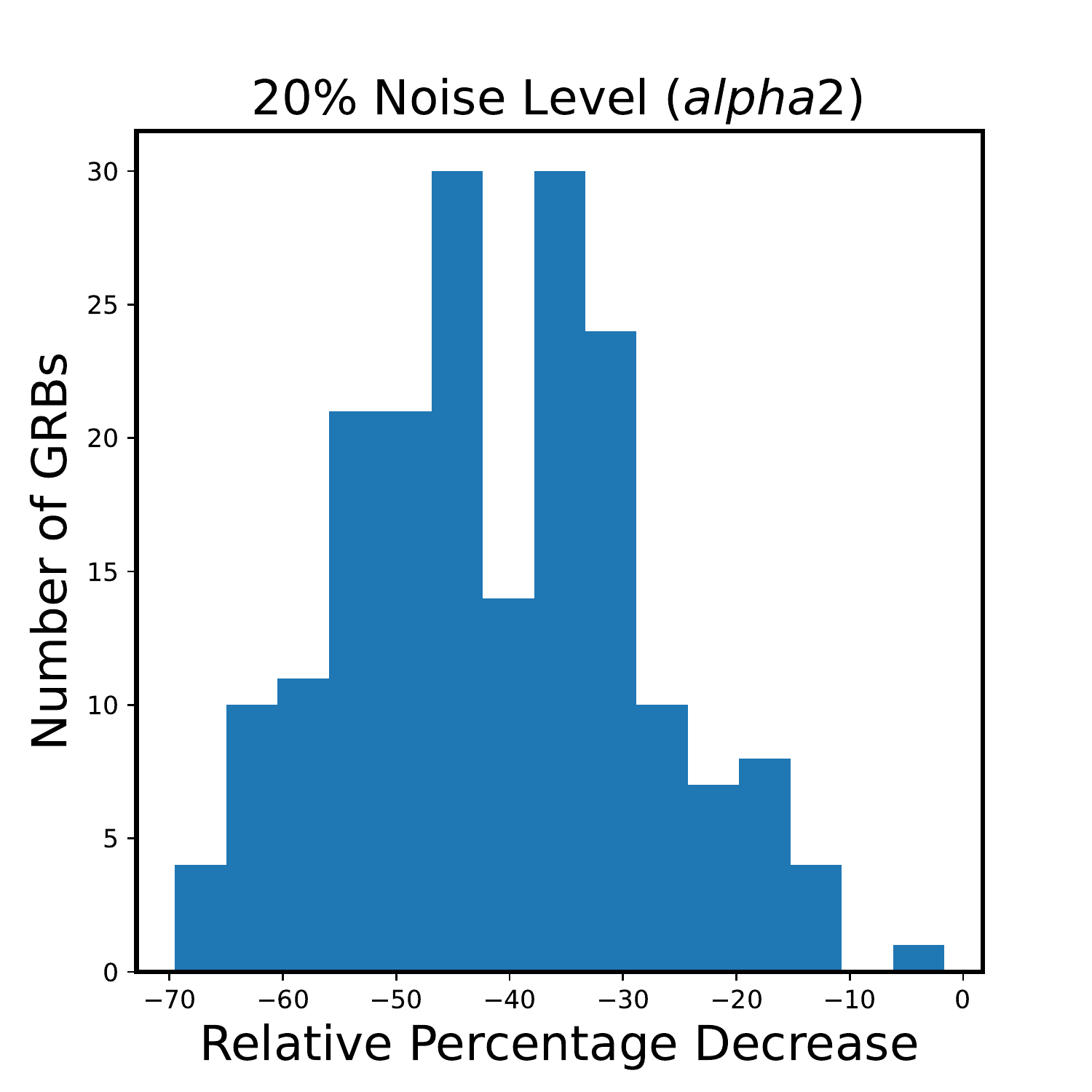}{0.22\textwidth}
{(h) Distribution of the 20\% noise level for the $\alpha_2$ with BPL}

\caption{Distribution of Relative Percentage Decrease for the parameters of the BPL fit when we apply the BPL for the reconstruction.}
    \label{fig:BPL_parameter_histo_BPL_RC}
\end{figure}

Overall, we see a similar trend to the W07 fitting. The results for the average change in error fraction for the BPL fitting are given in Table \ref{tab:table3}. The comparison between two noise levels for a sample GRB, GRB121217A, for 10\% and 20\% of the BPL fit is shown in the lower panels of Fig. \ref{fig:recon}. Figure \ref{fig:BPL_parameter_histo_BPL_RC} shows the distribution of the percentage decrease (upper panel shows the 10\% noise level, while the bottom panel shows the 20\% noise level) for all four of the BPL parameters. Again, the reconstruction leads to a decrease in the uncertainties on the BPL parameters for all the GRBs. The average percentage decrease on all parameters for both the W07 and BPL fits are summarized in Table \ref{tab:table6}.

We also check the differences between the Gaussian processes (a completely model-independent approach) and the BPL (a model-dependent approach) to assess the differences between these methods.


\subsection{Results from Gaussian process reconstruction}\label{sec:GPresults}

\begin{figure*}
\begin{center}
\includegraphics[width=.48\textwidth]{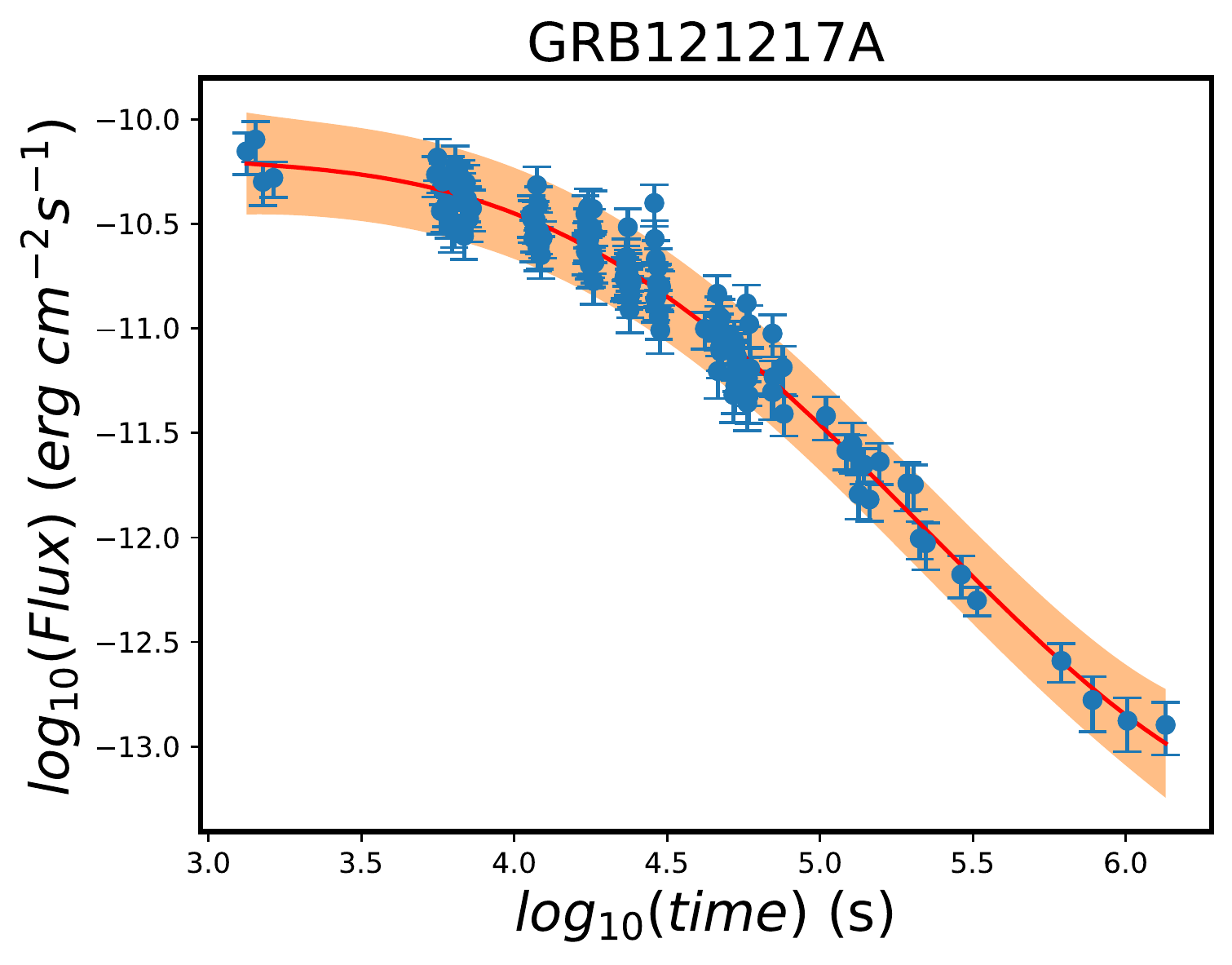}
\includegraphics[width=.48\textwidth]{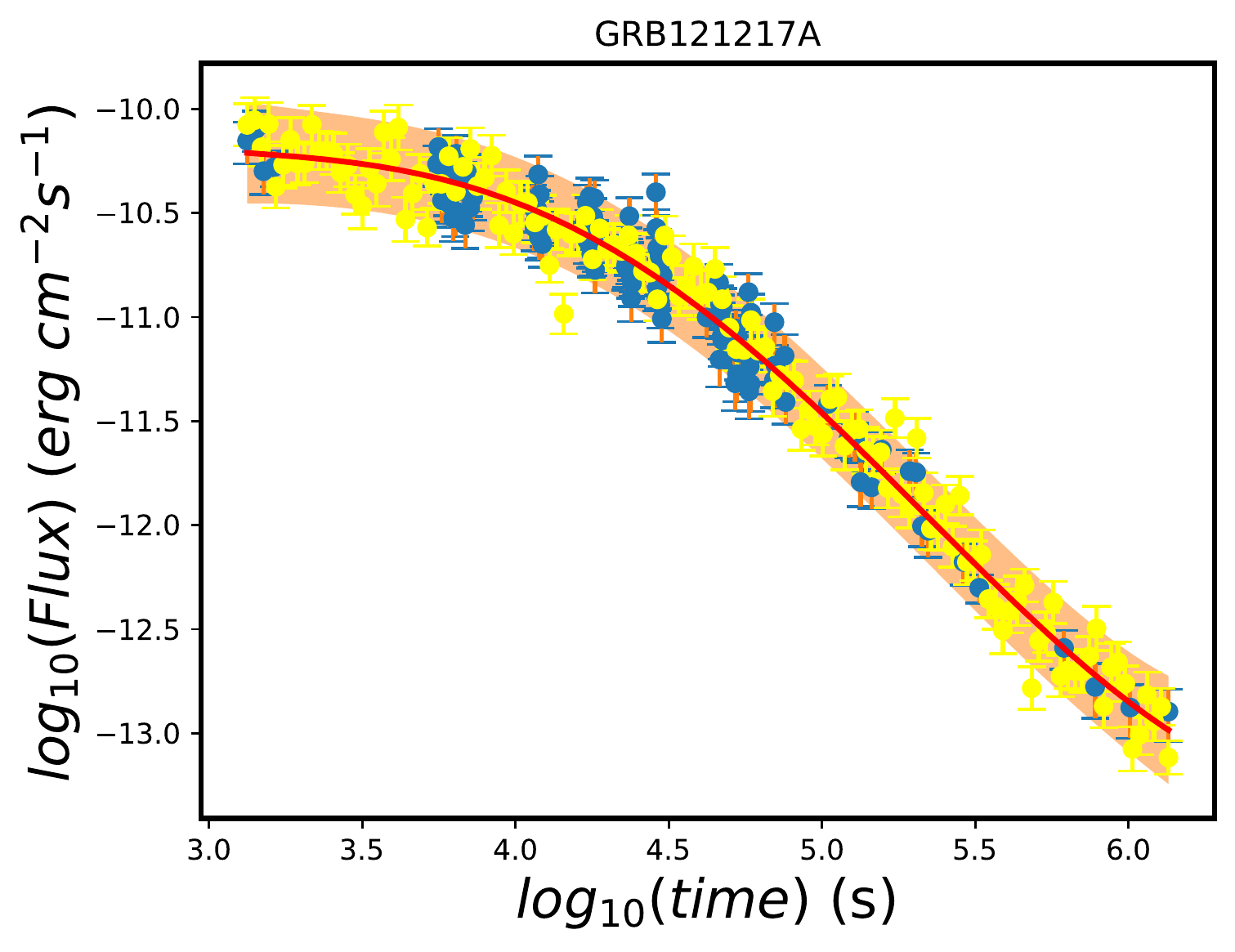} 
\end{center}

 \caption{Gaussian Process for the GRB 121217A. The left panel shows the Gaussian process fit, while the right panel shows the reconstructed LCs.} 
    \label{fig:GP}
\end{figure*}

Gaussian Process-based reconstruction provides a model-independent way of achieving our aim.
The reconstructed LC of GRB121217A is presented in Fig. \ref{fig:GP}. Following the reconstruction, the full LC is fitted with both the W07 and the BPL model to obtain new estimates of the parameters and their errors.

\begin{figure}
\fig{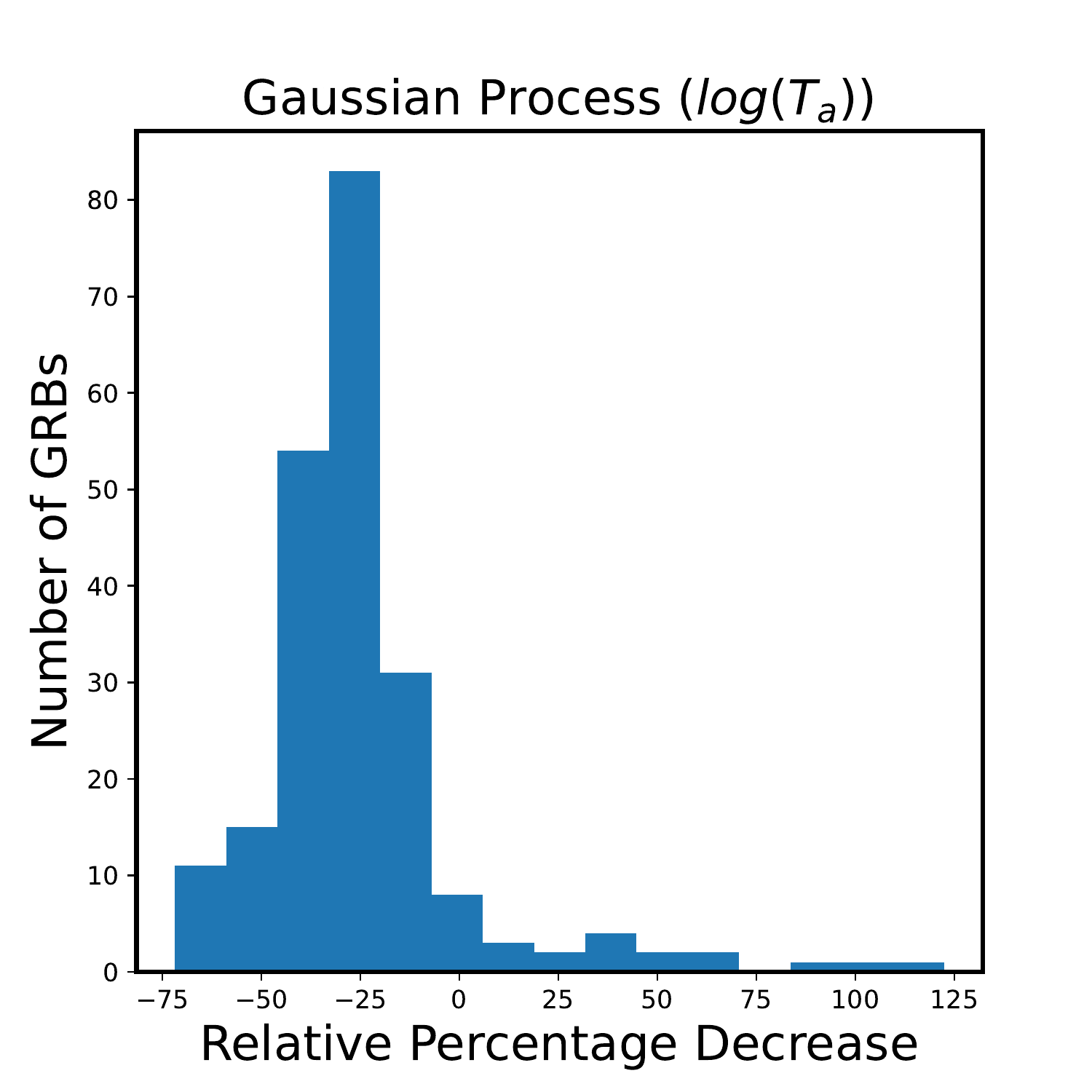}{0.27\textwidth}
{(a) Distribution of error fraction decrease for $\log F_a$.}
\fig{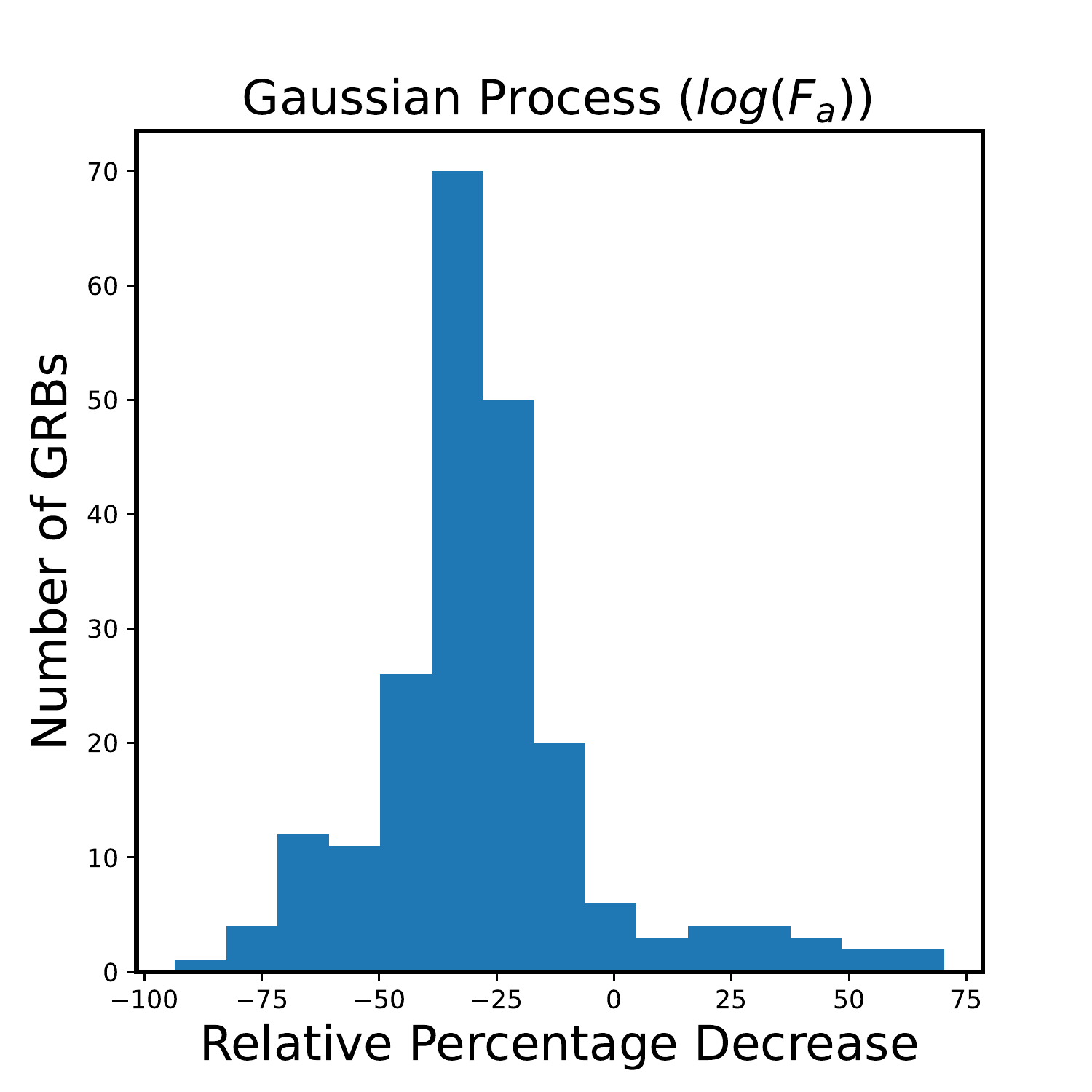}{0.27\textwidth}
{(b) Distribution of error fraction decrease for $\log T_a$.}
\fig{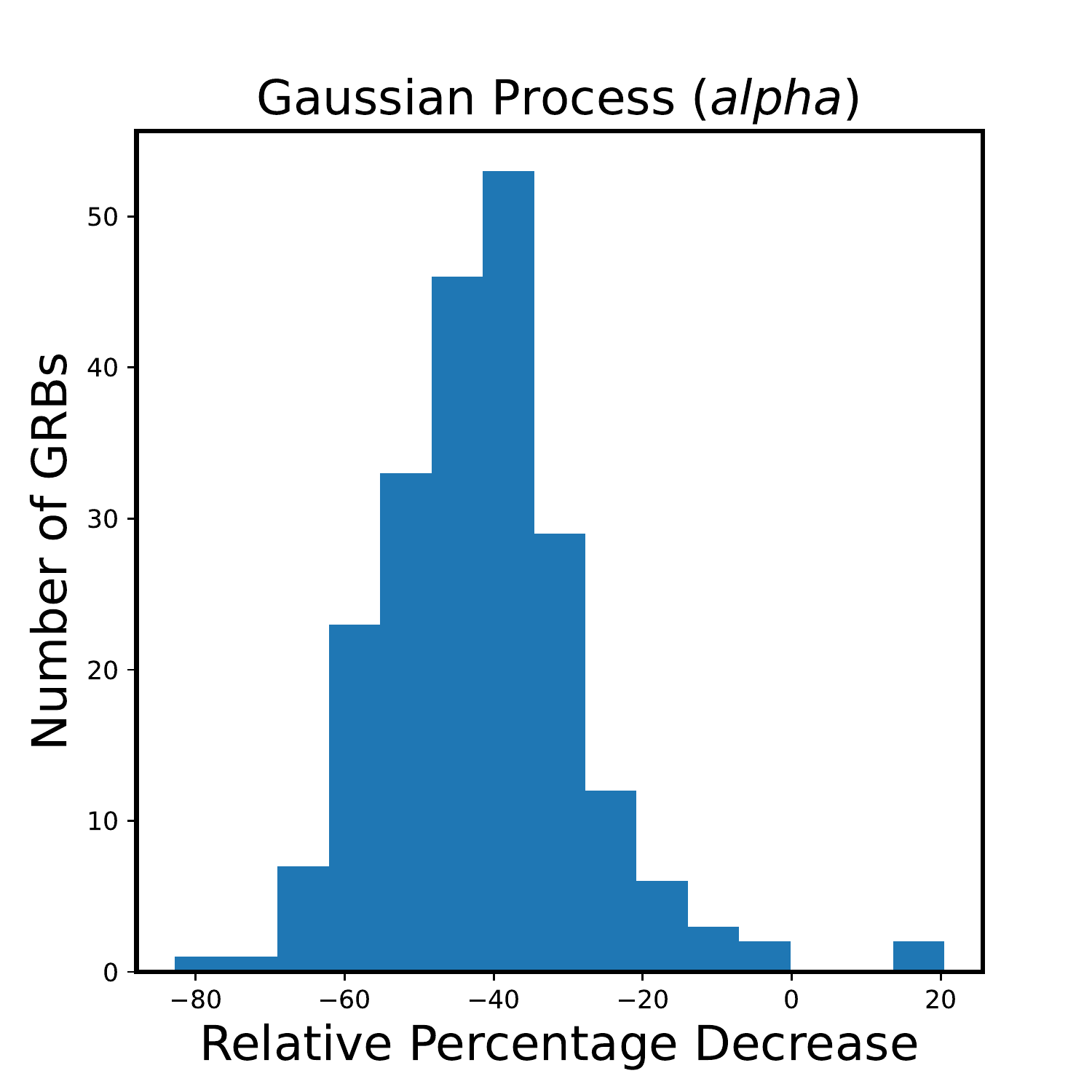}{0.27\textwidth}
{(c) Distribution of error fraction decrease for $\alpha_a$.}

\caption{Distribution of Relative Percentage Decrease for the parameters of the W07 function following Gaussian Process reconstruction.}

    \label{fig:GP_histo}
\end{figure}

For the parameter $\log(T_a)$ we observe an average decrease in the error fraction of 25\%. Similarly, for $\log(F_a)$ and $\alpha$ the average decreases in error fraction are 27.9\% and 41.5\%, respectively.
Note that in this case also the error fractions are calculated according to Eq. \ref{eqn8}.
The histogram distributions of Fig. \ref{fig:GP_histo} show the distribution of the relative percentage decrease of the three W07 parameters. The results of the GP for the W07 are given for each GRB in Table \ref{tab:table7}, while the average percentage decrease on the uncertainties of the parameters for all GRBs are presented in Table \ref{tab:table6}.

For the BPL fits, we see that for the $\alpha_1$ parameter, there is occasionally an increase in error fraction from the original parameters to the reconstructed parameters. This is because the slope of the LC before the break in the BPL fitting can be very flat, which will cause the uncertainty to be high compared to the $\alpha_1$ value and cause the error fraction to be $> 1$. Because of this, we remove the GRBs which exhibit a flat $\alpha_1$, or $|\alpha_1| < 0.1$. We see a percentage decrease in the parameters of 15.02\% for $\log(T_a)$, 11.91\% for $\log(F_a)$, 25.10\% for $\alpha_1$, and 35.92\% for $\alpha_2$. Again, the BPL fitting results show a similar trend to the W07 fits. The results of the GP for each GRB are given in Table \ref{tab:tableBPL}, while the average percentage decrease on the uncertainties of the parameters for all GRBs are given in Table \ref{tab:table6}.

\begin{figure}
\fig{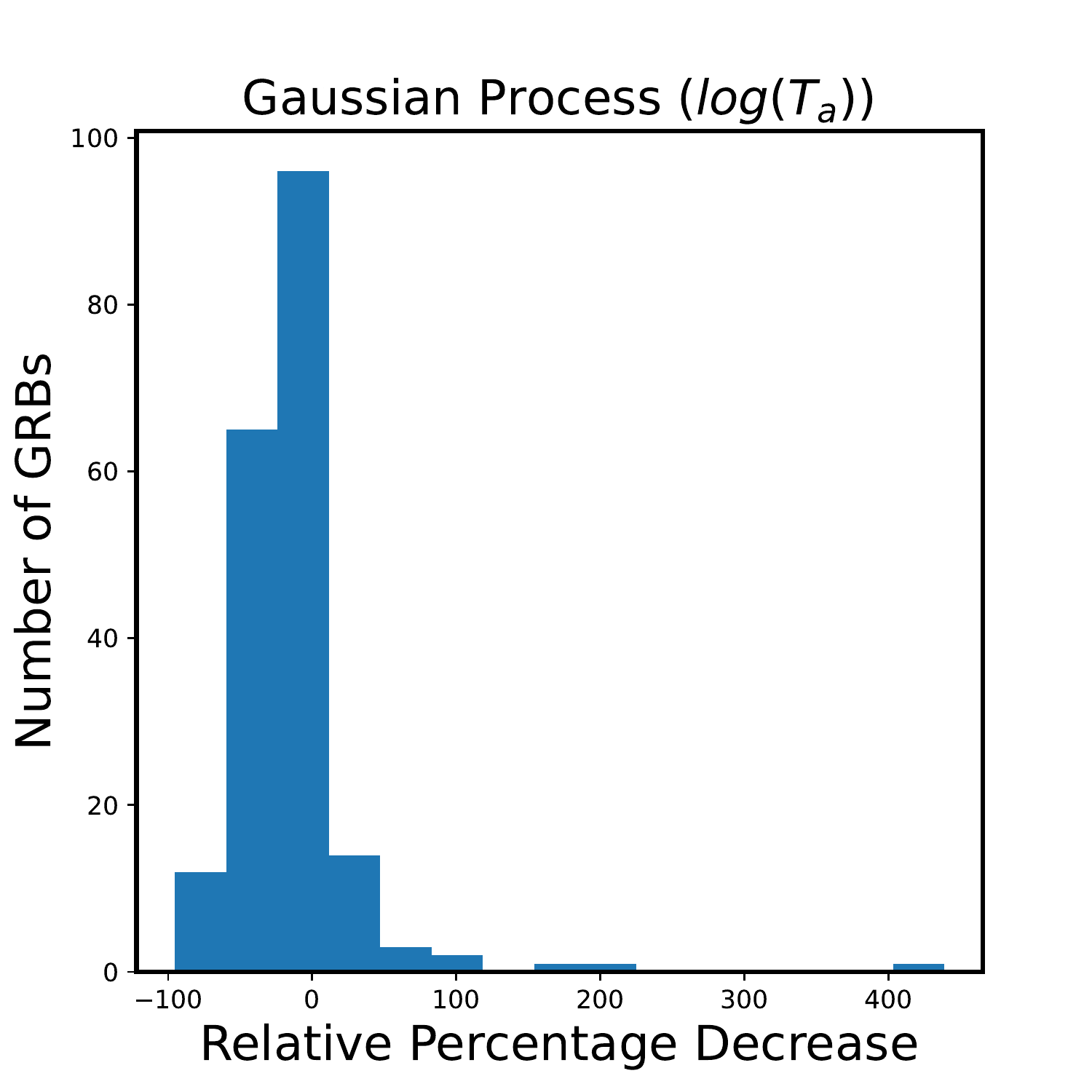}{0.22\textwidth}
{(a) Distribution of error fraction decrease for $\log(T_a)$}
\fig{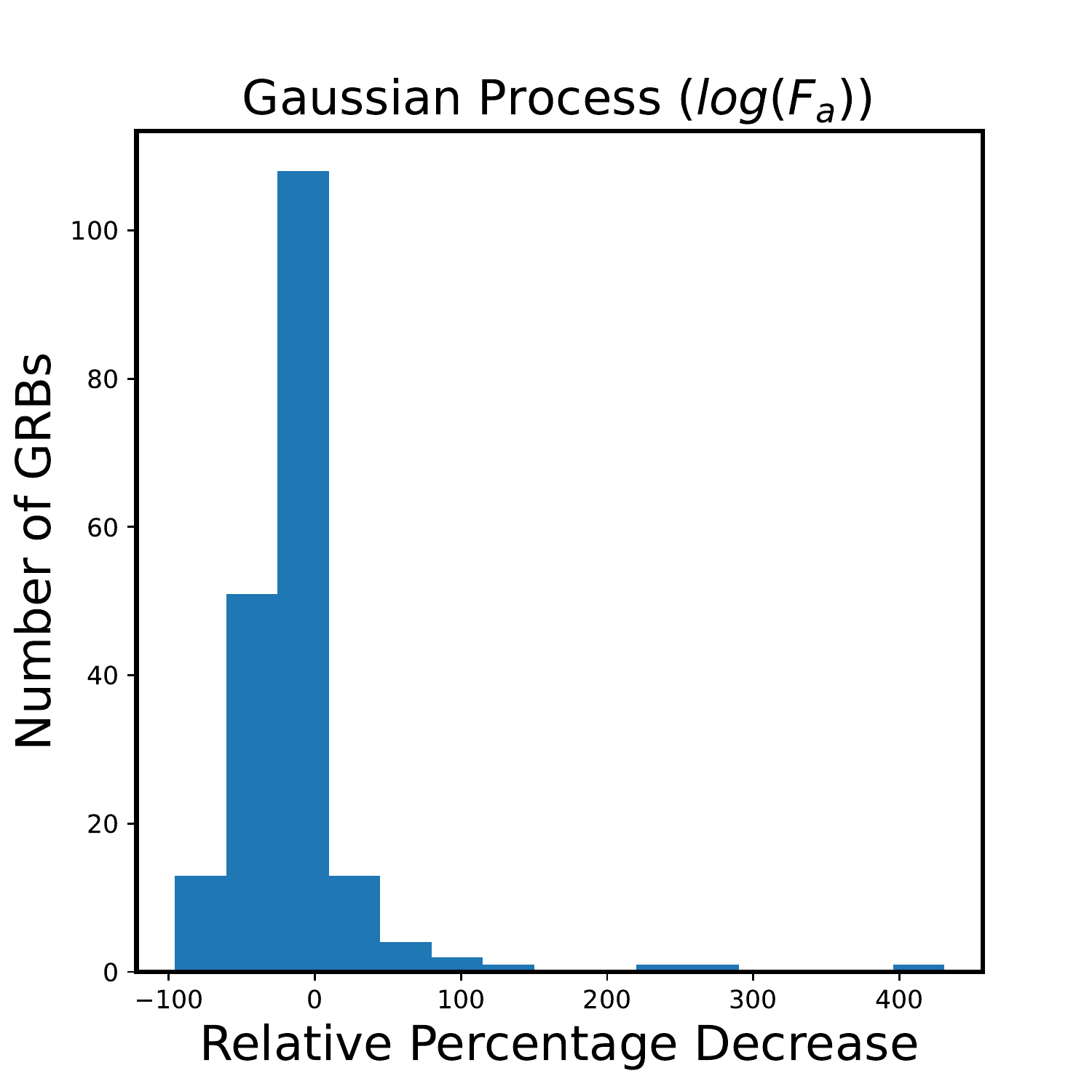}{0.22\textwidth}
{(b) Distribution of error fraction decrease for $\log(F_a)$}
\fig{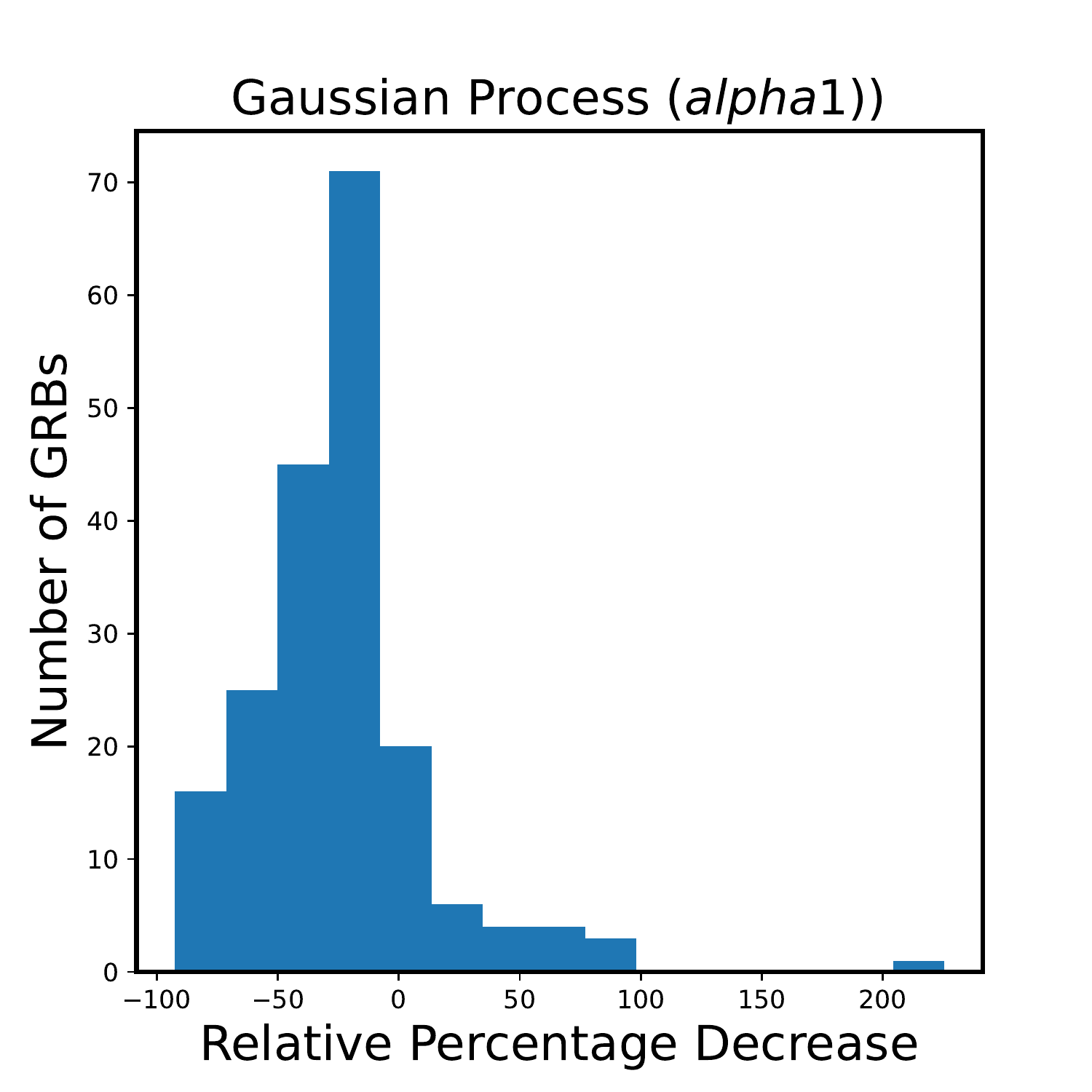}{0.22\textwidth}
{(c) Distribution of error fraction decrease for $\alpha_1$. }
\fig{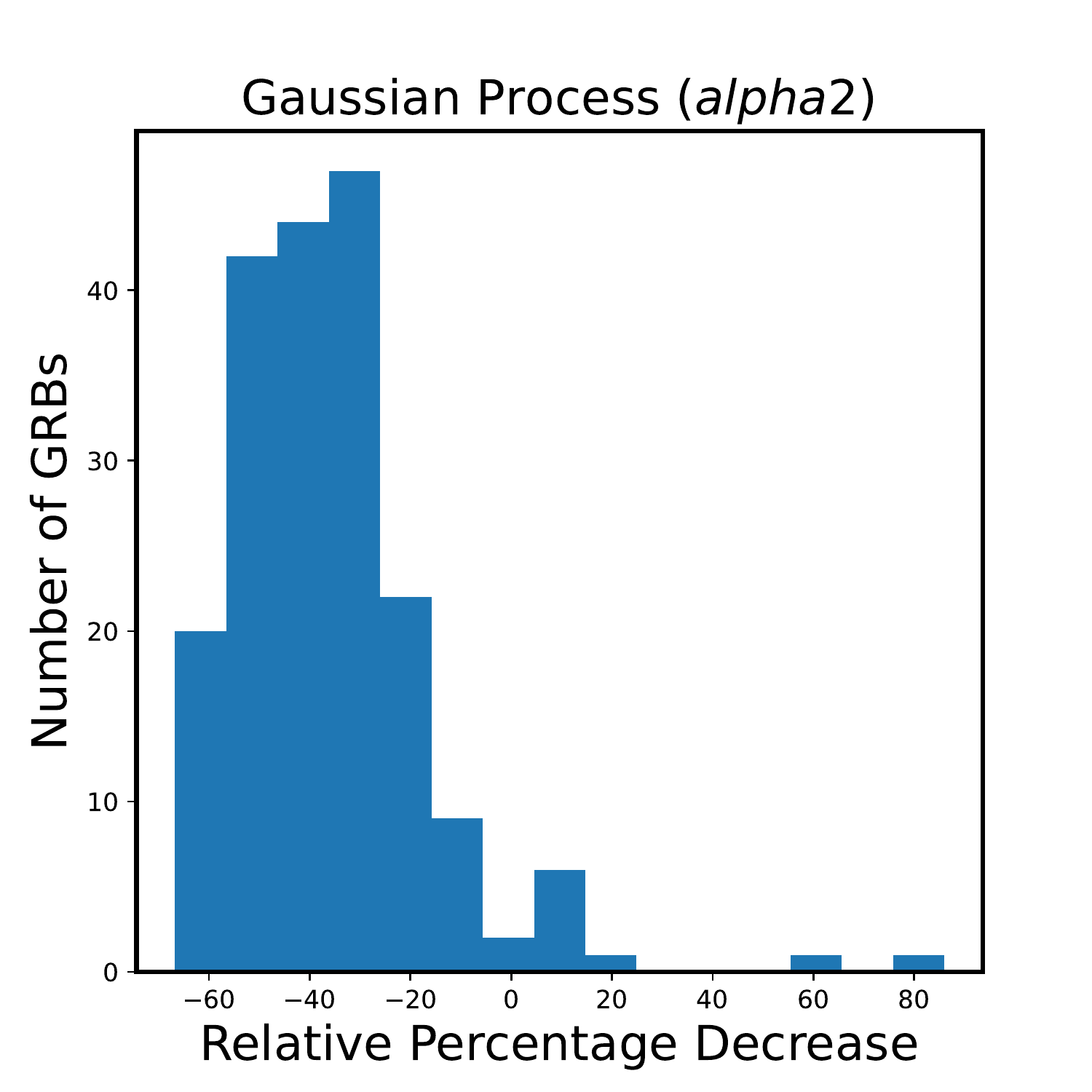}{0.22\textwidth}
{(d)  Distribution of error fraction decrease for $\alpha_2$.}

\caption{Distribution of Relative Percentage Decrease for the parameters of the BPL function following Gaussian Process
reconstruction.}
    \label{fig:BPL_GP_histogram}
\end{figure}

\section{Summary and Conclusion}\label{section:conclusion}

We propose a relatively simple methodology for GRB LCR for LCs with X-ray plateaus observed by Swift. 
This case study involves two functional form toy-models, the W07 and the BPL model and a limited GRB sample, as the main aim is to open the way to a new methodology that can be tested with more fitting functions and different statistics (e.g., Bayesian) in the future.

Our method assumes the W07 and the BPL models as the starting model for reconstruction, but the methodology presented here is versatile and can be extended to any other model (smoothly BPL, double BPL, etc).
Even models such as the pulse model, which includes the prompt emission, can be explored in the future. 
We also took a further step in the analysis adopting Gaussian processes with the aim of being completely model-independent.
We find that after testing our methodology on a sample of 218 Good GRBs, we obtain a reduction in the error bars of the plateau parameters of the W07 fitting - in this case, we have a reduction of 37.22\% on average for all parameters when the noise is set at the 10\% level, and a 33.77\% reduction, on average, for all parameters with a 20\% noise level. 
Regarding the BPL, we have a reduction of 30.69\% in average for the 10\% noise level and a reduction of 24.97\% for the 20\% noise level. In average for GP we have a decrease of 31.43\% for the W07 and 21.99\% for the BPL.
The reduction of uncertainties on plateau parameters can be crucial for many cosmological studies that benefit from precise measurements of these parameters, including the investigation of Population III stars or the use of GRBs as standard candles to study the early universe.

There are many advantages to the community from our study, as filling in the temporal gaps of observed LCs with a toy model solves the problem of the lack of data and thus allows us to use the reconstructed LC for:

\begin{itemize}
    \item discovering plateau features in GRB LCs, which otherwise may remain undetected,
    \item increasing the sample size for building GRB correlations among the plateau parameters,
    \item testing theoretical models which would use the reconstructed LCs, since the method is general and can be used with many different functions,
    \item cosmological studies to reduce the scatter on the cosmological parameters. This will be achieved using the enhanced GRB correlations involving the plateau emission built with LCR,
    \item estimating the redshift information with machine learning analysis which uses the new, more accurate estimates of the plateau parameters. This can allow us to determine the redshift of the high-z GRBs and thus further enhance the studies of Population III stars,
    \item classification of GRBs according to their morphology with increased accuracy.
\end{itemize}

As a final remark, this reconstruction is based on the LCs provided by Swift, but it can be extended to any current or future missions and across different wavelengths as well.

\begin{table*}
    \centering
    \begin{tabular}{|l|c|c|c|c|c|c|c|c|c|}
    \hline
        GRB ID& $EF_{\log_{10}(T_i)}$ & $EF_{\log_{10}(F_i)}$ & $EF_{\alpha_{i}}$ & $EF_{\log_{10}(T_i)}$ RC & $EF_{\log_{10}(F_i)}$ RC & $EF_{\alpha_{i}}$ RC & $\%_{\log_{10}(T_i)}$ & $\%_{\log_{10}(F_i)}$ & $\%_{\alpha_{i}}$  \\
        \hline
        
&&&&10\% noise & & & & & \\
\hline

050712 & 0.019 & 0.005 & 0.044 & 0.014 & 0.004 & 0.027 & -26.67 & -24.96 & -38.04
\\050318 & 0.011 & 0.006 & 0.046 & 0.008 & 0.004 & 0.033 & -22.91 & -24.05 & -28.62
\\050416A & 0.024 & 0.005 & 0.018 & 0.016 & 0.003 & 0.012 & -31.24 & -33.54 & -34.85
\\050607 & 0.021 & 0.005 & 0.044 & 0.016 & 0.004 & 0.027 & -22.62 & -22.79 & -39.18
\\050713A & 0.01 & 0.003 & 0.018 & 0.008 & 0.002 & 0.011 & -16.21 & -14.31 & -36.35
\\050822 & 0.011 & 0.003 & 0.026 & 0.007 & 0.002 & 0.015 & -31.06 & -35.46 & -43.07
\\050824 & 0.025 & 0.006 & 0.094 & 0.015 & 0.003 & 0.056 & -40.7 & -38.45 & -40.27
\\050826 & 0.029 & 0.019 & 0.131 & 0.026 & 0.016 & 0.196 & -9.54 & -15.21 & 49.99
\\050915B & 0.036 & 0.008 & 0.115 & 0.025 & 0.005 & 0.068 & -29.03 & -34.25 & -40.92
\\051016A & 0.033 & 0.006 & 0.051 & 0.021 & 0.004 & 0.024 & -36.11 & -28.49 & -53
\\051109A & 0.012 & 0.005 & 0.016 & 0.006 & 0.002 & 0.01 & -51.32 & -58.89 & -33.99
\\051221A & 0.02 & 0.005 & 0.051 & 0.015 & 0.004 & 0.033 & -27.39 & -28.33 & -34.6
\\060105 & 0.004 & 0.001 & 0.007 & 0.003 & 0.001 & 0.003 & -16.2 & -14.1 & -54.34
\\060108 & 0.024 & 0.006 & 0.071 & 0.018 & 0.004 & 0.047 & -26.39 & -30.58 & -33.59
\\060109 & 0.014 & 0.005 & 0.057 & 0.008 & 0.003 & 0.025 & -40.63 & -45.09 & -56.2
\\060124 & 0.008 & 0.004 & 0.012 & 0.006 & 0.003 & 0.007 & -28.27 & -29.34 & -42.7
\\060218 & 0.028 & 0.014 & 0.082 & 0.014 & 0.005 & 0.065 & -50.59 & -63 & -20.77
\\060306 & 0.012 & 0.003 & 0.024 & 0.009 & 0.002 & 0.017 & -22.76 & -28.7 & -29.23
\\060418 & 0.018 & 0.005 & 0.03 & 0.01 & 0.003 & 0.01 & -43.68 & -32.48 & -66.37
\\060421 & 0.041 & 0.01 & 0.087 & 0.022 & 0.006 & 0.039 & -46.03 & -40.58 & -55.27

\\
\hline
&&&&20\% noise & & & & & \\
\hline

050712 & 0.019 & 0.005 & 0.044 & 0.014 & 0.004 & 0.028 & -24.37 & -22.28 & -35.56
\\050318 & 0.011 & 0.006 & 0.046 & 0.009 & 0.004 & 0.034 & -18.58 & -19.74 & -24.97
\\050416A & 0.024 & 0.005 & 0.018 & 0.017 & 0.003 & 0.012 & -27.39 & -29.91 & -31.97
\\050607 & 0.021 & 0.005 & 0.044 & 0.017 & 0.004 & 0.028 & -17.04 & -17.37 & -36.04
\\050713A & 0.01 & 0.003 & 0.018 & 0.009 & 0.003 & 0.012 & -12.34 & -10.16 & -33.05
\\050822 & 0.011 & 0.003 & 0.026 & 0.008 & 0.002 & 0.015 & -27.43 & -32.13 & -40.11
\\050824 & 0.025 & 0.006 & 0.094 & 0.016 & 0.004 & 0.059 & -37.67 & -35.27 & -36.93
\\050826 & 0.029 & 0.019 & 0.131 & 0.025 & 0.014 & 0.172 & -13.33 & -28.04 & 31.46
\\050915B & 0.036 & 0.008 & 0.115 & 0.027 & 0.006 & 0.073 & -23.99 & -30.54 & -37.03
\\051016A & 0.033 & 0.006 & 0.051 & 0.022 & 0.004 & 0.025 & -32.78 & -24.82 & -50.61
\\051109A & 0.012 & 0.005 & 0.016 & 0.006 & 0.002 & 0.011 & -48.9 & -56.82 & -30.45
\\051221A & 0.02 & 0.005 & 0.051 & 0.016 & 0.004 & 0.035 & -23.51 & -24.37 & -31.17
\\060105 & 0.004 & 0.001 & 0.007 & 0.003 & 0.001 & 0.003 & -11.94 & -9.72 & -51.99
\\060108 & 0.024 & 0.006 & 0.071 & 0.018 & 0.004 & 0.049 & -23.82 & -27.76 & -30.82
\\060109 & 0.014 & 0.005 & 0.057 & 0.009 & 0.003 & 0.026 & -37.84 & -42.4 & -54.15
\\060124 & 0.008 & 0.004 & 0.012 & 0.006 & 0.003 & 0.007 & -23.7 & -24.55 & -40.13
\\060218 & 0.028 & 0.014 & 0.082 & 0.014 & 0.005 & 0.067 & -48.31 & -61.5 & -18.12
\\060306 & 0.012 & 0.003 & 0.024 & 0.01 & 0.002 & 0.018 & -20.13 & -25.98 & -26.29
\\060418 & 0.018 & 0.005 & 0.03 & 0.011 & 0.003 & 0.011 & -41.52 & -29.94 & -64.95
\\060421 & 0.041 & 0.01 & 0.087 & 0.024 & 0.006 & 0.041 & -41.96 & -36.37 & -52.87
\\

    \hline
    \end{tabular}
    \caption{The error fractions of $\log_{10}(T_{a})$, $
    \log_{10}(F_{a})$ and $\alpha_{a}$ before and after reconstruction (with relative percentage decrease in error for all three parameters) at 10\% noise ($n=0.1$; top) and 20\% noise ($n=0.2$; bottom). First three columns give error fraction for original W07 fit, while second three columns give error fraction for new W07 fit after reconstruction. Final three columns give percentage decrease in error fraction after reconstruction.  Full table is available online.}
    \label{tab:table2}
\end{table*}

\begin{sidewaystable*}
    \tiny
    \centering
    \begin{tabular}{|l|c|c|c|c|c|c|c|c|c|c|c|c|}
    \hline
        GRB ID& $EF_{\log_{10}(T_i)}$ & $EF_{\log_{10}(F_i)}$ & $EF_{\alpha_{1}}$ & $EF_{\alpha_{2}}$ & $EF_{\log_{10}(T_i)}$ RC & $EF_{\log_{10}(F_i)}$ RC & $EF_{\alpha_{1}}$ RC &  $EF_{\alpha_{2}}$ RC & $\%_{\log_{10}(T_i)}$ & $\%_{\log_{10}(F_i)}$ & $\%_{\alpha_{1}}$   & $\%_{\alpha_{2}}$\\
        \hline
        
&&&&10\% noise & & & & & \\
\hline 

GRB050712	&	0.031	&	0.031	&	0.155	&	0.056	&	0.022	&	0.008	&	0.158	&	0.033	&	-30.6	&	-28.48	&	2.15	&	-39.97	\\
GRB050318	&	0.011	&	-0.007	&	0.075	&	0.047	&	0.009	&	0.005	&	0.058	&	0.038	&	-20.09	&	-23.11	&	-23.14	&	-19.47	\\
GRB050416A	&	0.038	&	-0.01	&	0.078	&	0.023	&	0.029	&	0.007	&	0.056	&	0.013	&	-23.52	&	-22.5	&	-28.03	&	-40.54	\\
GRB050607	&	0.036	&	-0.011	&	0.15	&	0.053	&	0.03	&	0.01	&	0.166	&	0.03	&	-16.39	&	-11.78	&	10.48	&	-43.15	\\
GRB050713A	&	0.015	&	-0.006	&	0.043	&	0.019	&	0.012	&	0.004	&	0.039	&	0.013	&	-18.58	&	-18.89	&	-9.77	&	-34.12	\\
GRB050822	&	0.012	&	-0.004	&	0.223	&	0.024	&	0.01	&	0.004	&	0.164	&	0.014	&	-14.58	&	-14.2	&	-26.57	&	-42.26	\\
GRB050824	&	0.026	&	-0.006	&	0.276	&	0.105	&	0.015	&	0.004	&	0.223	&	0.057	&	-40.25	&	-30.63	&	-19.2	&	-46.1	\\
GRB050826	&	0.008	&	-0.004	&	0.328	&	0.064	&	0.005	&	0.002	&	0.226	&	0.039	&	-33.14	&	-35.02	&	-30.9	&	-38.46	\\
GRB050915B	&	0.021	&	-0.006	&	0.085	&	0.13	&	0.012	&	0.004	&	0.057	&	0.069	&	-43.28	&	-33.53	&	-32.4	&	-46.8	\\
GRB051016A	&	0.089	&	-0.025	&	0.327	&	0.048	&	0.047	&	0.013	&	0.2	&	0.024	&	-47.27	&	-48.27	&	-38.99	&	-49.86	\\
GRB051109A	&	0.011	&	-0.005	&	0.136	&	0.015	&	0.009	&	0.003	&	0.049	&	0.01	&	-21.7	&	-24.17	&	-63.81	&	-31.43	\\
GRB051221A	&	0.021	&	-0.007	&	0.082	&	0.058	&	0.016	&	0.006	&	0.067	&	0.038	&	-23.99	&	-23.82	&	-18.86	&	-34.93	\\
GRB060105	&	0.007	&	-0.002	&	0.011	&	0.011	&	0.003	&	0.002	&	0.011	&	0.003	&	-51.9	&	-34.29	&	8.85	&	-68.73	\\
GRB060108	&	0.026	&	-0.008	&	0.147	&	0.075	&	0.019	&	0.006	&	0.113	&	0.048	&	-26.83	&	-25.54	&	-23.34	&	-36.11	\\
GRB060109	&	0.01	&	-0.003	&	0.347	&	0.047	&	0.007	&	0.003	&	0.271	&	0.022	&	-26.18	&	-9.85	&	-22.03	&	-54.26	\\
GRB060124	&	0.014	&	-0.008	&	0.034	&	0.019	&	0.011	&	0.007	&	0.027	&	0.01	&	-17.7	&	-16.08	&	-21.58	&	-47.95	\\
GRB060218	&	0.013	&	-0.005	&	0.582	&	0.064	&	0.01	&	0.004	&	0.811	&	0.042	&	-22.2	&	-21.99	&	39.27	&	-34.22	\\
GRB060306	&	0.075	&	-0.026	&	0.395	&	0.024	&	0.018	&	0.006	&	0.107	&	0.018	&	-76.3	&	-78.27	&	-72.97	&	-24.73	\\
GRB060418	&	0.066	&	-0.034	&	0.11	&	0.029	&	0.029	&	0.014	&	0.054	&	0.011	&	-56.32	&	-57.75	&	-51.17	&	-64.14	\\
GRB060421	&	0.092	&	-0.038	&	0.557	&	0.082	&	0.04	&	0.015	&	0.249	&	0.041	&	-56.87	&	-59.92	&	-55.33	&	-50.46	\\

\hline
&&&&20\% noise & & & & & \\
\hline
GRB050712	&	0.031	&	-0.011	&	0.155	&	0.056	&	0.022	&	0.008	&	0.161	&	0.035	&	-28.73	&	-25.69	&	3.85	&	-37.46	\\																								
GRB050318	&	0.011	&	-0.007	&	0.075	&	0.047	&	0.009	&	0.006	&	0.061	&	0.039	&	-16.43	&	-19.69	&	-19.51	&	-17.34	\\																								
GRB050416A	&	0.038	&	-0.01	&	0.078	&	0.023	&	0.03	&	0.008	&	0.06	&	0.014	&	-20.36	&	-19.71	&	-23.32	&	-38.02	\\																								
GRB050607	&	0.036	&	-0.011	&	0.15	&	0.053	&	0.031	&	0.01	&	0.175	&	0.032	&	-13.92	&	-9.42	&	16.3	&	-40.4	\\																								
GRB050713A	&	0.015	&	-0.006	&	0.043	&	0.019	&	0.012	&	0.005	&	0.04	&	0.013	&	-15.14	&	-15.83	&	-7.65	&	-30.87	\\																								
GRB050822	&	0.012	&	-0.004	&	0.223	&	0.024	&	0.011	&	0.004	&	0.176	&	0.015	&	-10.2	&	-9.64	&	-21.26	&	-39.39	\\																								
GRB050824	&	0.026	&	-0.006	&	0.276	&	0.105	&	0.016	&	0.005	&	0.243	&	0.06	&	-36.94	&	-26.72	&	-11.73	&	-43.25	\\																								
GRB050826	&	0.008	&	-0.004	&	0.328	&	0.064	&	0.005	&	0.003	&	0.228	&	0.041	&	-29.66	&	-32.24	&	-30.38	&	-35.41	\\																								
GRB050915B	&	0.021	&	-0.006	&	0.085	&	0.13	&	0.012	&	0.004	&	0.059	&	0.071	&	-42.53	&	-31.23	&	-30.44	&	-45.47	\\																								
GRB051016A	&	0.089	&	-0.025	&	0.327	&	0.048	&	0.049	&	0.013	&	0.199	&	0.025	&	-44.35	&	-45.47	&	-39.16	&	-48.04	\\																								
GRB051109A	&	0.011	&	-0.005	&	0.136	&	0.015	&	0.009	&	0.004	&	0.051	&	0.011	&	-18.25	&	-20.91	&	-62.47	&	-28.61	\\																								
GRB051221A	&	0.021	&	-0.007	&	0.082	&	0.058	&	0.017	&	0.006	&	0.071	&	0.04	&	-19.03	&	-18.86	&	-14.41	&	-31.45	\\																								
GRB060105	&	0.007	&	-0.002	&	0.011	&	0.011	&	0.004	&	0.002	&	0.012	&	0.004	&	-49.46	&	-30.98	&	14.36	&	-67.13	\\																								
GRB060108	&	0.026	&	-0.008	&	0.147	&	0.075	&	0.02	&	0.006	&	0.12	&	0.05	&	-24.59	&	-22.87	&	-18.53	&	-33.57	\\																								
GRB060109	&	0.01	&	-0.003	&	0.347	&	0.047	&	0.008	&	0.003	&	0.291	&	0.023	&	-22.04	&	-4.53	&	-16.26	&	-51.69	\\																								
GRB060124	&	0.014	&	-0.008	&	0.034	&	0.019	&	0.012	&	0.007	&	0.029	&	0.01	&	-13.29	&	-11.75	&	-15.88	&	-45.85	\\																								
GRB060218	&	0.013	&	-0.005	&	0.582	&	0.064	&	0.01	&	0.004	&	0.876	&	0.045	&	-17.72	&	-17.19	&	50.51	&	-30.2	\\																								
GRB060306	&	0.075	&	-0.026	&	0.395	&	0.024	&	0.018	&	0.006	&	0.108	&	0.019	&	-75.59	&	-77.65	&	-72.61	&	-20.71	\\																								
GRB060418	&	0.066	&	-0.034	&	0.11	&	0.029	&	0.03	&	0.015	&	0.056	&	0.011	&	-53.89	&	-55.46	&	-49.07	&	-62.74	\\																								
GRB060421	&	0.092	&	-0.038	&	0.557	&	0.082	&	0.041	&	0.016	&	0.262	&	0.041	&	-55.29	&	-58.42	&	-53.06	&	-50.02	\\

    \hline
    \end{tabular}
    \caption{The error fractions of $\log_{10}(T_{a})$, $
    \log_{10}(F_{a})$, $\alpha_{1}$ and $\alpha_{2}$ before and after reconstruction (with relative percentage decrease in error for all three parameters) at 10\% noise ($n=0.1$; top) and 20\% noise ($n=0.2$; bottom). First three columns give error fraction for original BPL fit, while second three columns give error fraction for new W07 fit after reconstruction. Final three columns give percentage decrease in error fraction after reconstruction.  Full table is available online.}
    \label{tab:table3}
\end{sidewaystable*}

\begin{table*}
    \centering
    \begin{tabular}{|l|c|c|c|c|c|}
    \hline
       \textbf{Reconstruction process} & $\%^{avg}_{\log_{10}(T_a)}$ & $\%^{avg}_{\log_{10}(F_a)}$ & $\%^{avg}_{\alpha_{a}}$ & $\%^{avg}_{\alpha_{1}}$ & $\%^{avg}_{\alpha_{2}}$\\
       \hline
W07 reconstruction (10\%) & -33.33 & -35.03 & -43.32 & - & -
\\ W07 reconstruction (20\%)  & -29.49 & -31.24 & -40.57 & - & -
\\
\hline 

BPL reconstruction (10\%) & -33.3 & -30.78 & - & -14.76 & -43.9 \\
BPL reconstruction (20\%) & -29.88 & -27.2 & - & -1.7 & -41.1 \\

    \hline
Gaussian process (W07) & -24.9 & -27.9 & -41.5 & - & -\\
\hline

Gaussian process (BPL) & -15.02 & -11.91 & - & -25.10 & -35.92\\ 
\hline
    \end{tabular}
    \caption{The average percentage decrease in the error fractions for the various parameters, following the reconstruction.
    }
    \label{tab:table6}
\end{table*}

\begin{table*}
    \centering
    \begin{tabular}{|l|c|c|c|c|c|c|c|c|c|}
    \hline
        GRB ID& $EF_{\log_{10}(T_i)}$ & $EF_{\log_{10}(F_i)}$ & $EF_{\alpha_{i}}$ & $EF_{\log_{10}(T_i)}$ RC & $EF_{\log_{10}(F_i)}$ RC & $EF_{\alpha_{i}}$ RC & $\%_{\log_{10}(T_i)}$ & $\%_{\log_{10}(F_i)}$ & $\%_{\alpha_{i}}$  \\
        \hline

        &&&& GP (W07) & & & & & \\
\hline

050712 & 0.019 & 0.005 & 0.044 & 0.012 & 0.003 & 0.028 & -34.01 & -31.09 & -36.97
\\050318 & 0.011 & 0.006 & 0.046 & 0.008 & 0.004 & 0.031 & -26.66 & -27.68 & -32.19
\\050416A & 0.024 & 0.005 & 0.018 & 0.016 & 0.003 & 0.011 & -32.44 & -34.04 & -38.46
\\050607 & 0.021 & 0.005 & 0.044 & 0.017 & 0.004 & 0.028 & -18.58 & -19.62 & -36.98
\\050713A & 0.01 & 0.003 & 0.018 & 0.008 & 0.002 & 0.011 & -18.81 & -15.29 & -35.28
\\050822 & 0.011 & 0.003 & 0.026 & 0.008 & 0.002 & 0.015 & -29.98 & -33.14 & -43.56
\\050824 & 0.025 & 0.006 & 0.094 & 0.015 & 0.003 & 0.056 & -39.2 & -37.19 & -40.83
\\050826 & 0.029 & 0.019 & 0.131 & 0.02 & 0.012 & 0.156 & -32.22 & -36.64 & 19.63
\\050915B & 0.036 & 0.008 & 0.115 & 0.024 & 0.005 & 0.068 & -34.36 & -37.46 & -41.29
\\051016A & 0.033 & 0.006 & 0.051 & 0.024 & 0.005 & 0.025 & -25.01 & -17.02 & -51.08
\\051109A & 0.012 & 0.005 & 0.016 & 0.007 & 0.002 & 0.009 & -41.5 & -49.53 & -40.98
\\051221A & 0.02 & 0.005 & 0.051 & 0.013 & 0.004 & 0.033 & -33.79 & -32.68 & -34.76
\\060105 & 0.004 & 0.001 & 0.007 & 0.003 & 0.001 & 0.003 & -23.04 & -20.31 & -52.2
\\060108 & 0.024 & 0.006 & 0.071 & 0.018 & 0.004 & 0.047 & -23.29 & -28.84 & -33.57
\\060109 & 0.014 & 0.005 & 0.057 & 0.008 & 0.003 & 0.025 & -40.9 & -44.23 & -55.74
\\060124 & 0.008 & 0.004 & 0.012 & 0.005 & 0.003 & 0.007 & -34.32 & -34.36 & -45.13
\\060218 & 0.028 & 0.014 & 0.082 & 0.013 & 0.005 & 0.062 & -54.56 & -65.62 & -24.38
\\060306 & 0.012 & 0.003 & 0.024 & 0.01 & 0.002 & 0.018 & -19.95 & -26.66 & -26.92
\\060418 & 0.018 & 0.005 & 0.03 & 0.009 & 0.003 & 0.01 & -48.49 & -37.4 & -67.46
\\060421 & 0.041 & 0.01 & 0.087 & 0.032 & 0.008 & 0.046 & -21.65 & -20.75 & -46.81

\\
\hline

    \end{tabular}
    \caption{The W07 error fractions of $\log_{10}(T_{a})$, $
    \log_{10}(F_{a})$ and $\alpha_{a}$ before and after performing the Gaussian Process (GP) reconstruction (with relative percentage decrease in error for all three parameters). First three columns give error fraction for original W07 fit, while second three columns give error fraction for new W07 fit after GP reconstruction. Final three columns give percentage decrease in error fraction after GP reconstruction.  Full table is available online.}
    \label{tab:table7}
\end{table*}

    
\begin{longrotatetable}

\begin{deluxetable*}{|l|c|c|c|c|c|c|c|c|c|c|c|c|}


\tabletypesize{\small}


\tablecaption{Data for the BPL fit of the GP-based reconstruction}


\tablehead{
\multicolumn{13}{c}{GP (BPL)} \\
\colhead{GRB ID} & \colhead{$EF_{\log_{10}(T_i)}$} & \colhead{$EF_{\log_{10}(F_i)}$} & \colhead{$EF_{\alpha_{1}}$} & \colhead{$EF_{\alpha_{2}}$} & \colhead{$EF_{\log_{10}(T_i)}$ RC} & \colhead{$EF_{\log_{10}(F_i)}$ RC} & \colhead{$EF_{\alpha_{1}}$ RC} & \colhead{$EF_{\alpha_{2}}$ RC} & \colhead{$\%_{\log_{10}(T_i)}$} & \colhead{$\%_{\log_{10}(F_i)}$} & \colhead{$\%_{\alpha_{1}}$} & \colhead{$\%_{\alpha_{2}}$} \\ 
\colhead{} & \colhead{} & \colhead{} & \colhead{} & \colhead{} & \colhead{} & \colhead{} & \colhead{} & \colhead{} & \colhead{} & \colhead{} & \colhead{} & \colhead{} } 

\startdata
050712 & 0.031 & -0.011 & 0.155 & 0.056 & 0.02 & 0.008 & 0.114 & 0.035 & -36.96 & -31.16 & -26.56 & -36.7 \\
050318 & 0.011 & -0.007 & 0.075 & 0.047 & 0.009 & 0.005 & 0.057 & 0.037 & -21.44 & -24.84 & -25.07 & -22.08 \\
050416A & 0.038 & -0.01 & 0.078 & 0.023 & 0.03 & 0.008 & 0.054 & 0.013 & -22.17 & -20.55 & -30.49 & -41.1 \\
050607 & 0.036 & -0.011 & 0.15 & 0.053 & 0.033 & 0.01 & 0.265 & 0.03 & -9.03 & -6.18 & 76.48 & -44.07 \\
050713A & 0.015 & -0.006 & 0.043 & 0.019 & 0.012 & 0.005 & 0.038 & 0.013 & -19.09 & -17.92 & -10.82 & -33.81 \\
050822 & 0.012 & -0.004 & 0.223 & 0.024 & 0.011 & 0.004 & 0.126 & 0.014 & -13.12 & -10.96 & -43.37 & -40.74 \\
050824 & 0.026 & -0.006 & 0.276 & 0.105 & 0.016 & 0.004 & 0.257 & 0.056 & -36.37 & -29.54 & -6.94 & -47.22 \\
050826 & 0.008 & -0.004 & 0.328 & 0.064 & 0.009 & 0.004 & 0.387 & 0.068 & 18.19 & 10 & 18.16 & 6.79 \\
050915B & 0.021 & -0.006 & 0.085 & 0.13 & 0.017 & 0.005 & 0.068 & 0.085 & -19.3 & -12.5 & -19.55 & -34.08 \\
051016A & 0.089 & -0.025 & 0.327 & 0.048 & 0.055 & 0.016 & 0.123 & 0.029 & -38.03 & -36.85 & -62.33 & -38.73 \\
051109A & 0.011 & -0.005 & 0.136 & 0.015 & 0.009 & 0.004 & 0.043 & 0.01 & -17.46 & -18.84 & -67.92 & -33.14 \\
051221A & 0.021 & -0.007 & 0.082 & 0.058 & 0.015 & 0.005 & 0.062 & 0.038 & -29.47 & -27.45 & -24.21 & -34.37 \\
060105 & 0.007 & -0.002 & 0.011 & 0.011 & 0.004 & 0.002 & 0.012 & 0.004 & -47.21 & -29.55 & 16.34 & -66.73 \\
060108 & 0.026 & -0.008 & 0.147 & 0.075 & 0.02 & 0.006 & 0.13 & 0.05 & -21.43 & -19.39 & -11.85 & -33.82 \\
060109 & 0.01 & -0.003 & 0.347 & 0.047 & 0.008 & 0.003 & 0.249 & 0.022 & -22.93 & -1 & -28.48 & -52.56 \\
060124 & 0.014 & -0.008 & 0.034 & 0.019 & 0.012 & 0.007 & 0.026 & 0.009 & -15.57 & -14.13 & -24.78 & -49.47 \\
060218 & 0.013 & -0.005 & 0.582 & 0.064 & 0.011 & 0.004 & 0.551 & 0.045 & -16.69 & -15.08 & -5.28 & -29.79 \\
060306 & 0.075 & -0.026 & 0.395 & 0.024 & 0.017 & 0.005 & 0.092 & 0.019 & -77.62 & -79.69 & -76.75 & -20.34 \\
060418 & 0.066 & -0.034 & 0.11 & 0.029 & 0.031 & 0.015 & 0.068 & 0.01 & -52.96 & -55.91 & -38.76 & -64.8 \\
060421 & 0.092 & -0.038 & 0.557 & 0.082 & 0.053 & 0.02 & 0.177 & 0.051 & -43.01 & -46.22 & -68.23 & -37.96 \\
\enddata


\tablecomments{The BPL error fractions of $\log_{10}(T_{a})$, $\log_{10}(F_{a})$, $\alpha_{1}$ and $\alpha_{2}$ before and after Gaussian Process (GP) reconstruction (with relative percentage decrease in error for all three parameters). First four columns give error fraction for original BPL fit, while second four columns give error fraction for new BPL fit after GP reconstruction. Final four columns give percentage decrease in error fraction after reconstruction. Full table is available online.}


\end{deluxetable*}
\end{longrotatetable}

\newpage
\section{Acknowledgments}
This work was made possible in part by the United States Department of Energy, Office of Science, Office of Workforce Development for Teachers and Scientists (WDTS) under the Science Undergraduate Laboratory Internships (SULI) program who funded the initial help in creating the code for the reconstructed LCs of Luke Nearhood and Sarvesh Gharat.
We thank Malgorzata Bogdan for helpful comments and discussion of our analysis. We also thank Susrestha Paul, Anish Kalsi, Shubham Bhardwaj, and Sachin Venkatesh for their help in the GRB classification. 
We are particularly grateful to Artem Poliszczuk for the initial discussion on the manuscript. 
We would also like to thank Dr. J. Xavier Prochaska for his helpful discussion of the Gaussian Processes.
This research was also support by the Visibility and Mobility module of the Jagiellonian University (Grant number: WSPR.WSDNSP.2.5.2022.5) and the NAWA STER Mobility Grant (Number: PPI/STE/2020/1/00029/U/00001). 
A. N. and M.G.D are grateful to the Exploratory Research Grant (ERF) of the Division of Science at NAOJ for the financial support for the visit of A. Narendra.
A. N. is also extremely grateful to NAOJ for providing all the essential amenities during the visit to the NAOJ Mitaka Campus.
D. L. is also grateful for the support of the ERF grant.
This work was also supported by the Polish National Science Centre grant UMO-2018/30/M/ST9/00757 and by Polish Ministry of Science and Higher Education grant DIR/WK/2018/12.

\bibliography{LC_reconstruction}
\bibliographystyle{aasjournal}



\end{document}